%% file: main.tex
\documentclass[
]{article}

\usepackage{graphicx} 
\usepackage{adjustbox}
\usepackage{wrapfig}
\usepackage{amsfonts}
\usepackage{amsmath}
\usepackage{amssymb}
\usepackage{amsthm}
\usepackage{bm}
\usepackage[round]{natbib}
\usepackage[left=2cm, right=2cm, top=2cm, bottom=2.5cm]{geometry}
\usepackage{booktabs}
\usepackage{hyperref}
\usepackage[table]{xcolor}
\usepackage{multirow}
\usepackage{multicol}
\usepackage{tikz}
\usepackage{wrapfig}
\usepackage[section]{placeins} 
\usepackage{orcidlink}
\usetikzlibrary{positioning}
\usetikzlibrary{calc} 
\usetikzlibrary{patterns, positioning, decorations.pathreplacing,}

\usepackage{pdfpages}

\usepackage{colortbl}
\newcolumntype{C}[1]{>{\centering\arraybackslash}m{#1}}
\newcolumntype{R}[1]{>{\raggedleft\arraybackslash}m{#1}}

\usepackage{enumitem}
\setlist[enumerate]{topsep=3pt,itemsep=-1ex,partopsep=1ex,parsep=1ex}
\setlist[itemize]{topsep=3pt,itemsep=-1ex,partopsep=1ex,parsep=1ex}

\usepackage[ruled,vlined, linesnumbered]{algorithm2e}
\SetKwRepeat{Do}{do}{while}
\SetKw{Return}{Return}
\SetKwComment{Comment}{\#{} }{}
\SetKwInput{KwStore}{Store}

\renewcommand{\vec}[1]{\ensuremath{\bm{\mathrm{#1}}}}
\newcommand{\mat}[1]{\ensuremath{\bm{\mathrm{#1}}}}
\newcommand{\transpose}[1]{\ensuremath{#1^\top}}
\newcommand{\hdmul}[0]{\ensuremath{\odot}}

\newtheorem{prop}{Proposition}

\newtheorem{definition}{Definition}

\newtheorem{remark}{Remark}
\newtheorem{example}{Example}

\title{Probabilistic Forecasting for Day-ahead Electricity Prices, Battery Trading Strategies and the Economic Evaluation of Predictive Accuracy}
\author{
    Simon Hirsch\textsuperscript{(\orcidlink{0009-0008-1409-9677},1,2)}  and
    Florian Ziel\textsuperscript{(\orcidlink{0000-0002-2974-2660},1)} \\ [1ex]
    \small (1) Data Science in Energy and Environment, University of Duisburg-Essen, Germany \\
    \small (2) Statkraft Trading GmbH, Germany
}
\date{April 21, 2026}

\begin{document}

\maketitle

\begin{abstract}
    Electricity price forecasting supports decision-making in energy markets and asset operation. Probabilistic forecasts are increasingly adopted to explicitly quantify uncertainty, typically issued as quantile predictions or ensembles of the full predictive distribution. However, how improvements in statistical forecast quality translate into economic value remains unclear. Battery storage arbitrage in day-ahead markets is a popular application-based benchmark for this purpose.
    We analyze quantile-based trading strategies (QBTS) and identify two critical flaws: they do not incentivize honest probabilistic forecasting and they ignore the intertemporal dependence structure of electricity prices. We therefore frame battery optimization as a stochastic program based on fully probabilistic forecasts and examine decision quality measurement for risk-neutral and risk-averse settings under different uncertainty models. Our discussion touches both sides of the coin: How reliable is the economic evaluation of forecasting models though (simplified) application studies -- and how do improvements in statistical forecast quality for stochastic programs relate to the decision-quality and economic performance?
    We provide theoretical justification and empirical evidence from a case study on the German electricity market. Our results highlight the pitfalls of ranking forecasting models through battery trading strategies. We conclude with implications for evaluation practice and directions for future research in application-based forecast assessment.
\end{abstract}


\section{Introduction}

Forecasting electricity prices is crucial for informed decision-making in the daily operations of energy companies. Over the last years, both industry and research have focused on probabilistic forecasting, which ``has a lot to offer, in particular, improved assessment of future uncertainty, ability to plan different strategies for the range of possible outcomes, increased effectiveness of submitted bids'' \cite[][p. 1549]{nowotarski2018recent}. A pressing question in practice is whether improved forecasting performance, in terms of scoring rules or loss functions, improves decision-making, measured in monetary terms \citep{granger2000economic,yardley2021beyond}. To address this need, various application studies have been proposed in the energy forecasting literature, \citep{nitka2023combining, vogler2021event}. A common showcase for the economic benefits of better forecasts is the simplified operation of grid-scale batteries energy storage systems (BESS), which can buy electricity (and charge) when prices are low and sell electricity (and discharge) when prices are high. Better anticipation of prices should translate into higher profits. This storage arbitrage example has been adopted for both point and probabilistic forecasting studies \citep[see][and the references in Table \ref{tab:literature_review}]{sang2022electricity, chȩc2025extrapolating, serafin2025loss, maciejowska2025statistical, uniejewski2025smoothing, o2025conformal, o2025optimising}.

It is widely agreed on that forecast evaluation and model selection should be conducted using proper scoring rules. Proper scoring rules incentivize honest forecasting, that is, the optimal score is attained for predicting the true distribution. Scoring rules are called strictly proper if the optimum is unique. Strictly proper scoring rules are thus the gold standard for forecast evaluation and model selection \citep{gneiting2007strictly}.
However, the strict propriety alone does not guarantee that scoring rules penalize forecast errors symmetrically \cite[see][]{buchweitz2025asymmetric}, provide meaningful discrimination \cite[see][]{alexander2024evaluating,marcotte2023regions} or that they are aligned with the decision-making process at hand. To bridge this gap, the properties of scoring rules and their modification through weighting, thresholding and censoring is an active area of research \citep{allen2023weighted, de2025localizing, shahroudi2025aligning}.

Consequently, some questions arise: (How) can we infer the quality of forecasts based on the quality of the decisions? How can we assess the quality of decisions under competing models of uncertainty? Can battery trading strategies provide meaningful discrimination between forecasts of different quality? Are battery trading strategies aligned with the concept of proper scoring rules? How are statistical scores related to the economic performance? These questions will guide us throughout this paper.

The majority of works concerning probabilistic forecasting of electricity prices focuses on the marginal distribution of hourly prices. Based on this, \emph{quantile-based trading strategies} (QBTS), originally introduced by \cite[][first published 2023 as working paper]{uniejewski2025smoothing}, employ quantile-based limit orders on the day-ahead market. Briefly, a trader places limit buy and limit sell orders based on $\alpha$ and ($1-\alpha$) -- quantiles of the predicted distribution, thereby controlling a trade-off between frequency of trading and anticipated profit of each trade. The approach has been used in various subsequent studies (see Table \ref{tab:literature_review} for an overview) and extended by \cite{o2025optimising,o2025conformal,o2024conformal}. While instinctively intuitive, we provide theoretical and empirical evidence that the approach can be gamed by providing systematically biased forecasts and is not aligned with the concept of proper scoring rules. Moreover, we show that the approach does not anticipate diversification benefits by neglecting the dependence structure of electricity prices.

Keeping the battery example, a natural further starting point is to ask whether battery trading strategies based on a fully multivariate probabilistic forecasts, optimized for common risk-measures such as expected profits or the conditional value-at-risk (CVAR) can be used to evaluate the quality of forecasts. We show that even in this case, the resulting profits are not necessarily a \emph{strictly} proper scoring rule and have potentially low discriminatory power. 
For point forecasts, \cite{maciejowska2025statistical} empirically evaluate the relationship between the statistical quality of point forecasts and the economic performance of battery trading strategies. However, formal analysis and the evaluation of the probabilistic and risk-averse case is notably absent in the literature. We contribute to this gap by providing a theoretical and empirical analysis of the relationship between forecast quality and a notion of decision quality, measured by the predictive accuracy of the objective value of our stochastic program under competing forecasts. We relate decision quality, statistical forecast and economic forecast evaluation.

Our contribution can be summarized in the following points: \begin{itemize}
    \item We analyze the theoretical properties of popular QBTS and show that they do not promote honest forecasting, as they can be gamed by providing overdispersed forecasts. We provide formal justification and a simulation study for this claim in Section \ref{sec:qbts}.
    \item We discuss more generally the evaluation of battery trading strategies framed as stochastic programs based on fully multivariate probabilistic forecasts (Section \ref{sec:ebts}). In particular: \begin{itemize}
        \item We show that forecast evaluation using battery trading strategies is no \emph{strictly} proper scoring rule and has potentially low discriminatory power. Low discriminatory power is an issue, as it reduces robustness of results across changing application settings (Section \ref{sec:ebts-theory}).
        \item We develop options for the evaluation of decision quality under different probabilistic forecasts in a risk-averse battery optimization setting (Section \ref{sec:ebts-decision-quality}). 
    \end{itemize}
    \item As a by-product, we extend the use of association measures for forecast evaluation introduced by \cite{maciejowska2025statistical} to the probabilistic case and present a proper scoring rule based on Kendall's $\tau$ in Section \ref{sec:forecast-evaluation}.
    \item  We provide an extensive empirical evaluation in Sections along a wide range of statistical and economic metrics, as well as different battery configurations in Section \ref{sec:results}. 
\end{itemize}

We believe that our findings are of interest to multiple research communities, including the forecasting community, the energy economics community and the stochastic optimization/programming community. \begin{itemize}
    \item The primary audience of this paper is the (electricity price) forecasting community. However, we believe that understanding the downstream impact of forecast on realistic and complex decision-making processes is also crucial for other forecasting applications such as supply chain and inventory management.
    \item Our results are also of interest to the stochastic optimization/programming community. Much research has been done on scenario reduction methods to reduce computational complexity, but less work has been done on the influence of the underlying scenarios on the solution quality \citep{dupavcova2003scenario, lohndorf2016empirical, ziel2021energy}, especially with respect to different underlying models of the uncertainty.
    \item A better understanding of the relationship between scoring rules and economic performance can also inform the development of new loss functions for the integration of forecasting and optimization, as strong theory exist for learning using scoring rules \citep[see][]{dawid2016minimum,gneiting2007strictly}, while custom loss functions are often developed ad-hoc \citep{sang2022electricity, sbaraglia2024optimizing,serafin2025loss}. 
\end{itemize}
On the other hand, we acknowledge some limitations a priori: this work aims at deepening the understanding of the relationship between forecast quality and decision quality in the context of battery trading strategies. We do not focus on the absolute profitability of BESS investments in energy markets \citep{spodniak2021profitability, staffell2016maximising} or on sophisticated, multi-market optimization \citep{lohndorf2023value, kraft2023stochastic, ghadimi2024stochastic}. 

The remainder of this paper is structured as follows: Section \ref{sec:qbts-theory} formalizes our claim and analyzes the objective value of quantile-based optimization from a theoretical vantage point. Section \ref{sec:ebts-dp} presents and analyzes the proposed multivariate dynamic programming approach. Section \ref{sec:case_study} gives a case study. The last section concludes and identifies avenues for further research. 

\section{Quantile-based Battery Trading Strategy}\label{sec:qbts}

Quantile-based trading strategies (QBTS) are a popular approach to evaluate the economic value of probabilistic forecasts in the context of battery trading. The main idea is to place limit orders based on quantiles of the predicted distribution of electricity prices. By adjusting the quantile level $\alpha$, the trader can control the trade-off between the frequency of trading and the anticipated profit of each trade. Two main variants exist in the literature, the quantile-based trading strategies based on the works of \cite{uniejewski2025smoothing} and the TS-1/2/3 strategies introduced by \cite{o2025optimising,o2024conformal,o2025conformal}.

\begin{table}[htb]
    \adjustbox{width=\linewidth}{%
        \begin{tabular}{lllll}
        \toprule
            Reference & Forecasting & Style & Comments \\
        \midrule
            \cite{sang2022electricity} & Point & BESS Expected Profits & Focus on integration of forecasting and optimization \\
            \cite{sbaraglia2024optimizing} & Point & BESS Expected Profits & Focus on custom loss functions \\
            \cite{serafin2025loss} & Point & BESS Expected Profits & Focus on custom loss functions \\
            \cite{serafin2025data} & Point & BESS Expected Profits & Focus on choosing the calibration window though battery optimization \\
            \cite{maciejowska2025statistical} & Point & BESS Expected Profits & Introduction of rank-based scoring rules \\ 
        \midrule
            \cite{uniejewski2021regularized} & Probabilistic &  QBTS &   Non-accepted into balancing market \\
            \cite{maciejowska2024probabilistic} & Probabilistic & QBTS & Balance non-accepted by unlimited bids on $d+1$\\
            \cite{uniejewski2025smoothing} & Probabilistic & QBTS &  Balance non-accepted by unlimited bids on $d+1$ \\
            \cite{lebedev2025analyzing} & Probabilistic & QBTS &  Balance non-accepted by unlimited bids on $d+1$ \\
            \cite{serafin2024ranking} & Probabilistic & QBTS &  Balance non-accepted by unlimited bids on $d+1$ \\
            \cite{nitka2023combining} & Probabilistic & QBTS & Balance non-accepted by unlimited bids on $d+1$ \\
            \cite{marcjasz2023distributional} & Probabilistic & QBTS & Balance non-accepted by unlimited bids on $d+1$, only bid if $Q_s > Q_b$. \\
            \cite{o2025optimising} & Probabilistic & TS-1/2/3  & Multi-Market (Day-ahead and Balancing) and ramp rate considerations \\
            \cite{o2024conformal} & Probabilistic & TS-1/2/3  & Multi-Market (Day-ahead and Balancing) and ramp rate considerations \\
            \cite{o2025conformal} & Probabilistic & TS-1/2/3  & Multi-Market (Day-ahead and Balancing) and ramp rate considerations \\
        \midrule
            \cite{katholnigg2025impact} & Probabilistic & Industrial Energy System Bidding  & Focus on load forecasting and grid constraints. \\
            \cite{beykirch2022bidding,beykirch2024value} & Probabilistic & Smart Energy System Bidding  & Focus on analytical results and single hour bids. \\
        \bottomrule
        \end{tabular}
    }%
    \caption{Literature overview on papers employing battery optimization case studies based on \cite{uniejewski2025smoothing} and related approaches for day-ahead price forecasting. The first two blocks focus directly on price forecasting for BESS operations, while the latter two works consider adjacent settings and markets, but have a clear connection to our analysis.}
    \label{tab:literature_review}
\end{table}

\subsection{Optimization Algorithm}\label{qbts-algorithm}

\begin{figure}[htb]
    \centering
    \includegraphics[width=0.5\textwidth]{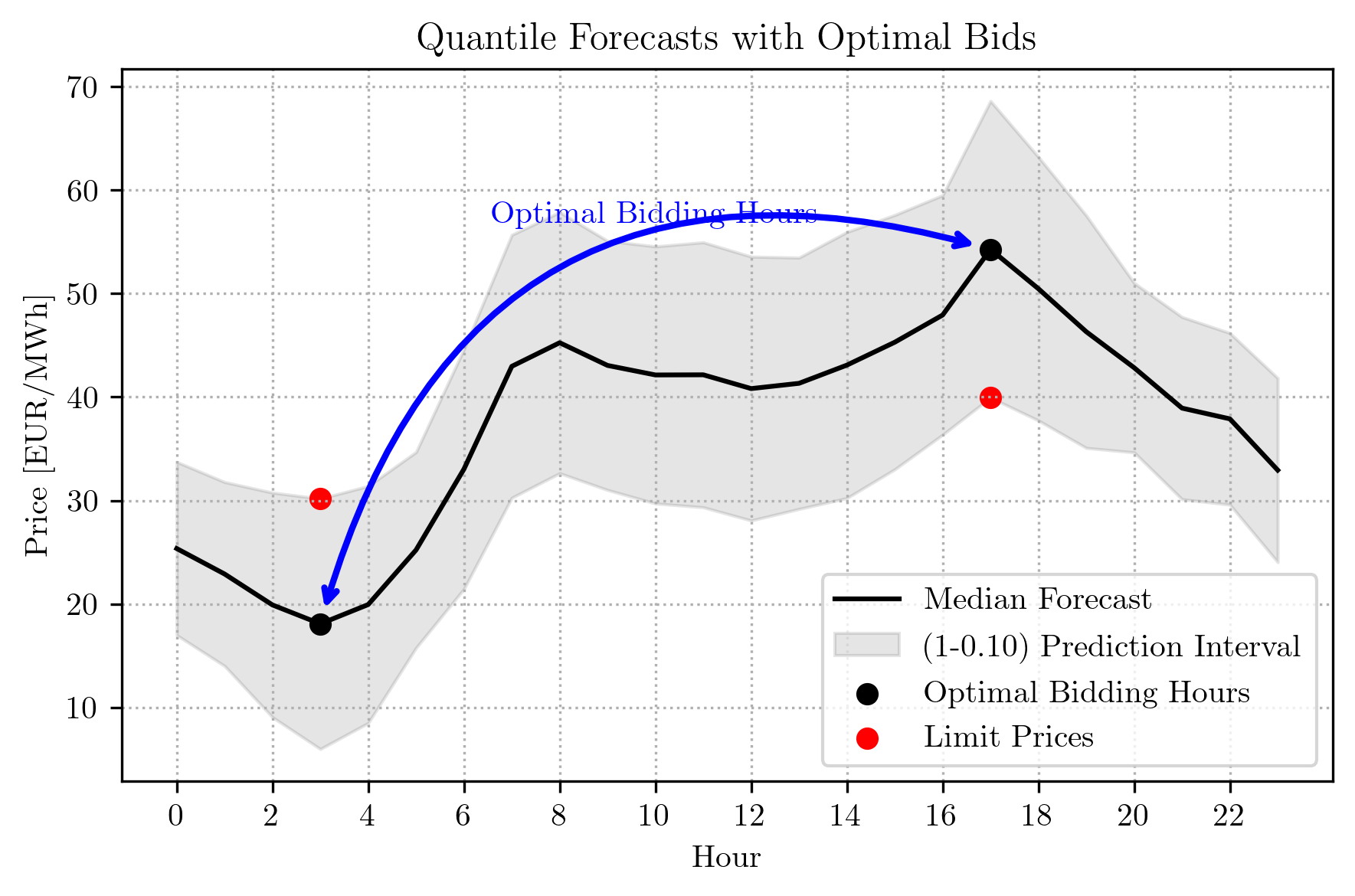}%
    \includegraphics[width=0.5\textwidth]{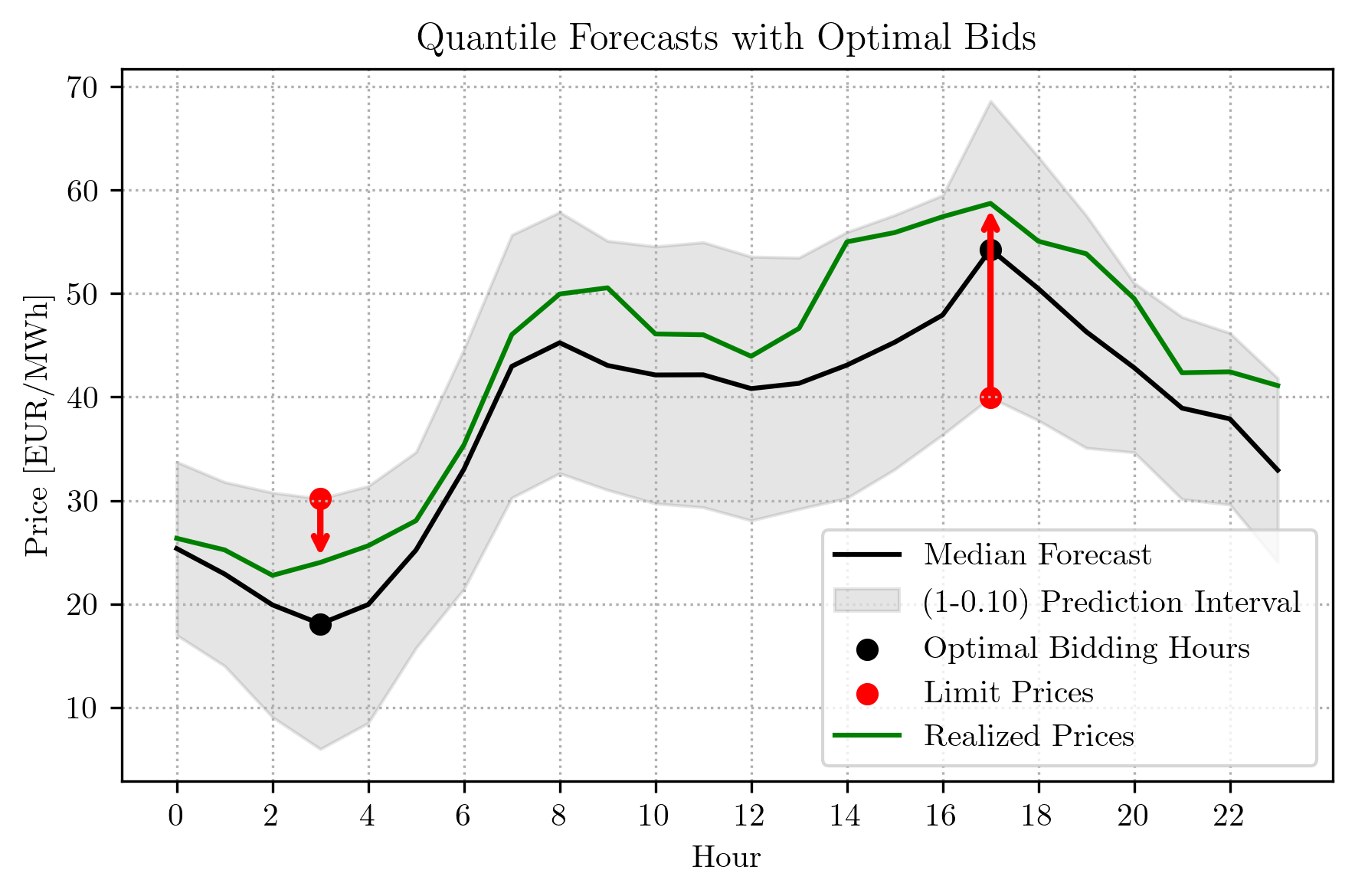}
    \caption{Example for quantile-based trading strategies. The left panel shows the selection of the optimal hours based on the median forecast and the placement of the limit orders based on the quantiles. The right panel shows the acceptance of the bids given the realized prices.}
    \label{fig:algorith-qtbs}
\end{figure}

We briefly describe QBTS and their variations in the literature. Assume the trader has access to a probabilistic forecast of the day-ahead electricity prices, which is issued as quantile predictions for each delivery hour. The trader aims to maximize the expected returns from trading a battery storage asset on the day-ahead market. The battery has a certain capacity $\kappa$ and efficiency $\eta$. The strategy consists of two steps: \begin{enumerate}
    \item Based on the median forecast, the trader identifies the hours with the lowest and highest predicted prices, denoted as $b$ and $s$. We ensure that $b \neq s$ to avoid buying and selling in the same hour.
    \item For a given level $\alpha$, the trader places a limit buy order at hour $b$ with a price based on the $(1-\alpha)$-quantile of the predicted distribution for hour $b$, and a limit sell order at hour $s$ with a price based on the $\alpha$-quantile of the predicted distribution for hour $s$.
\end{enumerate} 
In this case, we assume that the execution of the buy and sell bid is coupled, i.e. either both are executed or neither is executed. Thereby, the battery is always in a valid state and no additional orders need to be placed in other markets to balance the battery.\footnote{This assumption is closely related to the so-called loop bid, a product available at the EPEX power exchange that guarantees joint or no execution of coupled buy and sell bids. Contrary to individual directional limit prices however, loop bids are conditional on the total cashflow of the linked bids \citep{epexspot2025trading}. Framing as a loop bid would further move away from the original QBTS.} Alternatively, the trader might be faced with partial execution, that is, only the buy or only the sell trade is executed. There are multiple ways to handle this, the most prominent ones are described in the following and an overview is given in Table \ref{tab:literature_review}. \begin{itemize}
    \item The trader needs to reserve capacity in the battery to always remain in physically valid states or place additional orders in the balancing market to ensure that the battery is on half-full at the end of the day. This approach has been used by \cite{uniejewski2025smoothing}.
    \item The trader can place additional orders in the balancing market to balance the battery at the end of the day. This approach has been used by \cite{maciejowska2024probabilistic}.
\end{itemize}
Both options introduce additional risk and make the (predicted) distribution of battery returns intractable. Therefore, we focus on the simplified approach in the following theoretical analysis and empirical evaluation. Figure \ref{fig:algorith-qtbs} summarizes the QBTS approach. Algorithmic descriptions of all battery trading strategies are given in the supplementary material.

The TS-1 strategy as proposed by \cite{o2025optimising} is based on a similar algorithm. However, instead of placing limit orders based on quantiles, quantiles are used to choose the buying and selling hours and unlimited bids are placed. The objective function of the TS-1 strategy is given by \begin{equation}
    \max \sum_{h=1}^{H} \left( -\frac{1}{\eta}\widehat{Q}_h^{1-\alpha} + \eta \widehat{Q}_h^{\alpha}\right).
\end{equation}
The main difference to QBTS is that the TS-1 strategy does not place limit orders but unlimited orders at the selected hours and hence bids are always accepted. The TS-2 strategy extends the TS-1 strategy by placing additional orders in the balancing market to balance the battery at the end of the day, while the TS-3 strategy extends the TS-1 strategy by considering ramp rate constraints. We focus on the TS-1 strategy in the following, as it is the most similar to QBTS and allows for a more direct comparison.

\subsection{Theoretical Analysis}\label{sec:qbts-theory}

We begin the theoretical discussion by recapitulating the definition of a strictly proper scoring. 

\begin{definition}[Strictly Proper Scoring Rule, \citep{gneiting2007strictly}]
    For a predicted distribution $\mathcal{F}$ and the realized value $x$ taken from the true distribution $\mathcal{D}$, the scoring rule $S(\mathcal{F}, x) \rightarrow \mathbb{R}^+$ denotes the forecaster's reward. As a convention, we have lower scores corresponding to better forecasts. For a proper scoring rule we require: \begin{equation}
        \mathbb{E}[S(\mathcal{D}, x)] \leq  \mathbb{E}[S(\mathcal{F}, x)]
    \end{equation} 
    that is, the score is only minimized when forecasting the true distribution. A scoring rule is called strictly proper if the minimum is unique for the true distribution, i.e. 
    $\mathbb{E}[S(\mathcal{D}, x)] =  \mathbb{E}[S(\mathcal{F}, x)]$ iff $\mathcal{F} = \mathcal{D}$.
\end{definition}

We denote the unknown, true distribution of day-ahead electricity prices~$\boldsymbol{P}_d = (P_{d,0}, ..., P_{d, 23}) \sim \mathcal{D}_d$ on day $d$ as $\mathcal{D}_d$. Denote the realized prices as $\vec{p}_d = (p_{d,0}, ..., p_{d, 23})$. We describe the 24-dimensional vector of decisions as $\vec{a}(b, s) = (a_0, ..., a_h, ..., a_H)$, where $b$ and $s$ denote the hour(s) in which electricity is bought and sold. We will omit the $d$ in the following as the optimization is repeated every day. Denote the efficiency as $\eta$ and the storage capacity as $\kappa$ and hence we have 
\begin{equation}
    a_i = \begin{cases}
        (1/\eta)\kappa & \text{if } i = b, \\
        - \eta \kappa & \text{if } i = s, \\
        0 & \text{else.}
    \end{cases}
\end{equation}
For the first part we constrain ourselves to only one buy and one selling bid. We implicitly assume that the charge/discharge capacity $\xi = \kappa$. Without loss of generality we assume that $b < s$, hence we buy (and charge) before we sell (and discharge).
The revenues $R_d$ for an unconstrained bid $\vec{a}(b, s)$ are given by
\begin{equation}
    R (\vec{a}(b, s)) 
    = - \transpose{\vec{a}(b ,s)}\boldsymbol{P}
    = \sum_{h=0}^H -a_h P_{h}
\end{equation}  
and for a bid with limit prices $Q_b^{1-\alpha}$ and $Q_s^{\alpha}$, we have:
\begin{equation}
    R (b, s, \alpha) 
    = \begin{cases}
        - (1/\eta) \kappa P_{b} + \eta \kappa P_{s} & \text{if } P_{b} \leq Q_b^{1-\alpha} \text{ and } P_{s} \geq Q_s^{\alpha}, \\
        0 & \text{else.}
    \end{cases}
\end{equation}
This allows us to state our main result regarding quantile-based trading strategies.

\begin{prop}[QBTS profits as scoring rule]\label{res:qtbs-scoring}
For given risk-control parameter $\alpha$, the expected profits of the QBTS are given by 
    \begin{small}\begin{align}\label{eq:qbts-expected-profits}
    \mathbb{E}[R(b, s, \alpha)] 
        &= \underbrace{\mathbb{P}(P_{b} \leq {Q}_b^{1-\alpha}; \; P_{s} \geq Q_s^{\alpha})}_\text{Acceptance probability.} 
            \underbrace{\left(-\frac{1}{\eta} \kappa \mathbb{E}[P_b | P_{b} \leq {Q}_b^{1-\alpha}; P_s > Q_s^{\alpha}] + \eta\kappa\mathbb{E}[P_s |  P_{b} < Q_b^{1-\alpha}; P_{s} \geq Q_s^{\alpha}] \right)}_\text{Expected profits if accepted (EP).} \\ &+ \left(1 - \mathbb{P}(P_{b} \leq {Q}_b^{1-\alpha}; \; P_{s} \geq Q_s^{\alpha})\right) 0 \nonumber
    \end{align}\end{small}%
This expression is not uniquely maximized for the true forecast, but is potentially maximized by an overdispersed respectively underdispersed forecast. Providing an overdispersed forecast increases the acceptance probability (AP), but decreases the expected profits (EP) if accepted. Vice versa, providing an underdispersed forecast decreases the AP, but increases the expected profits if accepted. The relative strength of the effects depends on the underlying price distribution and the risk-control parameter $\alpha$.
\end{prop}

We show that an overdispersed forecast can lead to higher expected profits than the perfect forecast. 
Assume the price distribution is symmetric and we have an over-dispersed forecasts $\widetilde{\boldsymbol{P}} = 
b \cdot \boldsymbol{P}$, where $\boldsymbol{P} \sim \mathcal{D}$ is the true price distribution and $b > 1$ is a dispersion parameter. We have $\widetilde{Q}_h^q = b \cdot Q_h^q$, where $Q_h^q$ is the quantile function of the true distribution. For simplificity, we assume that delivery periods are independent. The mean and median forecast are unbiased, $\widetilde{Q}_h^{0.5} = Q_h^{0.5}$.
We therefore have: \begin{align}
    \mathbb{E}[\widetilde{R}(b, s, \alpha)] 
    &= \mathbb{P}(P_{b} \leq \widetilde{Q}_b^{1-\alpha}; \; P_{s} \geq \widetilde{Q}_s^{\alpha})
        \left(-\frac{1}{\eta} \kappa \mathbb{E}[P_b | P_{b} \leq \widetilde{Q}_b^{1-\alpha}; P_{s} > \widetilde{Q}_s^{\alpha}] + 
        \eta\kappa\mathbb{E}[P_s | P_{b} < \widetilde{Q}_b^{1-\alpha}; P_{s} \geq \widetilde{Q}_s^{\alpha}] \right) \\
    &+ \left(1 - \mathbb{P}(P_{b} \leq \widetilde{Q}_b^{1-\alpha}; \; P_{s} \geq \widetilde{Q}_s^{\alpha})\right) 0. \nonumber
\end{align}
We are interested in the difference between the expected profits of the over-dispersed forecast and the perfect forecast, that is, $\mathbb{E}[\widetilde{R}(b, s, \alpha)] - \mathbb{E}[R(b, s, \alpha)]$. We look at the probability of bid acceptance and the expected value of the returns separately. We assume independence of the delivery periods, hence we have $\text{AP} = \mathbb{P}(P_{b} \leq Q_b^{1-\alpha}; \; P_{s} \geq Q_s^{\alpha}) = (1 - \alpha)^2$ for the perfect forecast and 
$\mathbb{P}(P_{b} \leq \widetilde{Q}_b^{1-\alpha}) = 1 - \widetilde{\alpha} \geq 1 - \alpha$ and hence:
\begin{align}
    \Delta \text{AP} &= \mathbb{P}(P_{b} \leq \widetilde{Q}_b^{1-\alpha}; \; P_{s} \geq \widetilde{Q}_s^{\alpha}) - \mathbb{P}(P_{b} \leq Q_b^{1-\alpha}; \; P_{s} \geq Q_s^{\alpha})  
    =  (1 - \widetilde{\alpha})^2 - (1 - \alpha)^2  \geq 0
\end{align}
The intuition can be seen in the following Figure \ref{fig:no_scoring_rule}, which shows the true acceptance region, which corresponds to the acceptance region of a perfect forecast, and the acceptance region of an overdispersed forecast. We can also see that it is sufficient if one side (buy or sell) is provided with an overdispersed forecast.

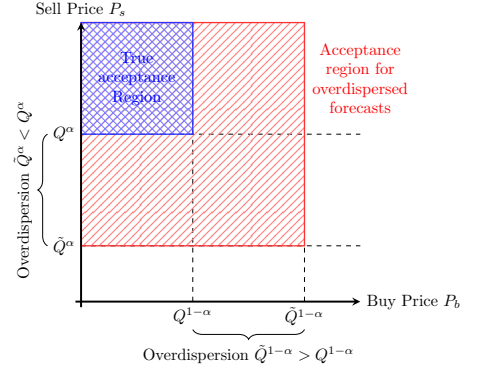
\begin{wrapfigure}[17]{r}{0.35\textwidth}
    \centering
    \resizebox{\linewidth}{!}{\input{tikz/no_proper_scoring_rule}}
    \caption{Visualization of the acceptance region of a perfect forecast (red) and an overdispersed forecast (blue). The overdispersed forecast has a larger acceptance region and hence, for the same parameter $\alpha$, a higher AP. }
    \label{fig:no_scoring_rule}
\end{wrapfigure}
\begin{figure}[htb]
    \centering
    \resizebox{0.8\linewidth}{!}{\includegraphics{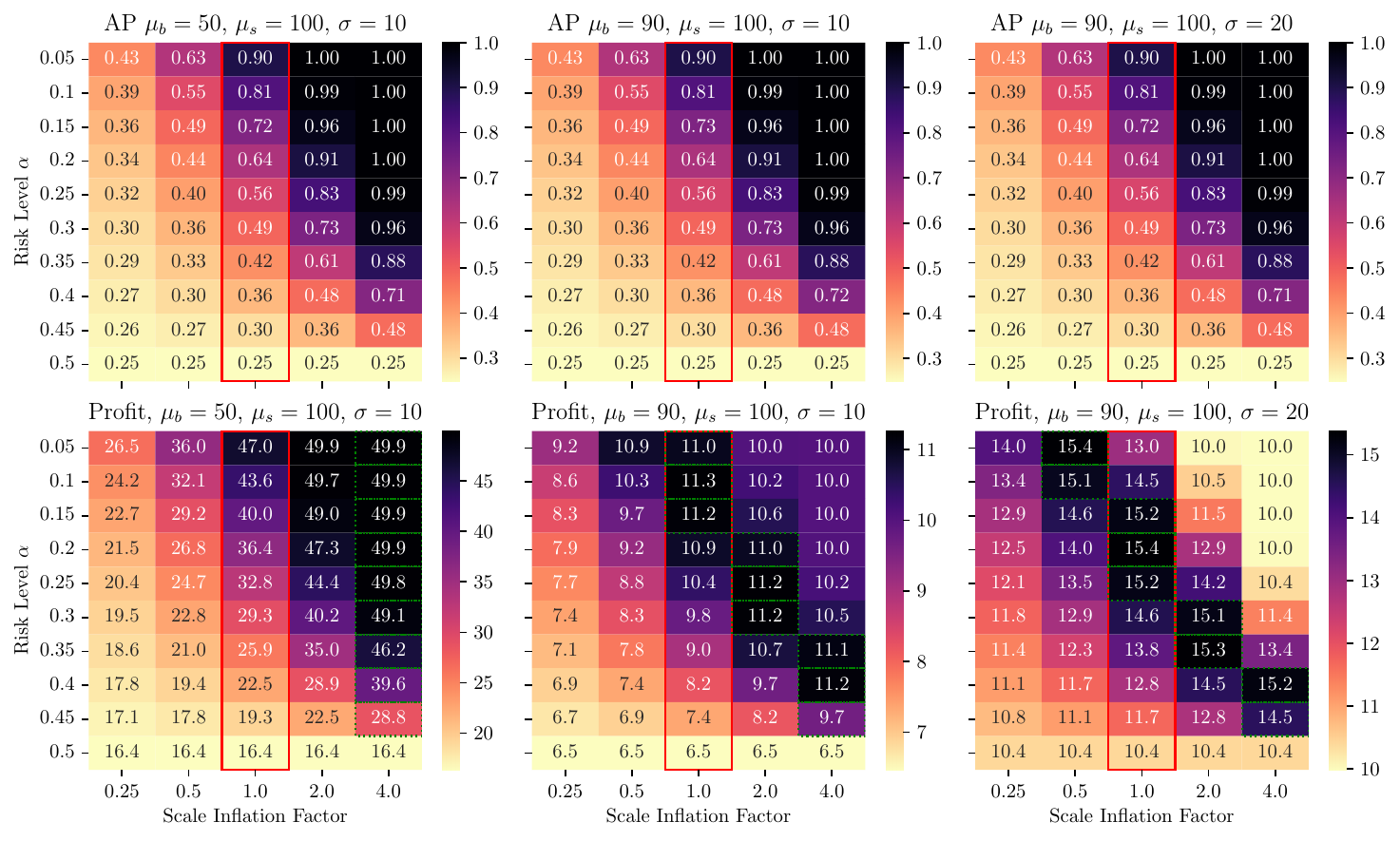}}
    \caption{Simulation study showing expected profits for QTBS for over-and underdispersed forecasts. True prices are simulated as Gaussian random variables with means $\mu_b$ and $\mu_s$. We over- and underdisperse the forecasts by multiplying the true prices with a dispersion parameter $b$. We then compute the expected profits for different values of $b$, $\alpha$ and vary the expected price $\mu_s$ and the standard deviation $\sigma$. The red box is the perfect forecast. The green boxes denote the fields with the highest (expected) profits.}
    \label{fig:qbts_simulation_1}
\end{figure}

The difference in the expected profits (EP) of the returns is given by:
\begin{small}\begin{align*}
    \Delta \text{EP} &=   \left(
            -\frac{1}{\eta} \mathbb{E}[P_b | P_{b} \leq \widetilde{Q}_b^{1-\alpha}; P_{s} \geq \widetilde{Q}_s^{\alpha}] + \eta\mathbb{E}[P_s | P_b \leq \widetilde{Q}_b^{1-\alpha}; P_{s} \geq \widetilde{Q}_s^{\alpha}] 
        \right) \\ &- 
        \left(
            -\frac{1}{\eta}  \mathbb{E}[P_b | P_{b} \leq {Q}_b^{1-\alpha}; P_{s} \geq Q_s^{\alpha}] + \eta\mathbb{E}[P_s | P_b \leq {Q}_b^{1-\alpha}; P_{s} \geq Q_s^{\alpha}] 
        \right) \\
    &= -\frac{1}{\eta} \left(
        \mathbb{E}[P_b | P_{b} \leq \widetilde{Q}_b^{1-\alpha}; P_{s} \geq \widetilde{Q}_s^{\alpha}] - \mathbb{E}[P_b | P_{b} \leq Q_b^{1-\alpha}; P_{s} \geq Q_s^{\alpha}]
        \right) \\ &+ 
        \eta\left(
            \mathbb{E}[P_s |  P_{b} \leq \widetilde{Q}_b^{1-\alpha}; P_{s} \geq \widetilde{Q}_s^{\alpha}] - \mathbb{E}[P_s | P_{b} \leq Q_b^{1-\alpha}; P_{s} \geq Q_s^{\alpha}]
        \right) \leq 0
\end{align*}\end{small}%
since we have \begin{small}\begin{align*}
    &\mathbb{E}[P_b | P_{b} \leq \widetilde{Q}_b^{1-\alpha}; P_{s} \geq \widetilde{Q}_s^{\alpha}] > \mathbb{E}[P_b | P_{b} \leq Q_b^{1-\alpha}; P_{s} \geq Q_s^{\alpha}] 
    \quad \text{and} \\
    &\mathbb{E}[P_s |  P_{s} \geq \widetilde{Q}_s^{\alpha}; P_{b} \leq \widetilde{Q}_b^{1-\alpha}] < \mathbb{E}[P_s |  P_{s} \geq Q_s^{\alpha}; P_{b} \leq Q_b^{1-\alpha}]
\end{align*}\end{small}%
as the conditioning sets of both expectations are larger for the over-dispersed forecast. Intuitively, we accept more trades and therefore more trades that are less profitable on average, i.e. higher buy prices and lower sell prices. Concluding, the difference in the expected value of the returns is negative. The overall effect therefore depends on the question whether the increase in the AP outweights the decrease in EP \begin{align}
    \mathbb{E}[\widetilde{R}(b, s, \alpha)] - \mathbb{E}[R(b, s, \alpha)] 
        = \underbrace{\text{AP} \cdot \Delta \text{EP}}_\text{\shortstack[c]{Negative effect \\ of lower EP.}}
        + \underbrace{\Delta \text{AP} \cdot \widetilde{\text{EP}}}_\text{\shortstack[c]{Positive effect \\ of higher AP.}}
\end{align}
For low values of $\alpha$, the AP is high and hence the first dominates. For high values of $\alpha$, the AP is low and hence the second dominates. Additionally, for larger spreads between the buy and sell price, the EP is higher and hence the first dominates. For intermediate values of $\alpha$, the effect depends on the underlying price distribution is hard to gauge. We show the implications of this result in a small simulation study: We simulate the true prices as Gaussian random variables with means $\mu_b$ and $\mu_s$. We over- and underdisperse the forecasts by multiplying the true $\sigma$ with a dispersion parameter $b$. We then compute the expected profits for different values of $b$ and $\alpha$. The results are shown in Figure \ref{fig:qbts_simulation_1}. First, we see that overdispersed forecasts lead mechanically to higher acceptance ratios (top row). However, the effect on the expected profits is not monotone. For high spreads ($\mu_b = 50, \mu_s = 100$) and low volatility ($\sigma = 10$), overdispersion leads to higher profits. For smaller spreads, and higher volatilty, profits decrease for overdispersion and low values of $\alpha$, since we start to accept more unprofitable trades. 

\begin{remark}[Bid acceptance of QBTS and the Dependence Structure of Electricity Prices] \label{result:dependence_structure}
Assume that the true and the predictive distribution of the day-ahead electricity prices are continuous and equal (perfect forecast). The probability of acceptance of the bid is crucially dependent on the dependence structure of the electricity prices. Denote $\mathcal{D}(p_0, ..., p_{23})$ as the joint CDF of the electricity prices and $\mathcal{D}_{h}(p_h)$ as the marginal distribution. We have:
\begin{align}
    \mathbb{P}(P_{b} \leq {Q}_b^{1-\alpha}; \; P_{s} \geq Q_s^{\alpha}) 
        & = \mathbb{P}\left(P_{b} \leq {Q}_b^{1-\alpha}\right) - \mathbb{P}\left(P_{b} \leq {Q}_b^{1-\alpha}; \; P_{s} \leq Q_s^{\alpha}\right) \\
        & = (1 - \alpha) - \mathcal{D} \left(\mathcal{D}_{b}^{-1}(1-\alpha), \mathcal{D}_{s}^{-1}(\alpha), ... \right) = (1 - \alpha) - \mathcal{C}_{b, s}(1 - \alpha, \alpha)
\end{align}
where $\mathcal{C}(\cdot, \cdot)$ is the Copula of the price distribution. Employing a copula-based argument here allows the argument to be margin-independent. It should be noted that $\mathcal{C}_{b, s}(1 - \alpha, \alpha)$ is not necessarily easy to compute for arbitrary dependence structures. For the Gaussian Copula $\mathcal{C}_{b, s}(1 - \alpha, \alpha)$ depends only on the pairwise Copula of $b$ and $s$. However, for other dependence structures, we can have an implicit dependence on all variables in between, since: $
    \mathcal{C}_{b, s}(1 - \alpha, \alpha) = \int_{[0, 1]^{H-2}} \frac{\partial \mathcal{C}(u)}{\partial u_0 \dots \partial u_H} \text{d}u_{h \neq b, h \neq s}.
$\end{remark}

\begin{figure}[htb]
    \centering
    \resizebox{0.8\linewidth}{!}{\includegraphics{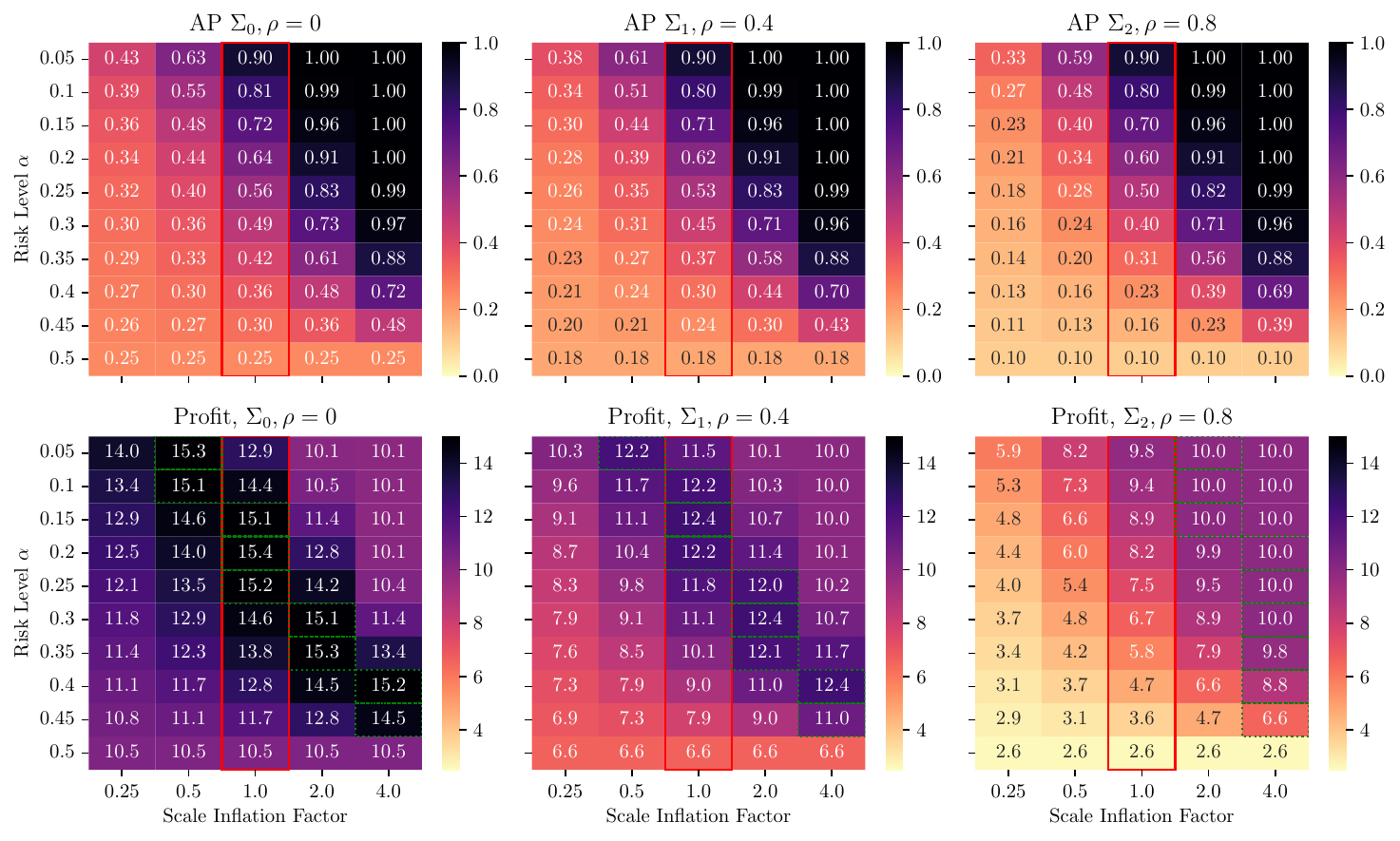}}
    \caption{Simulation study showing expected profits for QTBS for over-and underdispersed forecasts and varying levels of dependence between $P_b$ and $P_s$ with standard deviation $\sigma$.}
    \label{fig:qbts_simulation_2}
\end{figure}

In a second simulation study, we show the implications of this result. We simulate true prices as multivariate Gaussian random variables with means $\mu_b = 90$ and $\mu_s = 100$ and standard deviation $\sigma = 20$ (same as Panel 3 in Figure \ref{fig:qbts_simulation_1}). We vary the correlation between $b$ and $s$, taking $\rho = \{0, 0.4, 0.8\}$ in the according covariance matrices $(\Sigma_0, \Sigma_1, \mat{\Sigma}_2)$. Again, we show the AP and the expected profits for different values of $\alpha$ in Figure \ref{fig:qbts_simulation_2}. The AP decreases for higher correlation. Profits decrease for higher correlation and higher values of $\alpha$. For overdispersed forecasts, the effect is not as strong. This raises two important points: First, in practice, electricity prices are positively correlated across delivery hours, which further weakens the validity of QTBS for forecast evaluation. Second, the positive dependence should potentially decrease the risk of battery trading strategies, however, the QTBS approach does not capture this effect and hence can lead to suboptimal bids.

Although formally more involved to analyze due to the multi-market setting, this results highlights an important consideration for the TS-1/2/3 strategies \citep{o2024conformal,o2025conformal,o2025optimising}, which directly optimizes for the difference of the quantile forecasts: Treating the hourly quantile forecasts as independent neglects the dependence structure of electricity prices and hence can lead to suboptimal bids.

\section{Battery Trading using Multivariate Probabilistic Forecasts}\label{sec:ebts}

Let us formalize the notation. 
We denote with~$\vec{a} = (a_0, .., a_H)$ a (fixed) vector of bids.\footnote{
    Strictly speaking $\vec{a}$ is a random variable, since the trading decision depend on the forecasted distribution $\mathcal{F}$, which in turn depends on learned coefficients and parameters. Thus, $\vec{a}$ might be biased. We generally treat $\vec{a}$ as deterministic and return to the issue   briefly in the following Section~\ref{sec:ebts-dp}.
} 
We have $\transpose{\vec{a}}\boldsymbol{P} \sim \mathcal{R}$ as the distribution of the revenues for bid vector $\vec{a}$. 
Given the forecast $\boldsymbol{F} \sim \mathcal{F}$, usually issued as ensemble or scenario forecast $\widehat{\mat{F}}$, we have $\transpose{\vec{a}}\boldsymbol{F} \sim \mathcal{P}^{\vec{a}}$ as the forecasted distribution of the revenues
for bid vector $\vec{a}$. 
We can then write the battery optimization problem as $\max_{\vec{a}} \rho(\mathcal{P}^{\vec{a}})$, where $\rho(\cdot)$ is the risk measure we want to optimize. 
The realized revenues are given by $r_d = \transpose{\vec{a}}\vec{p}_d$.

\subsection{An Intuitive Optimization Algorithm}\label{sec:ebts-dp}

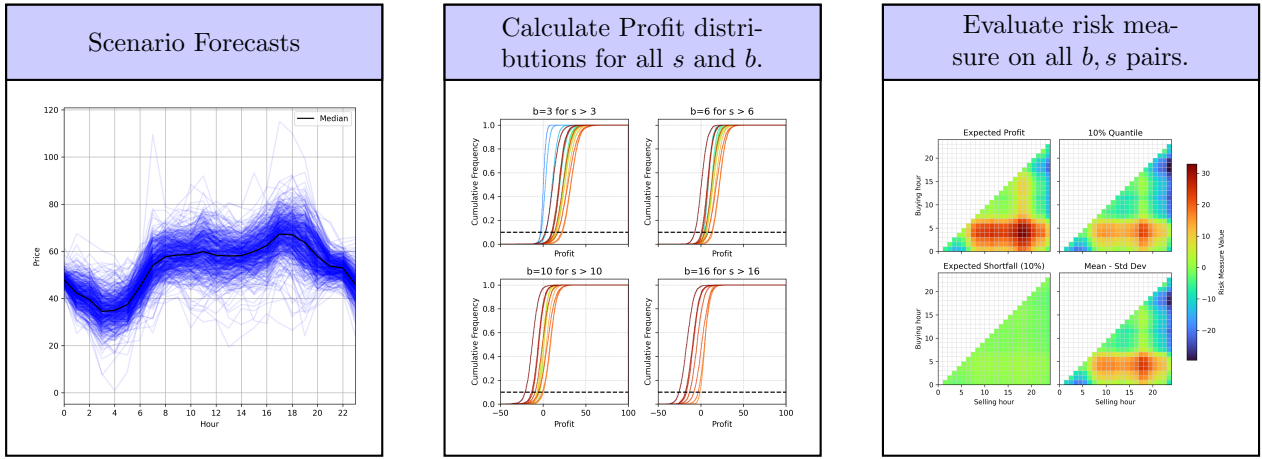
\begin{figure}[htb]
    \centering
    \input{tikz/ensemble_approach}
    \caption{Schematic overview of the dynamic programming approach using multivariate probabilistic forecasts for battery optimization.}
    \label{fig:schema_dp_approach}
\end{figure}

We approach the issue from the objective function. Commonly, we want to optimize some kind of risk measure~$\rho(\cdot)$ of the returns $$
    \max_{b,s} \quad \rho\left(R(\vec{a}(b, s), \mathcal{F})\right)
$$
this can be expected profits, or the (conditional) value-at-risk, a mixture of both or some specific functions. We assume that the forecaster equips the trader with a \emph{multivariate} predictive distribution $\mathcal{F}$, on which decisions are based. $\mathcal{F}$ is commonly represented through a (large) number of samples $\widehat{\mat{F}}$. Intuitively, we can use the following dynamic programming approach: \begin{enumerate}[noitemsep]
    \item For each scenario $m = 1, ..., M$ calculate returns for each buy/sell pair $\widehat{\mat{P}}_{bsm} = -\transpose{\vec{a}(b ,s)}\widehat{\mat{F}}_m$
    \item Calculate the discretized version of $\widehat{\mat{V}}_{b,s} = \rho(\widehat{\mat{P}}_{bsm})$ for all $b, s$ pairs, where $b < s$. Taking no action corresponds to $\widehat{\mat{V}}_{0,0} = \rho(0)$. 
    \item Choose $b, s = \arg \max (\widehat{\mat{V}}_{b,s})$, define the optimal decision $\vec{a}^* = \vec{a}(b^*, s^*)$ and place unrestricted bids.
\end{enumerate}
It is straightforward to see that, in this setting, we can directly optimize any risk measure. However, the approach is computationally demanding and does not scale well in the number of bids placed and the relative size of the bids. In the following Section \ref{sec:milp}, we discuss a more scalable approach based on mixed-integer linear programming. Nevertheless, this mental model provides a clear and intuitive understanding of how to employ multivariate probabilistic forecasts for battery optimization, can serve as a straightforward benchmark and highlights the importance of the dependence structure. Figure \ref{fig:schema_dp_approach} visualizes the approach.

\begin{remark}[Connection to Markowitz Portfolio Optimization]
    Note that the battery optimization can be seen as a long-short portfolio management problem with specific constraints on the weights, that is, the sum of the weights needs to be zero and the absolute weight of each asset is constrained by the battery capacity and depends on the order of assets (hours). In the classical Markowitz mean-variance optimization \citep{markowitz1952portfolio,markowitz1952utility}, the (random) returns are commonly denoted as $\boldsymbol{r} \sim \mathcal{N}(\vec{\mu}, \mat{\Sigma})$ and the portfolio weights as $\vec{w}$, which corresponds to the (random) prices $\boldsymbol{P}$ and the bid vector $\vec{a}$ in our setting \citep{kan2007optimal}.
\end{remark}
This connection to portfolio optimization allows us to discuss two interesting points. \begin{itemize}
    \item A somewhat counterintuitive property of QBTS is described in Result \ref{result:dependence_structure}: The expected return of QBTS decreases with increasing correlation between the day-ahead prices. On the contrary, for long-short portfolio optimization, positive correlation between the positions is desirable, since it reduces the variance of the portfolio returns and hence increases risk-adjusted returns. QBTS is not able to exploit this property.
    \item  It would be interesting to further analyze the distribution of the optimal bid vector $\vec{a}^*$, which itself is a random variable. This is analogous to the distribution of the optimal portfolio weights in the context of portfolio optimization, which has been analyzed in the context of mean-variance optimization \citep{kan2007optimal}. However, the distribution of the optimal bid vector $\vec{a}^*$ is more complex, since the optimization problem is non-convex and the solution space is not continuous. Hence, we leave this analysis for future research. 
\end{itemize}

\subsection{Formulation as Mixed-Integer Linear Problem}\label{sec:milp}

While the dynamic programming approach is intuitive, it does not scale well. Hence, we propose an alternative formulation as a mixed-integer linear programming problem. A drawback of this approach is that only the expected value and the conditional value at risk can be optimized directly. Mean-variance and value-at-risk objective functions cannot be reformulated as MILP problems \citep{rockafellar2000optimization, uryasev2001conditional}. We will discuss this difference in the simulation study in Section \ref{sec:case_study}.

We again assume that we have $M$ scenarios $\widehat{\mat{F}}_{h,m}$ for each hour $h$ and scenario $m$. We further relax some assumptions from the section before and allow for multiple buy and sell bids. Denote the number of buy bids as $N_b$ and the number of sell bids as $N_s$. We treat the bid volume as decision variable $a_h^b > 0$ and $a_h^s >0$. We have 
\begin{align}
    & \max_{a^b, a^s} && \rho(R_m) && && &&  \\
    & \text{where} && R_m = \textstyle\sum_{h=0}^H \left(- a_h^b (1/\eta) \widehat{\mat{F}}_{h,m} + a_h^s \eta \widehat{\mat{F}}_{h,m}\right) && m = 1, ..., M && \text{Revenues for each scenario.} \\
    & \text{s.t.} && 0 \leq a_h^b \leq (1/\eta) \xi \quad \text{and} \quad 0 \leq a_h^s \leq \eta \xi, && h = 0, ..., H && \text{Bids smaller than capacity.}\\ 
    & && \mathbf{1}(a_h^b > 0) + \mathbf{1}(a_h^s > 0) \leq 1 && h = 0, ..., H && \text{No simultaneous buy and sell.} \\
    & && \textstyle\sum_{h=0}^{H} \mathbf{1}(a_h^b > 0) \leq N_b && && \text{Number of buy bids} \\
    & && \textstyle\sum_{h=0}^{H} \mathbf{1}(a_h^s > 0) \leq N_s && && \text{Number of sell bids} \\
    & && 0 \leq \textstyle\sum_{h=0}^k \eta a_h^b - (1/\eta) a_h^s \leq \kappa && k = 0, ..., H && \text{Charge limits} \\
    & && \textstyle\sum_{h=0}^H (\eta a_h^b - (1/\eta) a_h^s) = 0 && && \text{Storage is empty in $h = H$.} \\
    & && \textstyle\sum_{h=0}^k \eta a_h^b \leq \kappa &&  && \text{Cycle limit.}
\end{align}
where $\kappa$ is the storage capacity and $\xi$ is the charge/discharge capacity (also called energy or rated capacity). In light of the recent market change to quarter-hourly trading in the single day-ahead coupling (SDAC), the use of such dynamic programming algorithms becomes computationally demanding and as such, the use of mixed-integer linear programming (MILP) solvers is advisable. Our implementation uses \texttt{pyomo} and the \texttt{HIGHS} solver \citep{hart2011pyomo,huangfu2018parallelizing}.

\subsection{Battery Optimization as Scoring Rules}\label{sec:ebts-theory}

In the following, we take the above formulation and discuss whether the multivariate probabilistic forecast can be scored through the battery optimization.  Using a simple example, we show that it is possible to construct different forecast distributions that lead to the same objective value $\rho(\cdot)$ for the battery optimization. This raises the issue of discriminatory power of battery optimization for the evaluation of multivariate probabilistic forecasts. 

The reason for this issue is the number of degrees of freedom within the battery optimization problem. In the classic forecast evaluation setting, we are interested in ranking a number $\mathcal{M}$ different models $h_m(\mat{X}) \rightarrow \mathcal{F}_m$ based on their forecasted distribution $\mathcal{F}_m$ and the realized values $y$ using scores $s_m = S(\mathcal{F}_m, \vec{p})$. In the battery optimization case, we rank different models based on the outcome of an optimization problem that takes the forecasted distribution as input. Therefore, we have \begin{equation}
    \vec{a}_m^* = \arg \max_{\vec{a}} \rho(\mathcal{P}^{\vec{a}}_m) \quad \text{and} \quad r_m = R(\vec{a}_m^*, \vec{p})
\end{equation}
where $\transpose{\vec{a}} \boldsymbol{F}_m \sim \mathcal{P}^{\vec{a}}_{m}$ is the forecasted distribution of the revenues for model $m$ and $R(\vec{a}_m^*, \vec{p})$ is the realized revenues for the optimal bid $\vec{a}_m^*$ and the realized prices $\vec{p}$. We can then rank the models based on $s_m$ and $r_m$, respectively, which raises the question of whether the ranking of the models based on $s_m$ and $r_m$ is equivalent. \begin{itemize}
    \item From a decision-maker's perspective, it is desirable that both rankings coincide, as this ensures that the best model according to the forecast evaluation is also the best model for the decision-making problem at hand. However, this equivalence is not guaranteed, as we will show in the following.
    \item Often, the ultimate goal is the application, so the ranking based on $r_m$ seems more relevant. However, ranking based on $s_m$ is more general, is based on well-established theory and allows for a more comprehensive evaluation of the forecast quality across different applications. It is reasonable to assume that a forecast that scores well in terms of $s_m$ will also perform well in terms of $r_m$ across a range of decision-making problems. 
\end{itemize}
In the following, we will show through a simple example that different forecast distributions can lead to the same ranking of bids in the battery optimization problem, hence placing a fundamental limit on forecast evaluation through battery optimization.

\begin{prop}[Battery Trading Strategies are not strictly proper scoring rules]\label{res:scoringrule_limit}
    Battery trading strategies compress the distributional information. Therefore, different forecasts can indistinguishably lead to the same optimal bids, given the risk measure $\rho(\cdot)$ employed in the battery optimization. Assume $\mathcal{F}_1$ is a perfect forecast, then there can exist different forecast distributions $\mathcal{F}_2$, where $\mathcal{F}_1 \neq \mathcal{F}_2$, such that\begin{equation}
        \rho(\mathcal{P}^{\vec{a}}_{1}) = \rho(\mathcal{P}^{\vec{a}}_{2})
     \end{equation}
    and the ranking of all pairs $(b,s)$ does not change. Hence, the battery optimization using multivariate probabilistic forecasts cannot be a strictly proper scoring rule $\square$
\end{prop}

From a practical perspective, this implies that economic backtest performance can be decision-relevant, but it is generally unsuitable as a stand-alone strictly proper scoring rule for model selection over full predictive distributions. We illustrate the issue of information compression through the battery optimization problem in the following examples.

\begin{figure}
    \centering
    \includegraphics[width=\linewidth]{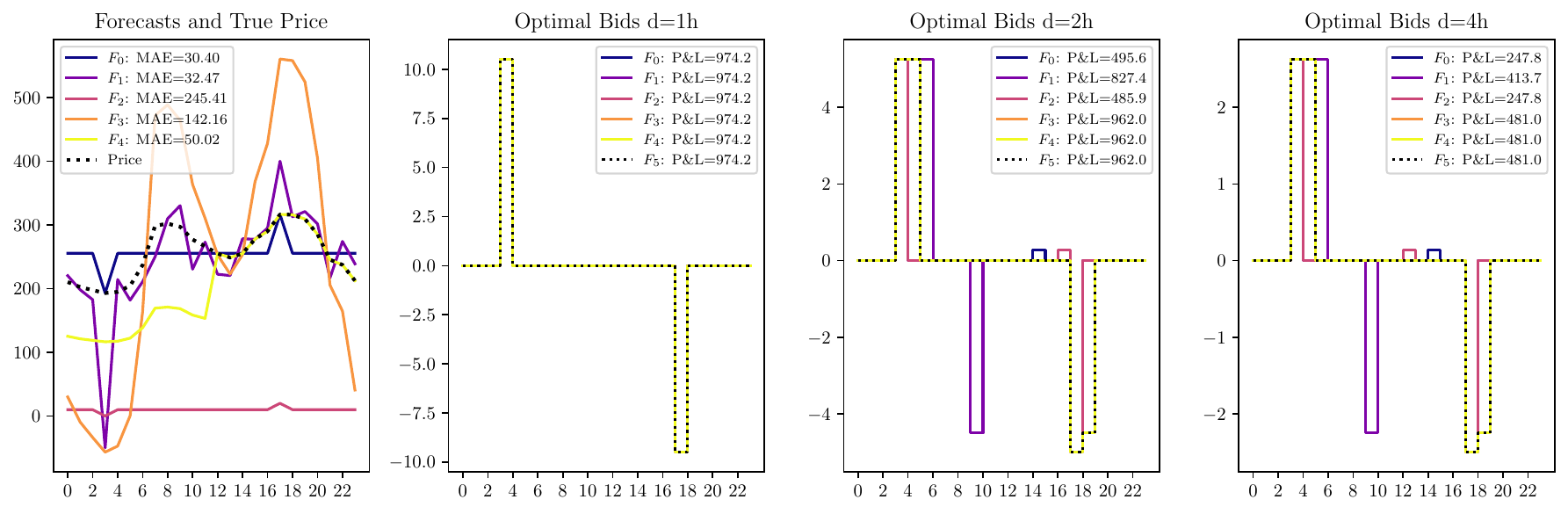}
    \caption{Exemplary forecasts that lead to the same expected revenues for a 1-hour battery despite being different forecasts. While some forecasts are close to the true prices, the forecast with the same expected revenues can be arbitrarily far away from the true distribution. The 3rd and 4th panel show the optimal bids for a 2-hour and 4-hour battery of the same capacity. These bids are far from optimal and emphasize the low discriminatory power of \emph{specific} battery trading strategies for forecast evaluation. A relatively small change in the asset portfolio can make a previously optimal forecast perform very poorly.}
    \label{fig:no_scoring_rule_mean}
\end{figure}

\begin{example}[Misspecified Mean Structure]\label{ex:mean_structure}
    A simple example can be constructed for optimizing the expected revenues $\rho(\cdot) = \mathbb{E}[\cdot]$ with relatively few assumptions. We have a perfect forecast $\mathcal{F}_1$ with $\vec{\mu}_1$ and a second forecast $\mathcal{F}_2$ with $\widehat{\vec{\mu}}$. For a 1-hour battery, we have the optimal bids $(b, s)$ where \begin{equation} 
        b = \arg \min_h \vec{\mu}_1 \quad \text{and} \quad s = \arg \max_h \vec{\mu}_1
    \end{equation}
    that is, we want to buy in the cheapest hour and sell in the most expensive hour. Conversely, we can have an arbitrary forecast $\mathcal{F}_2$ with \begin{equation}
        \vec{\mu}_{2,s} = \vec{\mu}_{1,s} \quad \text{and} \quad \vec{\mu}_{2,b} = \vec{\mu}_{1,b} 
    \end{equation}
    and arbitrary values $\vec{\mu}_{1,b} < \vec{\mu}_{2,h} < \vec{\mu}_{1,s} \; \forall h \in \{1, \ldots, 24\} \setminus \{b,s\}$. This forecast will lead to the same optimal bids and hence the same expected revenues, despite being a different forecast. The formulation can be extended to ranks, i.e. if the ranking of all hours is the same, the expected revenues will be the same. This allows us to construct forecasts that are arbitrarily far away from the true distribution, but still score perfectly and highlights the issue of information compression in the battery optimization example. This is further illustrated in Figure~\ref{fig:no_scoring_rule_mean}, which shows different forecasts that lead to the same expected revenues.
\end{example}

\begin{example}[Misspecified Covariance Structure]
    Assume the true and forecasted distribution of prices follow a multivariate normal distribution $\vec{p} \sim \mathcal{N}(\vec{\mu}, \vec{\Sigma})$ and $\mathcal{N}(\widehat{\vec{\mu}}, \widehat{\mat{\Sigma}})$ with potentially misspecified marginal variances and correlation structure. For a bid pair $(b, s)$, the profit distribution is given as \begin{eqnarray}
        \mathcal{P} = \mathcal{N}(\mu_R, \sigma^2_R) 
            & \mu_R = \eta^2 \kappa(\mu_s - \mu_b) 
            &  \sigma^2_R =                 
                (1/\eta)^2 \kappa^2\sigma^2_b + 
                \eta^2 \kappa^2 \sigma^2_s + 
                2 \kappa^2\rho_{b,s}\sigma_b \sigma_s\\
        \widehat{\mathcal{P}}= \mathcal{N}(\widehat{\mu}_R, \widehat{\sigma}^2_R)
            & \widehat{\mu}_R = \eta^2 \kappa(\widehat{\mu}_s - \widehat{\mu}_b) 
            & \widehat{\sigma}_R^2 = 
                (1/\eta)^2 \kappa^2\widehat{\sigma}^2_b + 
                \eta^2 \kappa^2 \widehat{\sigma}^2_s + 
                2 \kappa^2\widehat{\rho}_{b,s}\widehat{\sigma}_b \widehat{\sigma}_s
    \end{eqnarray}
    Trivially, for the perfect forecast, the forecasted and true revenue distribution are equal. For calibrated mean forecasts, we have $\widehat{\mu}_R = \mu_R$. Focusing on the variance, we have:
    \begin{align}\label{eq:difference_variance}
        \sigma_R^2 - \widehat{\sigma}_R^2 &= 
            (1/\eta)^2 \kappa^2\sigma^2_b +  \eta^2 \kappa^2 \sigma^2_s - 2 \kappa^2\rho_{b,s}\sigma_b \sigma_s 
            - \left((1/\eta)^2 \kappa^2\widehat{\sigma}^2_b - \eta^2 \kappa^2 \widehat{\sigma}^2_s - 2 \kappa^2\widehat{\rho}_{b,s}\widehat{\sigma}_b \widehat{\sigma}_s\right) \\ \nonumber
            &= (1/\eta)^2 \kappa^2(\sigma^2_b - \widehat{\sigma}^2_b) + \eta^2 \kappa^2 (\sigma^2_s - \widehat{\sigma}^2_s) - 2 \kappa^2 (\rho_{b,s}\sigma_b \sigma_s - \widehat{\rho}_{b,s}\widehat{\sigma}_b \widehat{\sigma}_s)
    \end{align}
    setting the right-hand side of Equation \ref{eq:difference_variance} to zero gives us a relationship between the forecasted variances and correlation such that the variance of the profit distribution is identical between true and forecasted distribution:
    \begin{equation}
            \widehat{\sigma}_b^2 = \frac{
                (1/\eta)^2\sigma^2_b 
                + 2 \sigma^2_s \widehat{\sigma}_s^2 \rho_{b,s}
                - \eta^2 (\sigma^2_s - \widehat{\sigma}^2_s) 
            }{
                (1/\eta)^2 - 2\rho_{b,s}\sigma^2_b
            }
    \end{equation}
    which allows us to construct different forecast distributions with identical risk measure outcomes for the battery optimization, assuming that the differences between the true and predicted variances are sufficiently small such that the ranking of all pairs $(b,s)$ does not change. 
\end{example}
These two examples already hint at potentially low discriminatory power of the battery optimization for forecast evaluation. It is important to be aware of this issue when evaluating forecasts through battery optimization. For optimizing the expected revenues, the issue has been discussed empirically by \cite{maciejowska2025statistical}. The main risk in the application is that, for slight changes in the application, a forecast that has previously performed well cannot be expected to perform well anymore. For example, a forecast that has performed well for a 1-hour battery might not perform well for a 2-hour or 4-hour battery, since the optimal bids can change and hence, the relevant part of the forecast distribution can change.

\begin{remark}[Relationship to Weighted Scoring Rules]
    The optimization can be seen as a censoring of the forecast distribution. \cite{allen2023weighted} and \cite{de2025localizing} discuss ways to create (strictly proper) scoring rules focused on specific parts of the distribution through weighted scoring rules. For example, \citet[][p. 4]{de2025localizing} define the region of interest as $A_w = \{y \in \mathcal{Y} : w(y) > 0\}$, where $w(\cdot)$ is a weighting function. However, the weighting function introduced by the battery optimization does not depend on the outcome variable alone, but also on the forecast distribution itself, which is not covered by the existing literature on weighted scoring rules.
\end{remark}

Based on these results, two extreme points of view can be taken:
\begin{itemize}
    \item One only cares about the decision-making problem at hand and hence, the ranking based on the application outcome is the only relevant one. In this case, one can even argue that the forecast ranking based on statistical measures is obsolete and the predictive modeling step can be fully integrated into the decision-making problem, e.g. through end-to-end learning approaches \citep{donti2017task, elmachtoub2022smart}. This argument is strongest if the application is stable. If the application is subject to change, e.g. due to changes in the asset portfolio, a general-purpose forecast is more desirable and flexible.
    \item One focuses on delivering the best general-purpose forecast in terms of a general scoring rule $S(\cdot, \cdot)$. There is a strong case that for strictly proper scoring rules, a forecast that performs well in terms of $s_m$ will also perform well in terms of $r_m$ across a range of decision-making problems and will be robust to changes in the application. 
\end{itemize}
There is a wide gray area in between these two extreme points of view. A plausible option is using weighted combinations of $s_m$ and $r_m$ to rank models. If rankings coincide, this gives us more confidence in the model choice and employ additional model diagnostics. Somewhat turning things around, it can be argued that this reliance on few parts of the distribution is advantageous. If we need only a certain part of the distribution to be (approximately) correct to create the optimal bids, we can utilize this for reducing the dimensionality of the problem. This can be hard to achieve a priori, as the optimal bids $\vec{a}_m^*$ are not known. In hindsight, we can analyze which parts of the forecast distribution were relevant for the decision-making problem and focus on improving these parts in future forecasts. 

\subsection{Evaluation of Decision Quality}\label{sec:ebts-decision-quality}

In the following, we discuss the evaluation of decision quality given competing forecasts. It is desirable that the evaluation of the decision quality is consistent with the objective function employed in the optimization. \begin{itemize}
    \item This ensures that the ranking is consistent with the decision-making problem at hand. However, this requires the risk measure to be elicitable itself. For example, the CVAR is not elicitable on its own, but only jointly with the value-at-risk \citep{fissler2016higher,ziegel2020robust,fissler2015expected}. 
    \item Under different competing forecasts $\mathcal{F}_m$, the optimal bids $\vec{a}_m^* = \arg \max_{\vec{a}} \rho(\mathcal{F}_m)$ might differ. This adds another layer of complexity to the evaluation of the forecasts based on battery optimization. This is in contrast to, e.g. the backtesting CVAR forecasts of GARCH models, where the implicit assumption is that the forecaster is long/short a fixed position in the asset. 
\end{itemize}
Therefore, the predicted objective value of the battery optimization for the forecast model $m$ as $\omega^*_m = \rho(\mathcal{P}^{\vec{a}_m^*}_m)$ and the realized returns of $r_m = \transpose{(\vec{a}_m^*)} \vec{p}$ cannot be evaluated across different models. 
For a strictly proper scoring rule $S(\cdot, \cdot)$ of $\rho(\cdot)$ we have 
\begin{align}
    S(\mathcal{P}^{\vec{a}_m^*}_m, \transpose{(\vec{a}_m^*)} \vec{p}) = S(\mathcal{P}^{\vec{a}_m^*}_m, r_m)
\end{align}
which shows that the forecast optimal bids of forecast $\mathcal{F}_m$, which we would like to evaluate, enters the scoring rule on both sides of the equation. As possible workaround, for a number of models $m \in \mathcal{M}$, we can evaluate the scoring rule for all optimal bids~$\vec{a}_m^*$ and all forecasts $\mathcal{F}_m$ and analyze the resulting scores. If the forecast $\mathcal{F}_m$ has the lowest score of all models for the risk measure for bids~$\vec{a}_m^*$, then this should a good indication that actually this decision is good. If many/all models $i = 1, ... ,M; i \ne{} m$ have better scores on the objective value~$\omega_m$ than~ $\mathcal{F}_m$, this is an indication that the decision~$\vec{a}_m^*$ is not globally optimal. This also gives rise to a formal test by comparing the scores of the models $i$ with the scores of model $m$ for the optimal bids $\vec{a}_m^*$, which can be done through a Diebold-Mariano test \citep[DM-test,][]{diebold2002comparing, diebold2015comparing}. We test for the hypothesis \begin{align}\label{eq:cross-score-h0}
    H_0: S(\mathcal{P}^{\vec{a}_m^*}_i, r_m) -  S(\mathcal{P}^{\vec{a}_m^*}_m, r_m) \geq 0 && \text{Model $m$ has lower (better) scores than model $i$.}
\end{align} 
If we can reject the null hypothesis, this implies that model $i$ was able to yield significantly better predictions of the objective value for this decision and hence an indication that the decision $\vec{a}_m^*$ based on forecast $\mathcal{F}_m$ should not be trusted. Importantly, this only addresses one side of the optimization problem, since we are only evaluating how well the forecast~$\mathcal{F}_m$ performs for the optimal bids~$\vec{a}_m^*$ relative to the other forecasts for the hours in which $m$ places bids, which is only a subset of hours. The predictive accuracy in the other hours is not evaluated, even though the predictive performance in these hours is relevant for the decision-making problem.

\section{Forecasting and Simulation Study}\label{sec:case_study}

In the following section we run a small forecasting experiment and analyze the impact of different forecast errors on the results of the optimization. We employ seven different forecasting models with different modeling strategies and distributional assumptions. These models range from the simple climatology and naive models, the lasso-estimated autoregressive model \citep{lago2021forecasting} and distributional regression approaches \citep{muniain2020probabilistic, hirsch2024online,hirsch2025online}. It is important to note that we have preferred models that provide either an option for bootstrapping or a parametric distributional assumption in order to draw samples for the stochastic program. For this reason, we have not included quantile regression or conformal prediction approaches \citep[see e.g.][]{uniejewski2025smoothing,o2025conformal}. As the main aim of the case study is to relate statistical measures and economic evaluation on a diverse pool of models, we acknowledge that we have not undertaken extensive feature engineering or hyperparameter optimization, but chose a configuration that has been shown to work well in previous work.

\subsection{Data and Market}\label{sec:case_study-data}

The day-ahead market is the primary exchange for electricity in Germany. The market is organized as a uniform price auction, where the price for each hour of the next day is determined by the intersection of the aggregated supply and demand curves. The market is cleared at 12:00 CET for the delivery of electricity on the next day. We use data from ENTSO-E and \url{investing.com} provided by \cite{lipiecki2024postprocessing}. The data covers the period from 2015 to 2023 and includes the day-ahead prices, the residual load, the marginal costs of gas, coal and oil fired production and the associated emission allowance costs. We use 2015-01-15 to 2018-12-26 for model training and 2018-12-27 to 2023-12-31 for testing, in line with other works using the same data set \citep{hirsch2024online,hirsch2025online,marcjasz2023distributional}.

\subsection{Forecasting Model}\label{sec:case_study-forecasting_model}

\newcommand{\gas}[0]{\ensuremath{\text{Gas}}}
\newcommand{\coal}[0]{\ensuremath{\text{Coal}}}
\newcommand{\eua}[0]{\ensuremath{\text{EUA}}}
\newcommand{\oil}[0]{\ensuremath{\text{Oil}}}
\newcommand{\resload}[0]{\ensuremath{\text{ResLoad}}}

We employ a distributional regression model or generalized additive model for location, scale and shape \citep[GAMLSS][]{rigby2005generalized} based on an extended expert model using Student's $t$-distribution for the marginal distribution of the electricity prices $$P_{d,h} \sim t(\mu_{d,h}, \sigma_{d,h}, \nu_{d,h})$$ Distributional models are popular for electricity price forecasting \citep{serinaldi2011distributional, abramova2020forecasting, muniain2020probabilistic, klein2023deep,marcjasz2023distributional}. We extend the well-known LASSO-estimated autoregressive model with exogenous variables \citep[LEAR, see e.g.][]{lago2021forecasting} model by non-linear effects representing the merit-order effects of residual load with the marginal costs~(MC) of gas, coal and oil fired production and the associated emission allowance costs. 
\begin{align}
    \mu_{d,h} 
        &= \beta_{d,h,0} 
        + \underbrace{\sum_{i=0}^{i=23}\beta_{d,h,i}P_{d,i}}_\text{Prices of previous day.} 
        + \underbrace{\sum_{l=2}^{l=14}\beta_{d,h,24+l}P_{d-l,h}}_\text{Time series effects.} 
        + \underbrace{\sum_{\text{WD} \in \mathcal{W}} \beta_{d,h,38+i} \operatorname{WD}(d)}_\text{Weekday effects.}
        + \underbrace{f_{d,h}(\resload_{d,h})}_\text{Residual load effects.} \\ \nonumber
        &+ \underbrace{f_{d,h}(\resload_{d,h}) \; \hdmul \; (\nu^\gas P^\gas_{d-2}+ \eta^\gas P^\eua_{d-2})}_\text{Interaction of residual load and the MC of gas plants.}
        + \underbrace{f_{d,h}(\resload_{d,h}) \; \hdmul \; (\nu^\coal P^\coal_{d-2}+ \eta^\coal P^\eua_{d-2})}_\text{Interaction of residual load and the MC of coal plants.} \\ \nonumber
        &+ \underbrace{f_{d,h}(\resload_{d,h}) \; \hdmul \; (\nu^\oil P^\oil_{d-2}+ \eta^\oil P^\eua_{d-2})}_\text{Interaction of residual load and the MC of oil plants.} 
\end{align}
where $\eta^\gas = 0.2, \eta^\coal = 0.35$ and $\eta^\oil = 0.3$ are emission factors and $\eta$ is a conversion factor between the fuel's unit and MWh thermal energy \citep[see][and Appendix \ref{app:equations} for details on the calculation of the marginal costs]{ghelasi2025data, ghelasi2025day} and $f(\cdot)$ denotes a b-Spline basis of degree 2 and 4 knots. Weekday effects are denoted with $\mathcal{W} = \{\text{Mon}, \text{Tue}, \text{Thu}, \text{Fri}, \text{Sat}, \text{Sun}, \text{Hol}\}$; holidays are encoded separately \citep{ziel2018modeling}. For the scale parameter $\sigma_{d,h}$ of the price distribution, we employ a similar structure, but with linear equations for residual load and the fundamental fuel prices. The degrees of freedom for the $t$-distribution are estimated as intercept. The equations can be found in the Appendix \ref{app:equations}. We name the model distributional LASSO-estimated non-linear autoregressive model (DLENAR). The model is estimated in a expanding window fashion using the online learning algorithm developed in \cite{hirsch2024online} to save computational time. We estimate separate models for each hour of the day. We model the dependence structure using a Gaussian Copula by the inference-for-margins approach. We convert the in-sample observations to the $\mathcal{U}(0, 1)$-space by the probability integral transformation (PIT) and subsequently to the $\mathcal{N}(0, 1)$ space by the inverse quantile transformation $\vec{g} = \Phi^{-1}(\vec{u})$. On the Gaussian space, we fit the correlation matrix.\footnote{Note that we additionally employ a ranking step to enforce strict uniformity on the residuals. This step does not alter the ordering of $\vec{u}$, but the conversion to ranks makes observations equally spaced.} We fit the three different models for the dependence structure:\begin{itemize}
    \item \textit{Independent (IND)}: We assume independence $\mat{\Omega} = \mat{I}$.
    \item \textit{Empirical (DEP)}: We use the empirical correlation matrix $\mat{\Omega} = \widehat{\operatorname{corr}}(\vec{g})$.
    \item \textit{Weekday (DWD)}: We estimate separate correlation matrices for each weekday $\mat{\Omega}_{\text{WD}} = \widehat{\operatorname{corr}}(\vec{g} | \text{WD})$.
\end{itemize}
We draw scenarios by sampling from the uniform distribution and applying a rank-reordering to ensure that the marginal distributions are \emph{exactly} equal across the different dependence structures. Additionally, we employ a number of benchmark models from the literature to obtain a diverse set of models. \begin{itemize}
    \item Climatology: We take a random sample from the in-sample observations as forecast for the next day. This is a common benchmark model for probabilistic forecasts \citep{gneiting2007strictly}.
    \item Naive Bootstrap (Naive-BS): The take the previous week, same weekday as the price forecast and sample from the in-sample residuals (of the same weekday). This is common benchmark model for seasonal time series such as electricity prices \citep{ziel2018modeling}.
    \item The widely used LASSO-estimated autoregressive model \cite[LEAR, see][]{lago2021forecasting} model with Gaussian errors $\mathcal{N}(0, \Sigma)$ estimated on the residuals of the mean model. The model is denoted as LEAR-N(0, $\Sigma$).
    \item The LEAR model with bootstrap residuals: we sample from the in-sample residuals of the mean model. We somewhat improve the bootstrap approach by clustering the days and predicting cluster membership by a $k$-nearest neighbor classifier \citep{cunningham2021k}
\end{itemize}
For all bootstrap approaches, we always sample full error trajectories to ensure that we (implicitly) capture the dependence structure of the errors. Model equations and further details can be found in the Appendix \ref{app:equations}.

\subsection{Battery Optimization Experiments}

We run a number of battery optimization experiments using different risk measures and correlation structures. Our experiments can largely be grouped in three categories. Generally, our experiments assume a storage of 10 MWh, a round-trip efficiency of $\eta^2 = 0.95^2$. The duration of the storage, that is, the ratio between storage capacity and maximum charge/discharge power is set to 1 hour unless otherwise stated. For larger battery capacities, we would need to take market impact into account \citep[see][]{narajewski2022optimal}, which is beyond the scope of this paper. 

\begin{table}[htb]
    \resizebox{\linewidth}{!}{%
    \begin{tabular}{ll}
        \toprule
        Optimization Method                 &   Dynamic Programming (DP), Mixed-Integer Linear Programming (MILP) \\
        Objectives                          &   Expected Profit (Risk-neutral), Conditional Value-at-Risk (CVAR, risk-averse, $\alpha \in 0.5, 0.75, 0.9$) \\
        Storage ($\kappa$)                  &   10 MWh \\
        Battery Duration ($d = \kappa / \xi$)   &   1, 2, or 4 hours (translates to $\xi \in 10, 5, 2.5$ MW) \\
        Number of Cycles ($c$)                    &   1 or 2 cycles per day \\
        \bottomrule
    \end{tabular}%
    }%
    \caption{Battery Configuration and Experimental Setup}
\end{table}

\paragraph{Multivariate Probabilistic Scoring Rules and Battery Revenues} In this experiment, we evaluate the relationship between multivariate probabilistic scoring rules and battery revenues. We discuss the results in light of Result \ref{res:scoringrule_limit} and analyze the relationship between the scoring rules, decision quality and battery revenues. We run the battery optimizations maximizing the expected profits and the CVAR at levels $\alpha \in \{0.90, 0.75, 0.50\}$ using the MILP approach for a 1, 2 and 4 hour battery. The results of this experiment are discussed in Section \ref{sec:results-economic}.

\paragraph{Multivariate Forecasts and Optimization Algorithms} In Sections \ref{sec:ebts-dp} and \ref{sec:milp} we have described an intuitive dynamic programming approach and a more scalable mixed-integer linear programming (MILP) approach to optimize battery trading strategies using multivariate probabilistic forecasts. Here, we briefly evaluate the impact of placing only one buy and sell pair compared to the MILP approach allowing for multiple bids. To this end, we compare results for a 1-hour battery, maximizing the expected profits and the CVAR at levels $\alpha \in \{0.90, 0.75, 0.50\}$ using both optimization approaches. The results of the experiment are discussed in Section \ref{sec:results-simplified}. 

\paragraph{Quantile-based Trading Strategies} We evaluate the quantile-based trading strategy (QBTS) as described in Section \ref{sec:qbts} using different forecast models. We evaluate the total profits, profits per MWh traded and the acceptance probability for different nominal prediction interval widths $\alpha$. Here, we focus providing an intuitive understanding of the Proposition \ref{res:qtbs-scoring} and Remark \ref{result:dependence_structure} through numerical examples and contrast the results with the stochastic programming approach. The results of this experiment are discussed in Section \ref{sec:results-qbts}.

\subsection{Economic Performance Measures}

We evaluate the economic performance of the different forecast models based on the daily returns from the battery optimization experiments. The daily return for day $d$ and model $m$ is defined as $r_{d,m} = \sum_{h=0}^H a_{d,h,m} P_{d,h}$ where $a_{d,h,m}$ are the optimal actions derived from the battery optimization using forecast model $m$ for day $d$. For daily returns $r_{d, m}$ of $m$ and day $d$ we calculate the total profits and the Sharpe ratio \citep{sharpe1966mutual} as \begin{equation}
    \text{Total Profits}_m = \sum_{d=1}^D r_{d,m} \quad \text{and} \quad
    \text{Sharpe Ratio}_m = \frac{\operatorname{E}[r_{d,m}]}{\operatorname{SD}[r_{d,m}]}.
\end{equation}
For the CVAR optimization, we also evaluate the ratio of Value-at-Risk (VaR) exceedances. The VaR at level $\alpha$ is defined as the ($1-\alpha$) -- quantile of the return distribution, that is, $\text{VaR}_\alpha = \inf \{x \in \mathbb{R} : P(r_{d,m} \le x) \ge \alpha\}$. The VaR exceedance ratio is then defined as \begin{equation}
    \text{VaR Exceedance Ratio}_m = \frac{1}{D} \sum_{d=1}^D \mathbb{I}(r_{d,m} < \text{VaR}_{\alpha, m}).
\end{equation} 
This can be both seen as a measure of the economic performance, and as a measure of the calibration of the CVAR forecasts, as for a well-calibrated CVAR forecast, the VaR exceedance ratio should be close to $1-\alpha$.

\section{Forecast Evaluation}\label{sec:forecast-evaluation}

We evaluate the forecast models using standard probabilistic scoring rules and economic metrics derived from the battery optimization experiments. 
We follow the general principle of sharpness subject to calibration \cite{gneiting2007probabilistic}. The following sections provide an overview of the measures of calibration and scoring rules. We denote the true value with $\vec{y}$ and the forecasted distribution with $\mathcal{F}$, represented by an ensemble of $M$ scenarios $\widehat{\mat{F}} \sim \mathcal{F}$ for $m = 1, \ldots, M$. We denote the mean and median forecasts as $\widehat{\vec{\mu}}$ and $\bar{\vec{\mu}}$. The data set of prices has $D$ days and $H$ hours per day and accordingly we aggregate by averaging. We test the statistical significance of differences in the scores by a Diebold-Mariano test \citep[DM-test,][]{diebold2002comparing, diebold2015comparing, nowotarski2018recent}.

\subsection{Scoring Rules for Continuous (Multivariate) Distributions}

A calibrated probabilistic forecast correctly represents the probabilities of the observed data. For univariate and quantile forecast, marginal calibration can be easily checked by comparing expected and observed frequencies. \begin{equation}
    \text{MC}_\alpha = \alpha - \frac{1}{T} \sum_{t=1}^T \mathbb{I}(\vec{y}_t < \hat{Q}_t^{\alpha})
\end{equation}
where $\hat{Q}_\alpha$ is the predicted quantile at level $\alpha$. For a well-calibrated forecast, the marginal calibration should be close to zero for all $\alpha$. Visually, this can be checked by plotting the marginal calibration curve, which plots the observed frequency of the event $\vec{y}_t < \hat{Q}_t^{\alpha}$ against the predicted quantile level $\alpha$. For a well-calibrated forecast, the marginal calibration curve should be close to the diagonal.

The following scoring rules are used throughout the empirical evaluation.
The MSE, MAE, CRPS and ES are strictly proper scoring rules for the mean, median, univariate and multivariate probabilistic forecasts. The DSS is only (strictly) proper for the mean-covariance structure of the predictive distribution. The VSS is not a strictly proper scoring rule. We keep the discussion of these scoring rules brief, as they are commonly used in PEPF. We have:
\begin{equation}
    \text{MSE} = \frac{1}{H} \sum_h^H (\hat{\vec{\mu}}_h - \vec{y}_h )^2 
    \quad \text{and} \quad \text{RMSE} = \sqrt{\text{MSE}} 
    \quad \text{and} \quad
    \text{MAE} =  \frac{1}{H} \sum_h^H | \bar{\vec{\mu}}_h - \vec{y}_h|
\end{equation}\begin{equation}
    \text{CRPS} = \frac{1}{M}\sum_{m=1}^M |\vec{y}_{h} - \widehat{\vec{F}}_{h,m}| - \frac{1}{2M^2} \sum_{m=1}^M \sum_{n=1}^M |\widehat{\vec{F}}_{h,m} - \widehat{\vec{F}}_{h,n}|
\end{equation}\begin{equation}
    \text{ES} = \frac{1}{M}\sum_{m=1}^M ||\vec{y} - \widehat{\vec{F}}_{m}||_2 - \frac{1}{2M^2} \sum_{m=1}^M \sum_{n=1}^M ||\widehat{\vec{F}}_{d,m} - \widehat{\vec{F}}_{d,n}||_2
\end{equation}\begin{equation}
    \text{VS} = \frac{1}{H^\frac{1}{p}} \sum_{i=0}^H \sum_{j=0}^H \left(|y_{i} - y_{j}|^p - \frac{1}{M} \sum_{m=1}^M |\widehat{\vec{F}}_{i,m} - \widehat{\vec{F}}_{j,m}|^p \right)^2
\end{equation}\begin{equation}
    \text{DSS} = (\vec{y} - \widehat{\vec{\mu}})^\top \widehat{\mat{\Sigma}}^{-1} (\vec{y} - \widehat{\vec{\mu}}) + \log |\widehat{\mat{\Sigma}}| 
\end{equation}
where $\widehat{\vec{\mu}}$ and $\widehat{\mat{\Sigma}}$ are the sample mean and covariance matrix of the ensemble forecasts $\widehat{\mat{F}}$.
The CRPS and the ES are estimated using the energy estimator \citep{gneiting2007strictly,zamo2018estimation}. For further information on the scoring rules we refer to the original papers \cite{scheuerer2015variogram,dawid1999coherent,gneiting2007strictly} and recent works on discriminatory power of scoring rules \citep[see][]{pinson2013discrimination,  marcotte2023regions, alexander2024evaluating} and potential asymmetric effects \citep{buchweitz2025asymmetric}. 

A particular issue is the evaluation of CVAR forecasts, as the CVAR is not elicitable on its own, but only jointly with the value-at-risk \citep{fissler2016higher, fissler2015expected}. For the evaluation of CVAR forecasts, we therefore use the strictly proper joint scoring rule for the (VaR, CVAR) proposed by \cite{fissler2015expected}. For a forecast of the VaR and CVAR at level $\alpha$, denoted as $(v, e)$, the scoring rule is defined as \begin{equation}
    S_\alpha((v,e), y) = \underbrace{(\mathbb{I}\{y \le v\} - \alpha) \left(G_1(v) - G_1(y)\right)}_\text{Pinball score for the VaR forecast.} + \frac{1}{\alpha}G_2(e)\mathbb{I}\{y \leq v\} (v - y) + G_2(e)(e - v) - \mathcal{G}_2(e)
\end{equation}
where $G_1(v) = v$, $G_2(e) = \exp(e)/(1 + \exp(e))$ and $\mathcal{G}'_2(e) = G_2(e)$ (see \citet[p. 1-3]{fissler2015expected} for details). The first term corresponds to the pinball score for the VaR forecast, while the second and third term correspond to the scoring rule for the CVAR forecast.

For the case of point forecasting and battery trading strategies, the use of correlation measures has been proposed as scoring rules by \cite{maciejowska2025statistical}. For probabilistic forecasts, we propose the using a association-based scoring rule based on Kendall's $\tau$ \citep{kendall1938new}, defined as 
\begin{equation}\label{eq:scores-ks}
    \text{KS} = 
        \frac{1}{2} \mathbb{E}
            [\tau(\widehat{\vec{F}}, \vec{y})] 
        - \mathbb{E}
            [\tau(\widehat{\vec{F}}, \widehat{\vec{F}})] 
        - \tau(\vec{y}, \vec{y})
\end{equation}%
where $\tau(\vec{v}, \vec{w})$ is the Kendall's $\tau$ correlation between two vectors $\vec{v}$ and $\vec{w}$. Note that we can calculate the score either on the price ensembles or on the rank ensembles, since Kendall's $\tau$ is invariant under strictly increasing transformations. Kendall's $\tau$ also relates to the dependence structure of the forecast's copula and thus provides a natural way to evaluate the dependence structure \citep{gijbels2011conditional,ziel2019multivariate}.

\begin{prop}
The scoring rule proposed in Equation \ref{eq:scores-ks} is proper. 
The construction follows the kernel score framework \cite[][Theorem 4]{gneiting2007strictly}, which requires that the kernel is a positive definite kernel.
\citet[][Theorem 1]{jiao2015kendall} show that Kendall's $\tau$ is a positive definite kernel $\square$
\end{prop}

\cite{maciejowska2025statistical} propose two BESS oriented scores for point forecasts: The mean profit deviation (MPD) and the mean hour deviation (MHD), defined as \begin{equation}
    \text{MHD} = |\hat{h}_\text{max} - h_\text{max}  | + |\hat{h}_\text{min}  - h_\text{min} | \quad \text{and} \quad
    \text{MPD} = |(\hat{p}_\text{max}  - p_{\text{max} })| - |(\hat{p}_\text{min}  - p_\text{min} )|    
\end{equation}
where $h_\text{min}$ and $h_\text{max}$ are the hours with the lowest and highest price. The MPD can be seen as a simplification of the returns of a 1-hour battery, relaxing the assumption of charging before discharging. The MHD is closely related to rank-based scoring rules, which we will treat in the following section.

\subsection{Scoring Rules for Rankings}

We denote the ranks of the observed prices $\vec{p}_d$ as $\vec{\pi}_d$ and the rank forecast derived from the predicted ensemble forecast $\widehat{\mat{F}}_d$ as $\widehat{\vec{\Pi}}_d$. We therefore have a predictive distribution over the ranks of the observed prices. Note that we have as many ranks as hours, i.e. $K = H + 1$. In the marginal case, rank based scoring rules can take two views: \begin{itemize}
    \item (Item view) For given hour $h$, did we correctly predict the rank $\pi_{h}$ of the observed price $\vec{y}_{h}$?
    \item (Rank view) For given rank $k$, did we correctly predict which hour $h$ corresponds to this rank?
\end{itemize} In the following, we discuss scoring rules for the item and rank view and propose a new scoring rule for the joint distribution of the ranks based on the kernel scoring rule framework.
The Brier score for the rank and item view is defined as \begin{align}
    \text{Brier}_k^\text{rank} &= 
        \frac{1}{K} 
        \sum_{h=0}^H 
        \left( 
            \frac{1}{M}\sum_{m=1}^M\mathbb{I}\{\mat{\Pi}_{d,h,m} = k\} - 
            \mathbb{I}\{\pi_{d,h} = k\} 
        \right)^2, \\
    \text{Brier}_h^\text{item} &= 
        \frac{1}{K} 
        \sum_{k=1}^K 
        \left( 
            \frac{1}{M}\sum_{m=1}^M\mathbb{I}\{\mat{\Pi}_{d,k,m} = h\} - 
            \mathbb{I}\{\pi_{d,k} = h\} 
        \right)^2.
\end{align} 
Note that is formulation corresponds to the multi-class Brier score, which is bounded in [0, 2]. For the marginal distribution of the ranks (item view), the ranked probability score (RPS) is a strictly proper scoring rule \citep{epstein1969scoring}. The RPS is defined as \begin{equation}
    \text{RPS} = 
        \frac{1}{DHK} \sum_{d=1}^D 
        \left(
                \frac{1}{M}\sum_{m=1}^M\mathbb{I}\{\mat{\Pi}_{d,h,m} \le k\} - 
                \mathbb{I}\{\pi_{d,h} \le k\}
        \right)^2.
\end{equation}
Again, we refer the reader to the original papers for further details on the scoring rules \cite{gneiting2007strictly,epstein1969scoring} and recents works on the properties of rank scoring rules \cite{constantinou2012solving,du2021beyond, wheatcroft2021evaluating}.

It is common to look at the performance of ranking measures for the top-$k$ and bottom-$k$ ranks. In the context of battery trading strategies, seems natural to define the relevant top-$k$ and bottom-$k$ as the product of the duration of the battery and the number of (ceiled) cycles per day. For example, for a 2-hour battery with a with one cycle per day, we would be interested in the top-2 and bottom-2 ranks. For a 4-hour battery, we would be interested in the top-4 and bottom-4 ranks -- the same as we would be for a 2-hour battery with two cycles per day. It is important to note that in this case, we have to consider the relevant structure of the battery optimization problem, that is, we can only discharge after charging and are thus interested in the top-$k$ ranks for the hours following the bottom-$k$ ranks. We therefore evaluate the top-$k$ and bottom-$k$ ordered by ranks, that is, we look at the top-$k$ which are preceded by at least $k$ hours with the bottom-$k$ ranks, which we denote as the BESS-$k$ scores.

\section{Results}\label{sec:results}

This section presents the results. We discuss the results of the statistical forecast evaluation and subsequently the results of the economic evaluation. To make matters easier for the reader, we consistently use the color scheme for the for tables/heatmaps and for the models.

\subsection{Statistical Forecast Evaluation}\label{sec:results-statistical}

\begin{figure}[htb]
    \centering
    \includegraphics[width=0.5\textwidth]{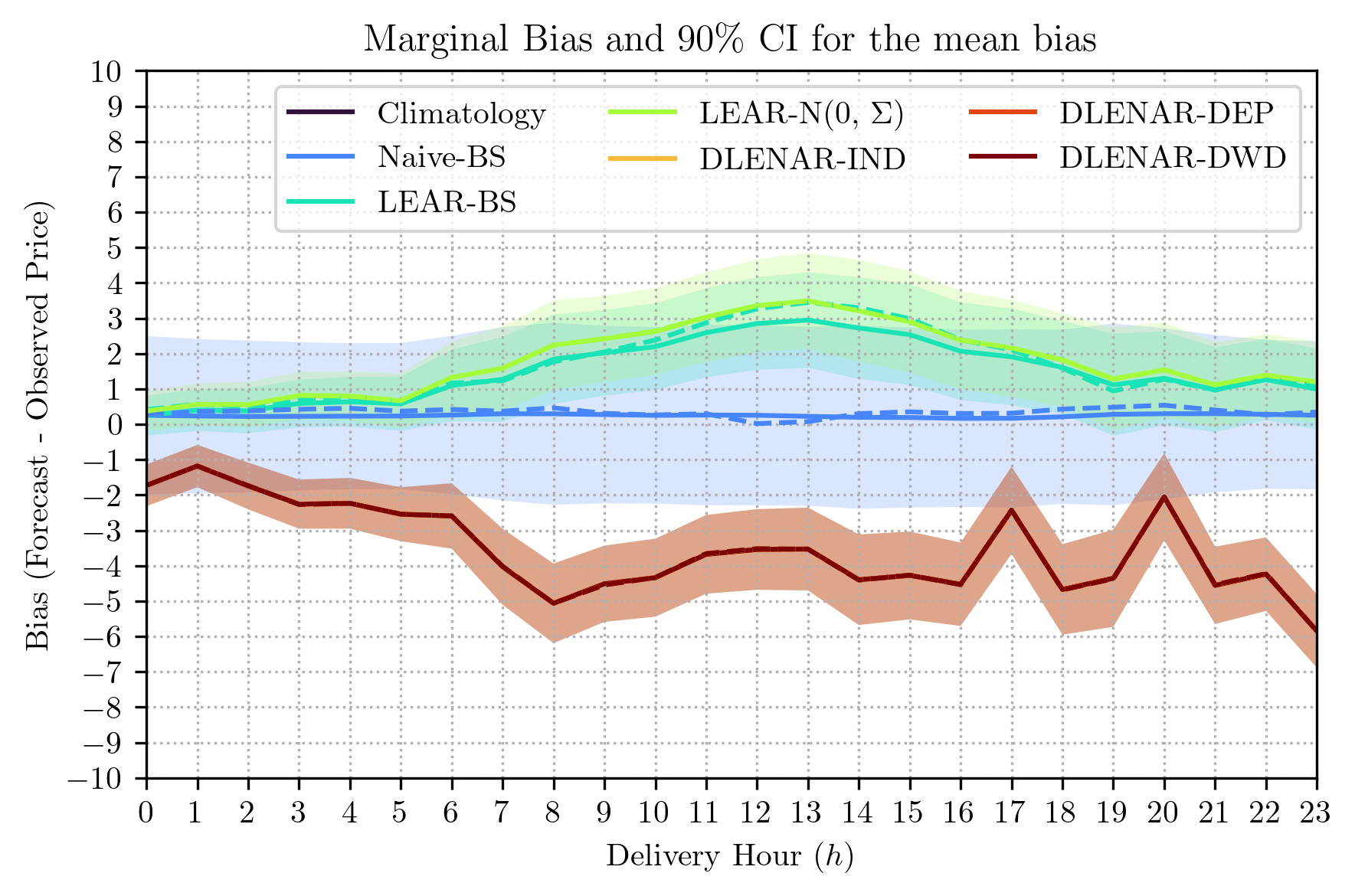}%
    \includegraphics[width=0.5\textwidth]{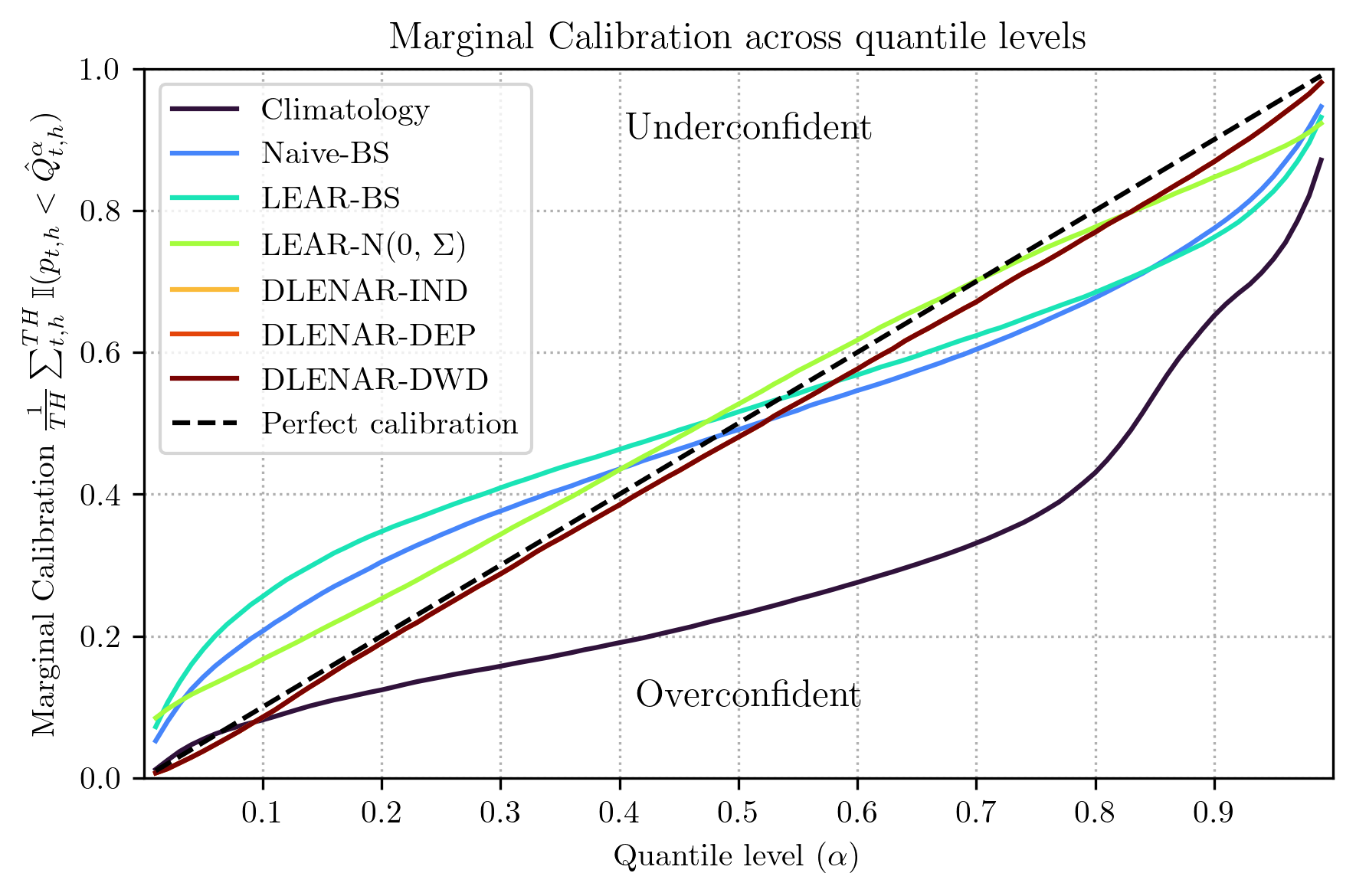}
    \caption{Marginal bias and marginal calibration for the different forecast models.}
    \label{fig:marginal_bias_calibration}
\end{figure}

\begin{figure}[htb]
    \centering
    \includegraphics[width=.8\linewidth]{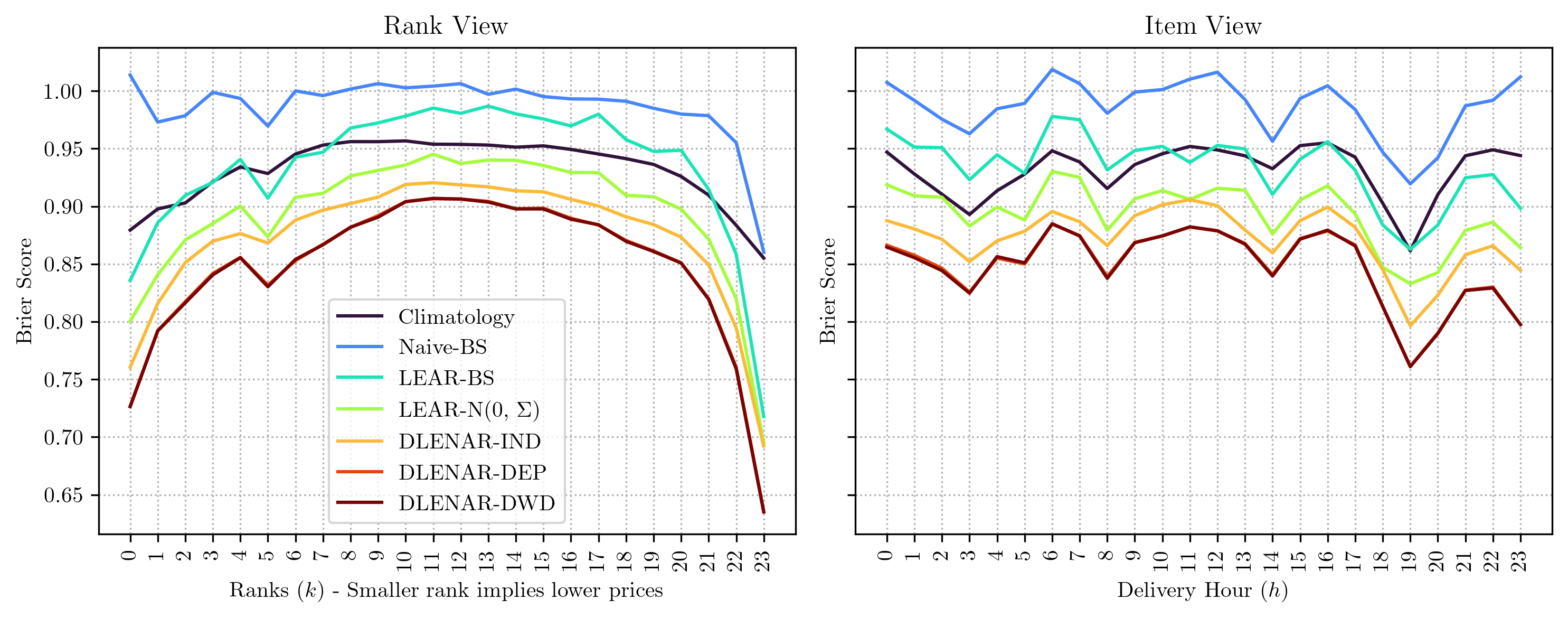}%
    \caption{Brier scores by rank and hour. Left panel shows the Brier score per rank, right panel shows the Brier score per hour.}
    \label{fig:brier_scores}
\end{figure}

This section briefly discusses the results of the forecast evaluation in terms of the scoring rules. Table \ref{tab:scoring_rules} shows the scores for the different forecast models and scoring rules. Table \ref{tab:top_k_scores} shows the top-$k$ Brier scores for the different forecast models and different definitions of top-$k$. Figure \ref{fig:brier_scores} shows the Brier scores by rank and hour. Marginal calibration is shown in the left panel of Figure \ref{fig:marginal_bias_calibration}, while the right panel of Figure \ref{fig:marginal_bias_calibration} shows the marginal calibration curve.

\begin{table}[htb]
    \centering
    \includegraphics[width=0.5\linewidth]{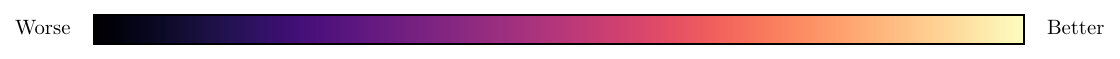} \\
    \resizebox{\linewidth}{!}{\input{tables/forecast_scores.tex}}%
    \caption{Scoring Rules for the different forecast models.}
    \label{tab:scoring_rules}  
\end{table}


Generally, the results follow an expected pattern, with the distributional regression model with the fitted dependence structure (DLENAR-DEP) performing best across all scoring rules, followed by the distributional regression model with the fitted dependence structure (DLENAR-DWD) and the distributional regression model with independence (DLENAR-IND). The LEAR-based benchmark models follow, with the LEAR-N(0, $\Sigma$) performing slightly better than the LEAR-BS. The climatology and Naive-BS perform worst. For the DLENAR-IND and the DLENAR-DWD and DLENAR-DEP, the difference between the ES is not as pronounced as for the other scoring rules, a potential indication that the ES is less sensitive to the dependence structure than other multivariate scoring rules \citep[see as well][on this issue]{alexander2024evaluating}, as the marginal model for these three is the same.

\begin{table}[htb]
    \centering
    \includegraphics[width=0.5\linewidth]{figures/legend.pdf} \\
    \resizebox{\linewidth}{!}{\input{tables/top_k_brier.tex}}%
    \caption{Top-$k$ Brier scores for the different forecast models and different definitions of top-$k$.} 
    \label{tab:top_k_scores}    
\end{table}

For the rank-based scores, the RPS and KS scores show a similar pattern as the ES, VS and CRPS. The Brier score shows a somewhat different pattern, as it yields worse scores for the Naive-BS than for the climatology model. This difference can be explained by the fact that the Naive-BS takes last weeks value as the forecast, while the climatology model yields an average over the training set. The predictions of the Naive-BS are hence somewhat more arbitrary, since the weather conditions between this week and last week can change drastically, while the climatology model captures at least the seasonal average. Figure \ref{fig:brier_scores} shows the Brier scores by rank and hour. The scores decrease quite steeply for the lowest ranks and then level off. For the hour/item view, we see that performance is rather stable over the hours, while hour 19 seems to be the easiest to predict, presumably since it is naturally the most expensive hour. For the top-$k$ scores in Table \ref{tab:top_k_scores}, we see that the BESS-$k$ scores are worse than the top-$k$ and bottom-$k$ scores, which is in line with the intuition that the middle ranks are harder to predict than the lowest and highest ranks. The differences between the different forecast models are not that pronounced, but we can see that the DLENAR-DEP-WD and DLENAR-DEP models perform better than the DLENAR-IND model, which is in line with the results for the other scoring rules.

\subsection{Economic Forecast Evaluation}\label{sec:results-economic}

\begin{figure}[htb]
    \centering
    \includegraphics[width=\linewidth]{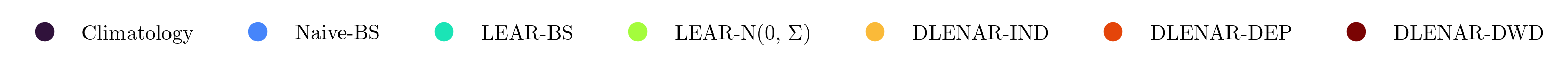}
    \includegraphics[width=\linewidth]{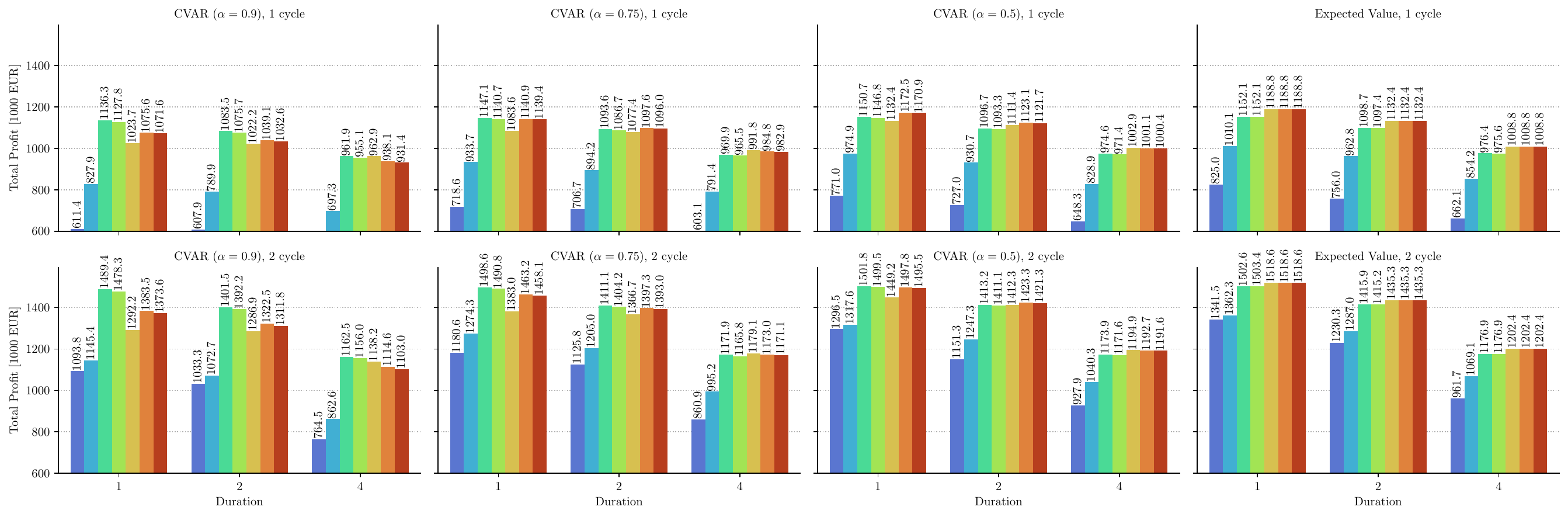}
    \caption{Total Profits for all models, durations and cycles.}
    \label{fig:ebts_milp_total_profits}
\end{figure}

\begin{figure}[htb]
    \centering
    \includegraphics[width=\linewidth]{figures/model_colormap_legend.png}
    \includegraphics[width=\linewidth]{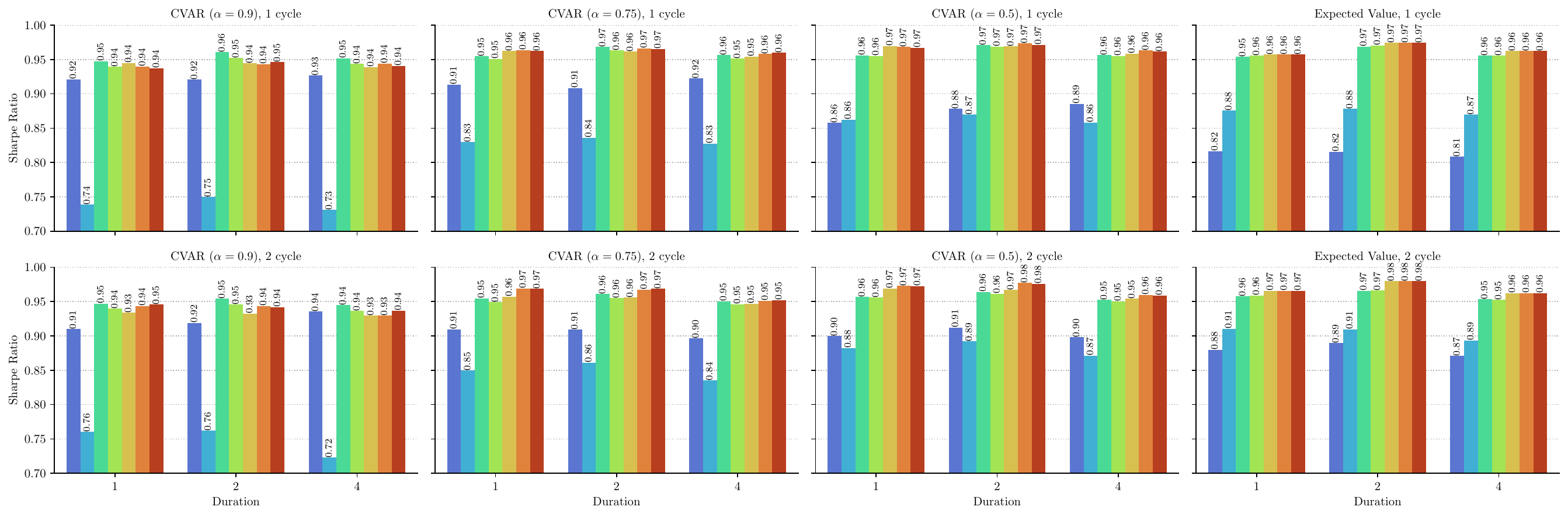}
    \caption{Sharpe ratios for all models, durations and cycles.}
    \label{fig:ebts_milp_sharpe_ratio}
\end{figure}

\begin{figure}
    \centering
    \includegraphics[width=\linewidth]{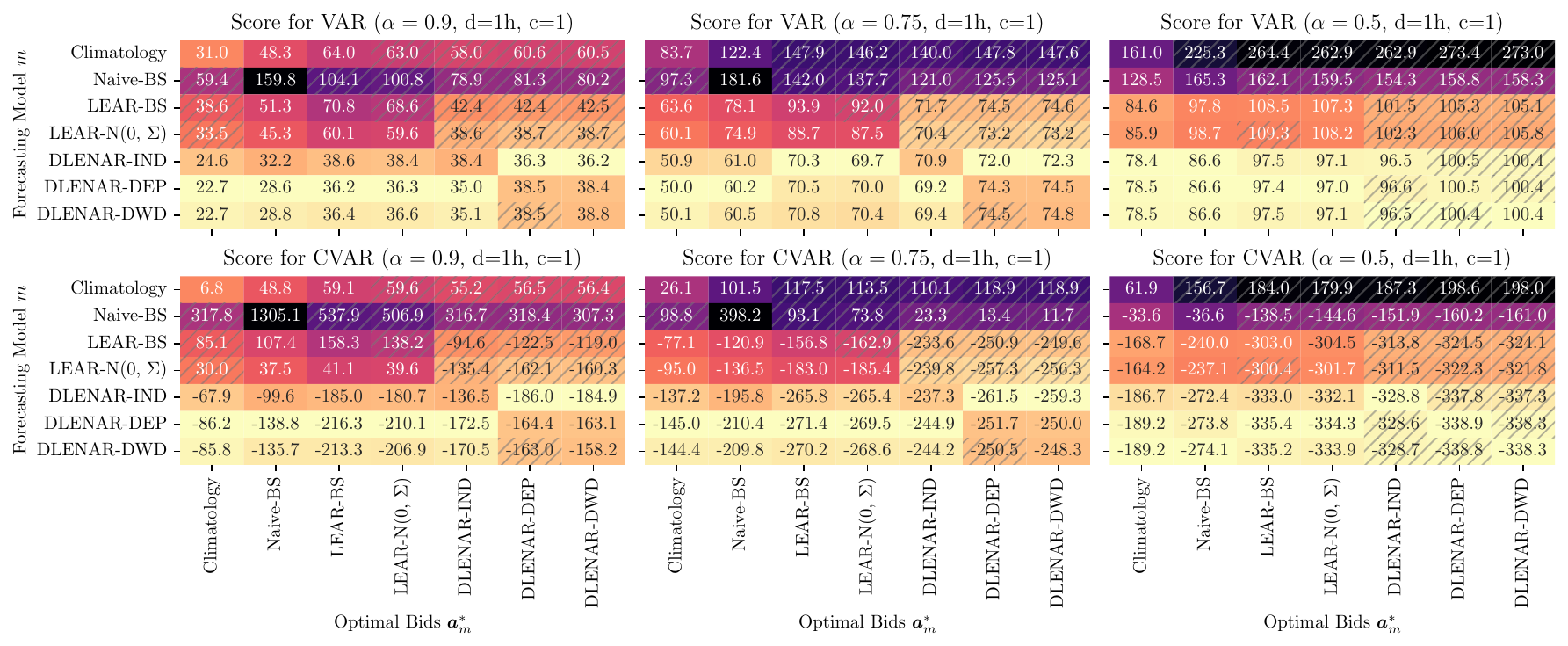}
    \caption{Cross-scoring for optimal bids on the objective function. Duration $d=1$, Number of cycles $c=1$. The hatched area denotes that the objective value forecasts from model (row, $i$) are not significantly better these of model which the optimal bids are derived (column, $m$), i.e. we cannot reject $H_0$ in Equation 
    \ref{eq:cross-score-h0}.}  
    \label{fig:ebts_cross_scores_d1_c1}
\end{figure}

\begin{figure}
    \centering
    \includegraphics[width=\linewidth]{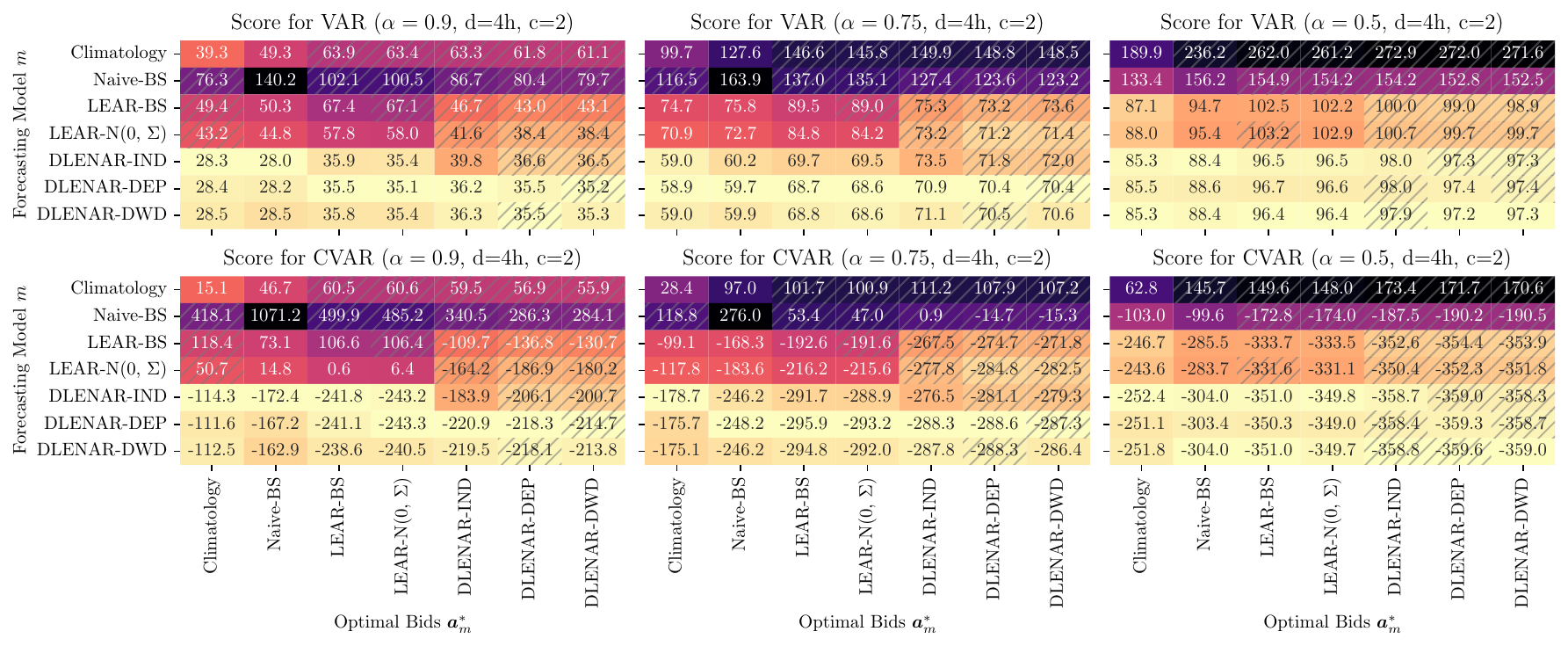}
    \caption{Cross-scoring for optimal bids on the objective function. Duration $d=4$, Number of cycles $c=2$. The hatched area denotes that the objective value forecasts from model $i$ (row) are not significantly better these of model which the optimal bids are derived (column), i.e. we cannot reject $H_0$ in Equation 
    \ref{eq:cross-score-h0}.}  
    \label{fig:ebts_cross_scores_d4_c2}
\end{figure}

In the following, we present an economic evaluation based on multivariate probabilistic forecasts. Given the fact that risk-neutral optimization approaches do not profit from probabilistic forecasting, we focus our discussion on the risk-averse case. As in the discussion of simplified BESS optimization before, we report the total profits and the Sharpe ratio for the combination of the battery configurations and forecast models. These results are shown in Figures \ref{fig:ebts_milp_total_profits} and \ref{fig:ebts_milp_sharpe_ratio}. We also report the Value-at-Risk (VaR) exceedance rates for the different models, which are shown in Table \ref{tab:exceedance_var}. Figures \ref{fig:ebts_cross_scores_d1_c1} and \ref{fig:ebts_cross_scores_d4_c2} give the cross-evaluation of the optimal bids' objective values for the different forecast models for the 1-hour, 1 cycle case and the 4-hour, 2 cycle case. The results for the other cases are similar and can be found in the supplementary material. The results for the economic performance measures are as follows.
\begin{itemize}
    \item For the optimization with respect to the expected profits, the DLENAR models yield the highest profits. This is in line with all the scoring rules, which rank the DLENAR models highest.
    \item For the risk-averse optimization, the LEAR-BS yields the highest profits for the CVAR optimization at the 90\% and 75\% level. For the CVAR at the 50\% level, the DLENAR-DEP model yields the highest profits. The results are relatively similar for the 1 and 2 hour duration.
    \item The climatology forecast yields surprisingly high Sharpe ratios for the risk-averse optimization. This is likely due to the fact that the optimal bids derived from the climatology forecast are very conservative, which leads to low profits but also low volatility of the profits, which results in a high Sharpe ratio. 
\end{itemize}

\begin{table}[htb]
    \centering
    \includegraphics[width=0.75\linewidth]{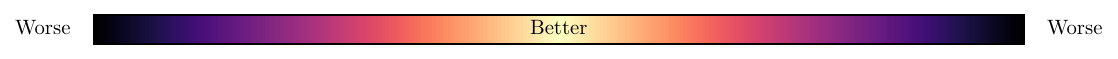} \\
    \resizebox{\linewidth}{!}{\input{tables/exceedance_value_at_risk_wide.tex}}
    \caption{Value-at-Risk (VaR) exceedance rates. Negative values imply that the losses exceed the VaR more frequently than the nominal level.}\label{tab:exceedance_var}
\end{table}

For the risk-averse optimization, we discuss the results in more detail in the following. \begin{itemize}
    \item The VaR exceedance rates for the CVAR optimization are closest to the nominal level for the DLENAR model. Interestingly, we see that for the rather low risk aversion of $\alpha =0.5$ and small battery configuration ($d \leq 2, c \leq 2$), the LEAR-based optimization yields the best calibrated VaR forecasts. For higher risk aversion, and larger battery configurations, modelling the dependence structure becomes more important and hence the DLENAR-DEP and DLENAR-DWD yield the best calibrated VaR forecasts. 
    \item The LEAR-BS model, which yields the highest profits for the CVAR optimization at the 90\% and 75\% level (see Figure \ref{fig:ebts_milp_total_profits}), also has exceedance rates of 11\% to 23\% above the nominal level. The LEAR-N(0, $\Sigma$) model, which yields the second highest profits for the CVAR optimization at the 90\% and 75\% level, has exceedance rates of 5\% to 15\% above the nominal level. This stark difference is striking, as the two models are rated similar in terms of the CRPS and the ES, and the LEAR-BS model even has better scores in terms of the VS and DSS (see Table \ref{tab:scoring_rules}). 
    \item The climatology model has VaR exceedance rates below the nominal level and, for $\alpha = 0.9$, also very close to the nominal level, which gives an indication that the derived bids are very conservative. This is in line with its high Sharpe ratios and relatively high profits considering the simplicity of the model.
\end{itemize}
Additionally, we evaluate the decision quality by cross-scoring of the optimal bids derived from the different models. Figures \ref{fig:ebts_cross_scores_d1_c1} and \ref{fig:ebts_cross_scores_d4_c2} show the results for the 1-hour, 1 cycle battery configuration and the 4-hour, 2 cycle battery configuration. We choose these configurations, as the first one represents a common reference case and the latter can be seen as a reference for hydro-pumped storage \cite[see e.g.][and the references therein]{lohndorf2023value} and as a shorter-duration battery with additional commitments to balancing markets \cite{kraft2023stochastic}. The upper panel show the pinball loss on the VaR forecast, the lower joint the joint score on the (VaR, CVAR) forecast. The columns denote the model whose optimal bids are evaluated and the rows denote the model whose forecasts are used for the evaluation. The diagonal hence shows the scores for the optimal bids derived from model $m$ evaluated on the forecasts of model $m$. The off-diagonal elements show the scores for the optimal bids derived from model $m$ evaluated on the forecasts of model $m' \neq m$. Figures for the remaining configurations are shown in the supplementary material and show similar results. The main conclusions from these results are as follows.
\begin{itemize}
    \item The results for the pinball scores for the VaR forecasts and the joint scores for the (VaR, CVAR) forecasts are similar, which is an indication that the results are not driven by a specific choice of the score function, but rather reflect a more general pattern.
    \item For decreasing risk aversion, the differences between the different models become smaller, which is in line with the fact that for risk-neutral optimization, the optimal bids are not derived from the probabilistic forecasts, but rather from the point forecasts. Hence, the scores of LEAR-BS and LEAR-N(0, $\Sigma$) and the scores within the DLENAR models converge. We also see less statistically significant differences.
    \item The DLENAR models have the lowest scores for the optimal bids derived from all models. Within the group of DLENAR models, for the 1 and 2 hour battery, the models are somewhat over-cross, the DLENAR-IND has the best forecasts for the bids derived from the DLENAR-DEP and vice versa. For the 4 hour battery, we see the DLENAR models have the best decision quality for the optimal bids derived from themselves, which is in line with the fact that for higher risk aversion and larger battery configurations, modelling the dependence structure becomes more important.
    \item The Naive-BS model has the worst scores for the optimal bids derived from itself, but also for all other models' optimal bids. The climatology model has surprisingly good scores not only for its own optimal bids, but also for the optimal bids derived from the LEAR-based models, especially for the very risk-averse optimization. This is in contrast to the statistical scoring rules, where the climatology model performs worst. 
\end{itemize}


Concluding, our economic evaluation of different forecasts in a risk-averse setting yields some interesting conclusions. While delivering the best profits and high Sharpe ratios, the LEAR-based models fail to deliver well-calibrated VaR forecasts and show shortcomings in the evaluation of the objective value of the optimal bids, when compared to simple benchmarks such as the climatology model. This is a stark reminder that total achieved profits are no suitable measure to compare model performance in a risk-averse setting. The stark differences in the VaR exceedance rates and the cross-scoring are not reflected in the total-profit ranking alone.

\subsection{Economic Evaluation based on simplified BESS}\label{sec:results-simplified}

Simplified case studies are popular for the monetary evaluation of forecasts through battery trading strategies. The predominant battery configuration is a 1-hour battery with one cycle per day, allowing one trade per day \citep[][and the literature on QBTS]{maciejowska2025statistical, chȩc2025extrapolating,serafin2025data}. This last constraint is crucial, as it allows us to use a simple dynamic programming approach to optimize the battery trading strategy. Arguably, for risk-neutral optimization, this constraint does not lead to suboptimal strategies, as we just select the hours with the highest spreads. However, for risk-averse optimization, this constraint can lead to suboptimal strategies, as we might want to place multiple bids to diversify the risk. 

\begin{figure}[htb]
    \centering
    \includegraphics[width=\linewidth]{figures/model_colormap_legend.png}
    \includegraphics[width=\linewidth]{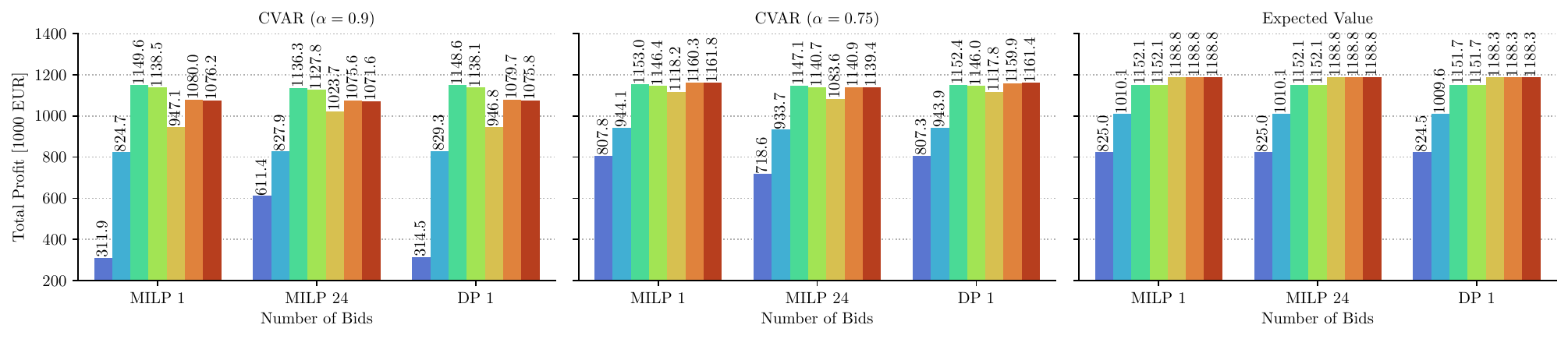}
    \includegraphics[width=\linewidth]{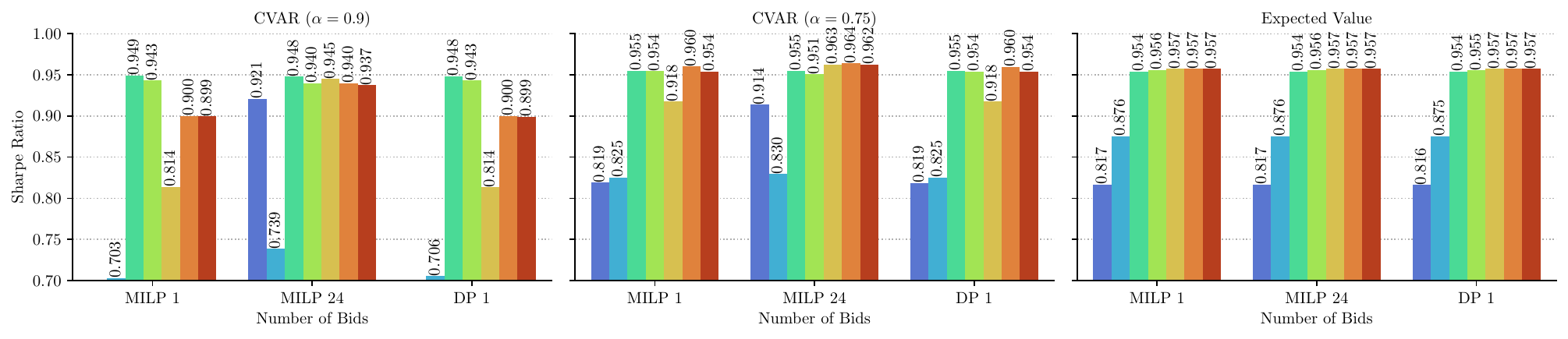} 
    \includegraphics[width=\linewidth]{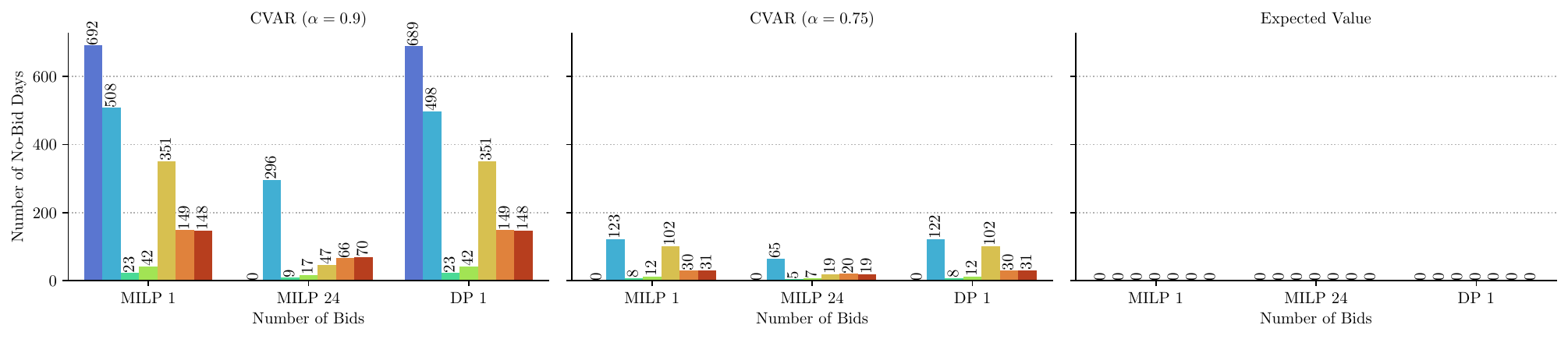}
    \caption{Total Profits, Sharpe ratios and the number of no-bid days for optimizing with the stylized DP optimization or the MILP formulation.}
    \label{fig:dp_vs_milp_evaluation}
\end{figure}

Figure \ref{fig:dp_vs_milp_evaluation} shows the total profits, Sharpe ratio and the number of no-bid days for the DP approach and the MILP approach for a 1-hour battery with one cycle per day. The results are shown for the expected profit optimization and the CVAR optimization at levels $\alpha \in \{0.90, 0.75, 0.50\}$. In the following, we discuss some key insights from these results. We denote with DP-1 the stylized optimization, with MILP-1 the MILP optimization with the same constraint and with MILP-24 the MILP optimization allowing for up to 24 bids per day.
\begin{itemize}
    \item For the expected profit optimization, we see that the profits are exactly the same for the simplified BESS and the MILP approach. This is in line with the theory, as for a 1-hour battery, we just select the highest spread hours.
    \item For the risk-averse optimization, we see that the profits are higher for the MILP-24 approach. The effect is most prominent for the CVAR ($\alpha=0.9$) optimization and the DLENAR-IND and the climatology model. The simplified approach does not allow for diversification gains on the marginal, but only on the dependence structure. The MILP-24 approach also rewards diversification along the marginal. Intuitively, the simplified approach only allows us to trade the largest spread, or not at all, while the MILP-24 approach allows us to trade the other spreads as well, which can yield diversification gains even if the dependence structure is not captured well. This effect is also visible in the number of no-bid days, where the simplified approach has more no-bid days than the MILP-24 approach.
\end{itemize}
In this case, the simplified approach overstates the effect of modelling the dependence structure for the DLENAR models, and, at the same time, understates the performance of naive strategies such as the climatology model, which can still yield good diversification gains on the marginal. This is an indication that the simplified approach might lead to spurious conclusions about the performance of the forecasts, especially for risk-averse optimization.

\subsection{Economic Evaluation based on QBTS}\label{sec:results-qbts}

\begin{figure}
    \centering
    \includegraphics[width=0.33\linewidth]{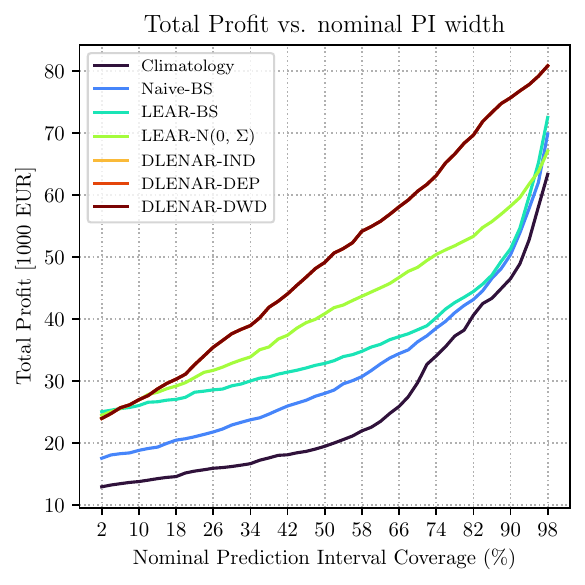}%
    \includegraphics[width=0.33\linewidth]{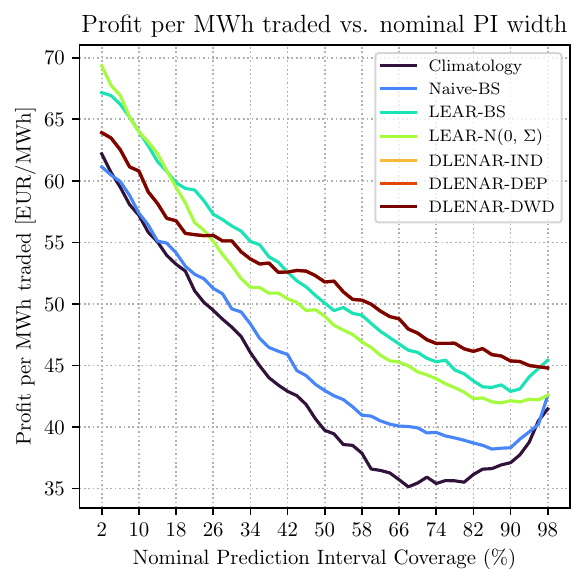}%
    \includegraphics[width=0.33\linewidth]{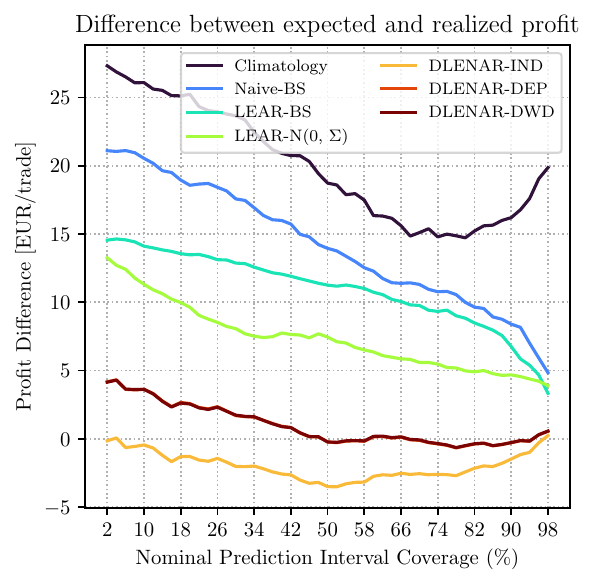}
    \caption{Total profits and profits per trade for the QBTS strategy.}
    \label{fig:qbts-profits}
\end{figure}

\begin{figure}
    \centering
    \includegraphics[width=0.5\linewidth]{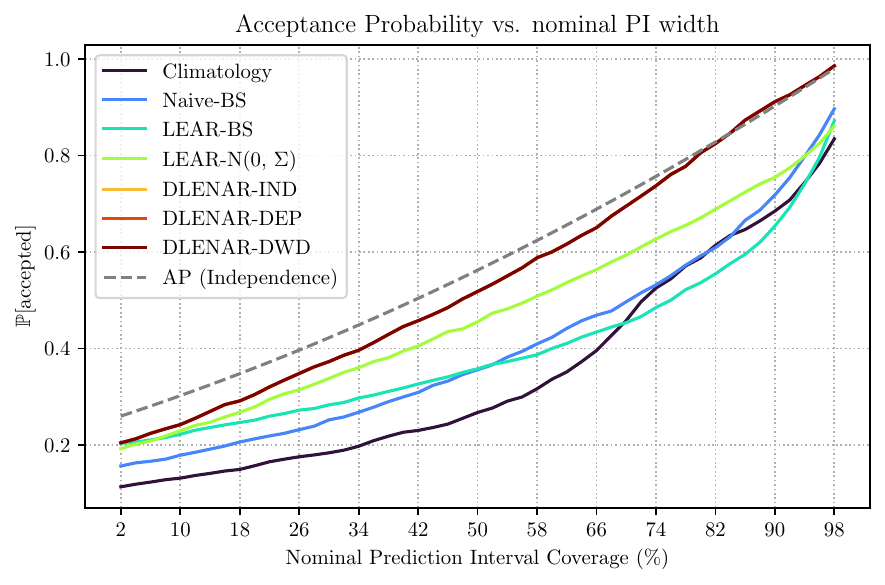}%
    \includegraphics[width=0.5\linewidth]{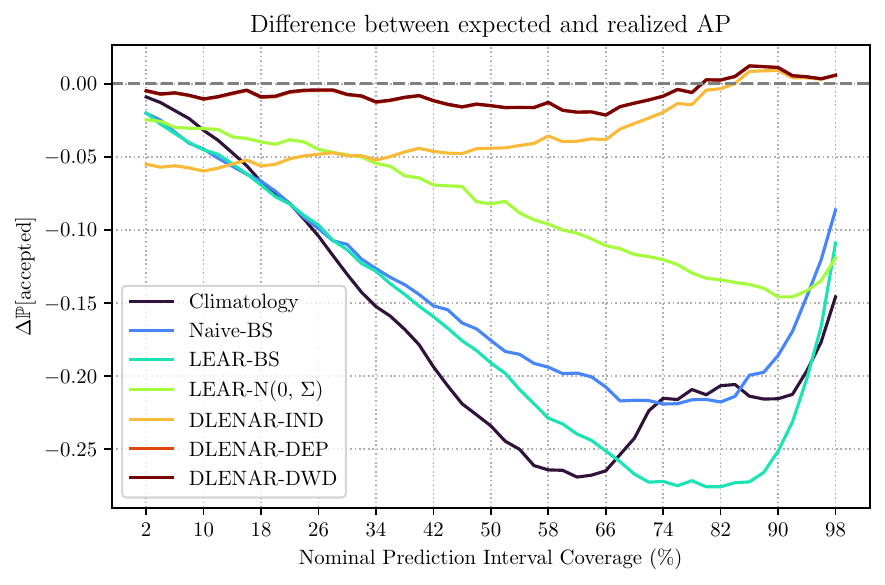}
    \caption{Acceptance probabilities for the QBTS strategy.}
    \label{fig:qbts-accept-prob}
\end{figure}

In this section, we briefly discuss the results of the QBTS for the different forecast models and potentially spurious conclusions. The results are shown in Figures \ref{fig:qbts-profits} and \ref{fig:qbts-accept-prob}, giving the total and per MWh profits and the acceptance probabilities (AP) for the different forecast models. We derive the AP based on the ensemble forecasts, taking the dependence structure into account: $
    \widehat{\text{AP}} = \frac{1}{M} \sum_{m=1}^M 
        \mathbb{I}\{
            \hat{Q}^{1-\alpha}_s \leq \widehat{\mat{F}}_{s,m} \cap 
            \hat{Q}^{\alpha}_b \geq \widehat{\mat{F}}_{b,m}
        \}.
$ We denote the nominal prediction interval coverage as $\lambda = 1 - 2 \alpha$, as we have $Q_b^{1-\alpha}$ and $Q_s^{\alpha}$ as upper and lower bounds for the QBTS.
The DLENAR models yield the highest total profits for the QBTS, followed by the LEAR-based models. The LEAR-BS and the Naive-BS are close, while the climatology model yields the lowest profits across all values of $\alpha$. In terms of the per-MWh profits, the LEAR-based models yield higher profits for small nominal prediction intervals, while the DLENAR models yield higher profits for larger nominal prediction intervals. The last panel gives the difference between expected profits and realized profits. Here, most models yield higher than expected profits. This effect is, however, defined by three effects: the level accuracy, the over- and underdispersion of the forecasts and the specification of the dependence structure. The difference is the largest for low nominal coverage and decreases with higher coverage. Interestingly, for the climatology model, the difference increases again for $\lambda > 0.8$, potentially due to the insufficient point accuracy of the model.

In Figure \ref{fig:qbts-accept-prob} we compare the realized acceptance probabilities with the expected AP under independence, and the expected AP under the dependence structure of the forecasts. The realized AP are below the expected AP under independence, which is in line with the fact that the QBTS does not take the dependence structure into account and hence overestimates the AP. This is in line with the results from Section \ref{sec:ebts-theory}, especially the Remark \ref{result:dependence_structure} and Figure \ref{fig:qbts_simulation_2}. Especially for the Climatology model, we can also see the clear correlation between the AP and the profits. Around $\lambda = 0.65$, the empirical AP increases and the profits increase as well, but due to overdispersion, also more loss-making trades are accepted and the profits per MWh decrease. The difference between the expected AP under the forecast $\widehat{\mat{F}}$ and the realized AP in the second panel of Figure \ref{fig:qbts-accept-prob} is defined by three sources of error. For the DLENAR-IND model, we see that the difference is close to the theoretical AP in the left panel, since the model assumes independence. The DLENAR-DWD and DLENAR-DEP have the lowest difference between expected and empirical AP. The largest differences appear for the Climatology, the LEAR-BS and the Naive-BS, while the LEAR-N(0, $\Sigma$) is somewhat in the middle. Generally, we see that for higher nominal prediction interval coverage, the difference increases.

These results emphasize that the QBTS can lead to hard-to-interpret and potentially spurious conclusions about the performance of probabilistic forecasts. Disentangling the different effects of level accuracy, over- and underdispersion and the dependence structure is not straightforward, but crucial for a proper interpretation of the results.

\section{Conclusion}\label{sec:conclusion}

Probabilistic forecasts are often proposed to enhance decision making, especially under risk. The economic evaluation of (probabilistic) forecasts is increasingly popular in electricity price forecasting and battery storage arbitrage has emerged as natural showcase application for the economic gains of better forecasts. However, the economic evaluation of forecasts, especially of probabilistic forecasts in a risk-averse setting, is not straightforward. This paper addresses two issues: \begin{itemize}
    \item We analyze the properties of the popular quantile-based trading strategy \cite[QBTS, see][]{uniejewski2025smoothing,o2025optimising} and show that it is not a proper scoring rule for the evaluation of forecasts, as systematic distortions of the forecasts can lead to higher profits. We further show that QBTS cannot anticipate diversification benefits through the correlation structure of the forecasts, which can lead to suboptimal decisions and thus to spurious conclusions about the performance of forecasts.
    \item Based on this insight, we discuss the evaluation of fully multivariate, probabilistic forecasts through battery trading strategies, framed as stochastic programming problems. We show that battery trading strategies can have low discriminatory power for the evaluation of forecasts. We discuss measures to analyze the decision quality in a risk-averse setting and use these to analyze the performance of different multivariate forecasts in a large-scale simulation study across various battery configurations.
 \end{itemize}
Our results shed light on the connection between statistical scoring rules for probabilistic forecasts and measures of decision quality for stochastic, risk-averse optimization problems. Similar to previous work on point forecasting \citep[see e.g][]{nitka2023combining,serafin2025data}, we see that measures of decision quality and scoring rules disagree also in the probabilistic setting. While we acknowledge that the results of this paper are based on a specific case study, we believe that the insights are more general and can be applied to other settings as well. Our research touches upon a number of potential avenues for future research. \begin{itemize}
    \item Strictly proper scoring rules give a principled way for the estimation of statistical models and better scores should translate into better decision quality. One avenue of further research is the development of new, objective-aligned evaluation measures, such as proposed by \cite{maciejowska2025statistical} for point forecasts. On the other side, the properties of established scoring rules such as the CRPS, DSS, and ES with respect to locality, symmetry and discrimination are continuously studied \citep{buchweitz2025asymmetric,marcotte2023regions,alexander2024evaluating} and their connection to decision quality is not yet fully understood.
    \item The evaluation of the decision is based on trades and their profits. However, the decision not to trade (in a certain hour) can also be a valuable decision, but is not as prominently captured in the evaluation. This hides part of the forecaster's skill, as one needs a good forecast to decide not to trade. The development of diagnostic tools to capture the value-add of improved forecasts in this area is an interesting avenue for future research.
    \item Our findings have further implications for the integration of decision and forecasting, e.g. in the setting of end-to-end learning \citep{donti2017task,elmachtoub2022smart, wen2025value}, where the low discriminatory power of the decision-based evaluation can hint at potentially demanding learning surfaces. On the other hand, the fact that only a small part of the full price distribution is relevant for the decision can be used to develop more efficient learning algorithms if the relevant part of the distribution can be identified a priori. The connection between the rank- and association-based scoring rules and the relationship between battery optimization and portfolio management suggest that the application of learning-to-rank algorithms could be an interesting avenue for future research as well \citep[see e.g.][for applications in financial portfolio management]{song2017stock, zhang2022constructing}.
\end{itemize}
We conclude by summarizing the key findings of this paper in a recommendation: the economic evaluation of probabilistic forecasts should be treated as an additional layer, not only as forecast evaluation, but as decision quality evaluation. A robust procedure therefore combines (a) forecast evaluation using (strictly) proper scoring rules, and (b) objective-aligned decision diagnostics, and (c) the evaluation of economic performance measures. Agreement across all measures gives strong evidence for model quality, while disagreement can give insights into the strengths and weaknesses of the different models.

\section*{Acknowledgments}

Simon Hirsch is employed as industrial Ph.D. student with Statkraft Trading GmbH and gratefully acknowledges the funding and support received. Florian Ziel acknowledges funding in the course of TRR 391 Spatio-temporal Statistics for the Transition of Energy and Transport (520388526) by the Deutsche Forschungsgemeinschaft (DFG, German Research Foundation). The authors are grateful for interesting discussions with Daniel Gruhlke and David Wozabal.

\section*{Declaration of generative AI and AI-assisted technologies in the manuscript preparation process}

During the preparation of this work the authors used Github Copilot (multiple LLM models) in order to improve the quality of the code, figures, tables and text. After using this tool/service, the authors reviewed and edited the content as needed and take full responsibility for the content of the published article.

\section*{CRediT Authorship Contribution Statement}

Simon Hirsch: Conceptualization; Data curation; Formal analysis; Investigation; Methodology; Software; Validation; Visualization; Roles/Writing - original draft. 

Florian Ziel: Conceptualization; Funding acquisition; Methodology; Project administration; Resources; Supervision; Validation; Writing - review \&{} editing.

\bibliographystyle{abbrvnat}
\setlength{\bibsep}{0pt}
\renewcommand*{\bibfont}{\footnotesize}
\bibliography{refs}

\appendix

\section{Further Information on Forecasting Models}\label{app:equations}

We discuss the derivation for the marginal cost calculation. We have the following $\text{CO}_2$ factors and unit conversion factors given the in the following table \citep[][p. 37]{ghelasi2025data,ghelasi2025day, department2024government}.

\newcommand{\EUR}{\text{EUR}}
\newcommand{\MWht}{\ensuremath{{\text{MWh}_\text{t}}}}
\newcommand{\MWhe}{\ensuremath{{\text{MWh}_\text{e}}}}
\newcommand{\bbl}{\ensuremath{{\text{bbl}}}}
\newcommand{\tCO}{\ensuremath{{\text{tCO}_2}}}
\newcommand{\ton}{\ensuremath{{\text{t}}}}
\newcommand{\fuel}{\ensuremath{{\text{Fuel}}}}
\newcommand{\unit}{\ensuremath{{\text{Unit}}}}

\begin{table}[thb]
    \centering
    \begin{tabular}{lccccccccc}
        \toprule
        Fuel & Price & Unit & $\text{CO}_2$ Price & Unit & Conversion ($\nu$) & Unit & $\text{CO}_2$ Factor ($\eta$)  & Unit \\
        \midrule
        Gas &   $P^\gas_{d-2} $ & $\frac{\EUR}{\MWht}$ & $P^\eua_{d-2} $ & $\frac{\EUR}{\tCO}$ 
            & - & - & 0.2  & $\frac{\tCO}{\MWht}$ \\ [0.5em]
        Coal & $P^\coal_{d-2}$ & $\frac{\EUR}{\ton}$ & $P^\eua_{d-2} $ & $\frac{\EUR}{\tCO}$ 
            & $\frac{1}{8.144}$ & $\frac{\ton}{\MWht}$ & 0.3  & $\frac{\tCO}{\MWht}$ \\ [0.5em]
        Oil & $P^\oil_{d-2} $ & $\frac{\EUR}{\bbl}$ & $P^\eua_{d-2} $ & $\frac{\EUR}{\tCO}$ 
            & $\frac{1}{1.17}$ & $\frac{\bbl}{\MWht}$ & 0.27  & $\frac{\tCO}{\MWht}$ \\
        \bottomrule
    \end{tabular}
\end{table}
and the marginal cost is calculated as \begin{alignat}{6}
    \beta \cdot \text{MC}_{d,h} &= 
        & \underbrace{\nu^\fuel} \cdot
        & \underbrace{P^\fuel} + 
        & \underbrace{\eta^\fuel} \cdot
        & \underbrace{P^\eua} \\
        & & {\dfrac{\unit}{\MWht}} \cdot & {\dfrac{\EUR}{\unit}} +
        & {\dfrac{\tCO}{\MWht}} \cdot & {\dfrac{\EUR}{\tCO}}
        = {\dfrac{\EUR}{\MWht}}
\end{alignat}
and given in $\frac{\EUR}{\MWht}$. The conversion to $\frac{\EUR}{\MWhe}$ is implicitly accounted for in the regression coefficients.
In the following, we present and discuss the equations for the remaining benchmark models, the LEAR model and the scale and degrees of freedom parameters for the distributional regression model. The equations for the point forecasts of the DLENAR models are given in the main text in Section \ref{sec:case_study-forecasting_model}. The climatology model is given by
$
    P_{d,h} = \frac{1}{|T|}\sum_{t \in T} P_{t,h} + \epsilon_{d,h}
$
where $\epsilon_{d}$ is a bootstrap sample from the training sample.

The Naive-BS \citep{ziel2018day,nowotarski2018recent} model is given by $
    P_{d,h} = P_{d-7,h} + \epsilon_{d,h} $
where $\epsilon_{d}$ is a bootstrap sample from the training sample. The LEAR model \citep[introduced by][for point forecasting]{lago2021forecasting} is given by
\begin{align}
    \mu_{d,h} 
        &= \beta_{d,h,0} 
        + \underbrace{\sum_{i=0}^{i=23}\beta_{d,h,i}P_{d,i}}_\text{Prices of previous day.} 
        + \underbrace{\sum_{l=2}^{l=14}\beta_{d,h,24+l}P_{d-l,h}}_\text{Time series effects.} 
        + \underbrace{\sum_{\text{WD} \in \mathcal{W}} \beta_{d,h,38+i} \operatorname{WD}(d)}_\text{Weekday effects.}
        + \underbrace{\beta_{d,h,46} \resload_{d,h}}_\text{Residual load effects.} \\ \nonumber
        &+ \underbrace{\beta_{d,h,47} P^\gas_{d-2} + \beta_{d,h,48} P^\coal_{d-2}  + \beta_{d,h,49} P^\oil_{d-2} + \beta_{d,h,50} P^\eua_{d-2}}_\text{Fundamental Fuel Prices} \nonumber
\end{align}
where use the multivariate Normal for sampling, estimated on the training set (LEAR-N(0, $\Sigma$)) or use a bootstrap sample from the training set (LEAR-BS), where we sample residual price paths using a simple importance sampling. We cluster the in-sample residuals using a k-means clustering approach and use a n-nearst neigher classifier to predict the cluster for the out-of-sample day and sample from the corresponding cluster of residuals.

For the distributional regression model, the scale parameter $\sigma_{d,h}$ is modelled as follows:
\begin{align}\label{eq:model_sigma}
    g_\sigma(\sigma_{d,h}) 
        &= \beta_{d,h,0} 
        + \sum_{i=0}^{i=23}\beta_{d,h,i}P_{d,i} 
        + \sum_{l=2}^{l=14}\beta_{d,h,24+l}P_{d-l,h} 
        + \sum_{\text{WD} \in \mathcal{W}} \beta_{d,h,38+i} \operatorname{WD}(d) \\
        &+ \beta_{d,h,46} \resload_{d,h}
        + \beta_{d,h,47} P^\gas_{d-2}
        + \beta_{d,h,48} P^\coal_{d-2}
        + \beta_{d,h,49} P^\oil_{d-2}
        + \beta_{d,h,50} P^\eua_{d-2} \nonumber
\end{align}
and the degrees of freedom $\nu_{d,h}$ are modelled as intercept only:
$g_\nu(\nu_{d,h}) = \beta_{d,h,0}$
with the inverse softplus link function $g_\sigma^{-1}(x) = \log(1 + \exp(x))$ and the inverse softplus shifted by two link function $g_\nu^{-1}(x) = \log(1 + \exp(x)) + 2$ to ensure positivity of the scale and degrees of freedom parameters \citep{hirsch2024online,hirsch2025online,marcjasz2023distributional}.

\clearpage


\section*{Supplementary material}

# 2 Generation of Scenarios

For the conducted experiments we needed to generate scenarios with different dependence structures, but otherwise as similar features as possible, especially with respect to empirical sample means, as well as sample quantiles/minima to ensure that the differences in objective value are due to the changed distributional property rather than random noise. Note that we draw $M = 2500$ scenarios for $H = 24$ hours and $T = 1831$ out-of-sample days, which gives close to 109 Mio random numbers and with highly volatile energy markets and heavy tailed distributions such as the $t$-distribution, outliers are expected. However, for a CVAR optimization at the $\alpha = 0.1$ level, the optimization maximizes the returns on the worst $2500 \times 0.1 = 250$ samples and hence, the impact of outliers on the potential decisions can be sizeable. To ensure we have the exact same marginal samples, we follow the following steps

1. Sample from the *independent* uniform normal $\mathbf{U}_h^{M \times H} \sim \mathcal{U}(0,1)$.
2. Sample from the *dependent* multivariate normal $\mathbf{N}^{M \times H} \sim N(\boldsymbol{\mu}, \boldsymbol{\Sigma})$.
3. Transform $\mathbf{N}$ to the Copula space via the PIT $\mathbf{U}^{[\mathbf{N}]} = \mathcal{N}^{-1}(\mathbf{N})$
4. Calculate the ranks of each column of $\mathbf{U}^{[\mathbf{N}]}$, i.e. $\mathbf{R}_h^{[\mathbf{N}]} = \text{Rank}(\mathbf{U}_h^{[\mathbf{N}]})$.
5. Reorder the rows of each column of $\widehat{\mathbf{U}}$ according to the ranks in $\mathbf{R}_h^{[\mathbf{N}]}$, i.e. $\widehat{\mathbf{U}}_h = \text{Order}(\mathbf{U}_h, \mathbf{R}_h^{[\mathbf{N}]})$.
6. Use the inverse PIT to transform back to the original scale, i.e. $\widehat{\mathbf{F}}_h = \mathcal{F}^{-1}(\widehat{\mathbf{U}}_h)$.

This procedure ensures that the generated scenarios have the same marginal samples, but different dependence structures, which allows us to isolate the effect of dependence on the performance of the trading strategies.

# 3 Algorithms

---
**Algorithm 1:** Quantile-based trading strategy (QBTS).
---
**Input:** Quantile forecasts $\widehat{Q}_h$, parameter $\alpha$.
1 Choose $b = \arg\min(\widehat{Q}_h^{0.5})$ and $s = \arg\max(\widehat{Q}_h^{0.5})$ with $b < s$.
2 Place a limit buy order at hour $b$ with price $\widehat{Q}_b^{1-\alpha}$ and a limit sell order at hour $s$ with price $\widehat{Q}_s^{\alpha}$.
3 Accept bid if $P_b \leq \widehat{Q}_b^{1-\alpha}$ and $P_s \geq \widehat{Q}_s^{\alpha}$.
**Output:** Return battery profits (adjusted for efficiency) if bid is accepted, 0 else.
---

---
**Algorithm 2:** TS-1 strategy (O'Connor et al., 2024, 2025a,b).
---
**Input:** Quantile forecasts $\widehat{Q}_h$, parameter $\alpha$.
1 Choose $b = \arg\min(\widehat{Q}_h^{\alpha})$ and $s = \arg\max(\widehat{Q}_h^{1-\alpha})$ with $b < s$.
2 Place an unlimited buy order at hour $b$ and an unlimited sell order at hour $s$.
**Output:** Return battery profits (adjusted for efficiency).
---

---
**Algorithm 3:** Dynamic Programming for Battery Trading using Multivariate Probabilistic Forecasts
---
**Input:** Multivariate probabilistic forecast $\widehat{\mathbf{F}}$ with $M$ scenarios, battery parameters $\kappa, \eta$, risk measure $\rho(\cdot)$.
1 For each scenario $m \in 1, ..., M$ calculate $\widehat{R}_{bsm}^{[DPS]} = -\mathbf{a}(b,s)^\top \widehat{\mathbf{F}}_m$ ;
2 Calculate the discretized version of $\widehat{V}_{b,s}^{[DPS]} = \rho(\widehat{R}_{bsm}^{[DPS]})$ ;
3 Choose $\arg\max(V_{b,s}^{[DPS]})$ ;
**Output:** Optimal buy and sell hour $b^*, s^*$.
---


\end{document}

%% file: tikz/no_proper_scoring_rule.tex
\begin{tikzpicture}[scale=1.2,>=stealth]

  \def\xone{2}
  \def\xtwo{4}
  \def\yone{1}
  \def\ytwo{3}

  \def\axislength{5}

  \tikzset{
    areaA/.style={
      fill=red!10,
      pattern=north east lines,
      pattern color=red!50,
      draw=red!80,
      thick
    },
    areaB/.style={
      fill=blue!10,
      pattern=crosshatch,
      pattern color=blue!50,
      draw=blue!80,
      thick
    },
  }

  \draw[dashed] (0,\yone) -- (\axislength,\yone);
  \draw[dashed] (0,\ytwo) -- (\axislength,\ytwo);
  \draw[dashed] (\xone,0) -- (\xone,\axislength);
  \draw[dashed] (\xtwo,0) -- (\xtwo,\axislength);

  \filldraw[areaA] (0,\yone) rectangle (\xtwo,\axislength);
  \filldraw[areaB] (0,\ytwo) rectangle (\xone,\axislength);

  \node[below] at (\xone,0) {$Q^{1-\alpha}$};
  \node[below] at (\xtwo,0) {$\tilde{Q}^{1-\alpha}$};
  \node[left] at (0,\yone) {$\tilde{Q}^{\alpha}$};
  \node[left] at (0,\ytwo) {$Q^{\alpha}$};

  \node[red, align=center] at (\xtwo + 1, \ytwo + 1) {Acceptance \\ region for \\ overdispersed \\ forecasts};
  \node[blue, align=center] at (1, \ytwo + 1) {True \\ acceptance \\ Region};
  \draw[decorate,decoration={brace, amplitude=6pt, mirror}]
    (\xone,-0.5) -- (\xtwo,-0.5)
    node[midway,below=6pt] {Overdispersion $\tilde{Q}^{1-\alpha} > Q^{1-\alpha}$};
  \draw[decorate,decoration={brace, amplitude=6pt}]
    (-0.6, \yone) -- (-0.6,\ytwo)
      node[midway,left=15pt,rotate=90,anchor=center] {Overdispersion $\tilde{Q}^{\alpha} < Q^{\alpha}$};
    
  \draw[very thick, ->] (-0.1,0) -- (\axislength,0) node[right] {Buy Price $P_b$};
  \draw[very thick, ->] (0,-0.1) -- (0,\axislength) node[above] {Sell Price $P_s$};

\end{tikzpicture}

%% file: tikz/ensemble_approach.tex
\tikzset{
  boxwidth/.store in=\boxwidth,
  boxheight/.store in=\boxheight,
  headerheight/.store in=\headerheight,
}

\tikzset{
  boxwidth=5cm,
  boxheight=6cm,
  headerheight=1cm,
}

\tikzset{
  boxwithheader/.style={
    draw,
    thick,
    minimum width=\boxwidth,
    minimum height=\boxheight,
    inner sep=0pt,
    font=\sffamily,
  },
  header/.style={
    draw, 
    thick,
    fill=blue!20,
    text centered,
    font=\normalfont,
    text width=\boxwidth,
    align=center,
    minimum width=\boxwidth,
    minimum height=\headerheight,
    inner sep=0pt,
  },
}

\begin{tikzpicture}[node distance=0.8cm and 0.8cm]

    \pgfmathsetlengthmacro{\figwidth}{0.9*\boxwidth}
  \node[boxwithheader] (b1) {};
  \node[header, anchor=north west] at (b1.north west)
    {Scenario Forecasts};

  \path let \p1 = (b1.north), \p2 = (b1.south) in
    coordinate (center1) at ($(b1.south)!0.5!(b1.north)+(0,-\headerheight/2)$);
  \node[anchor=center] at (center1)
    {\includegraphics[width=\figwidth, keepaspectratio]{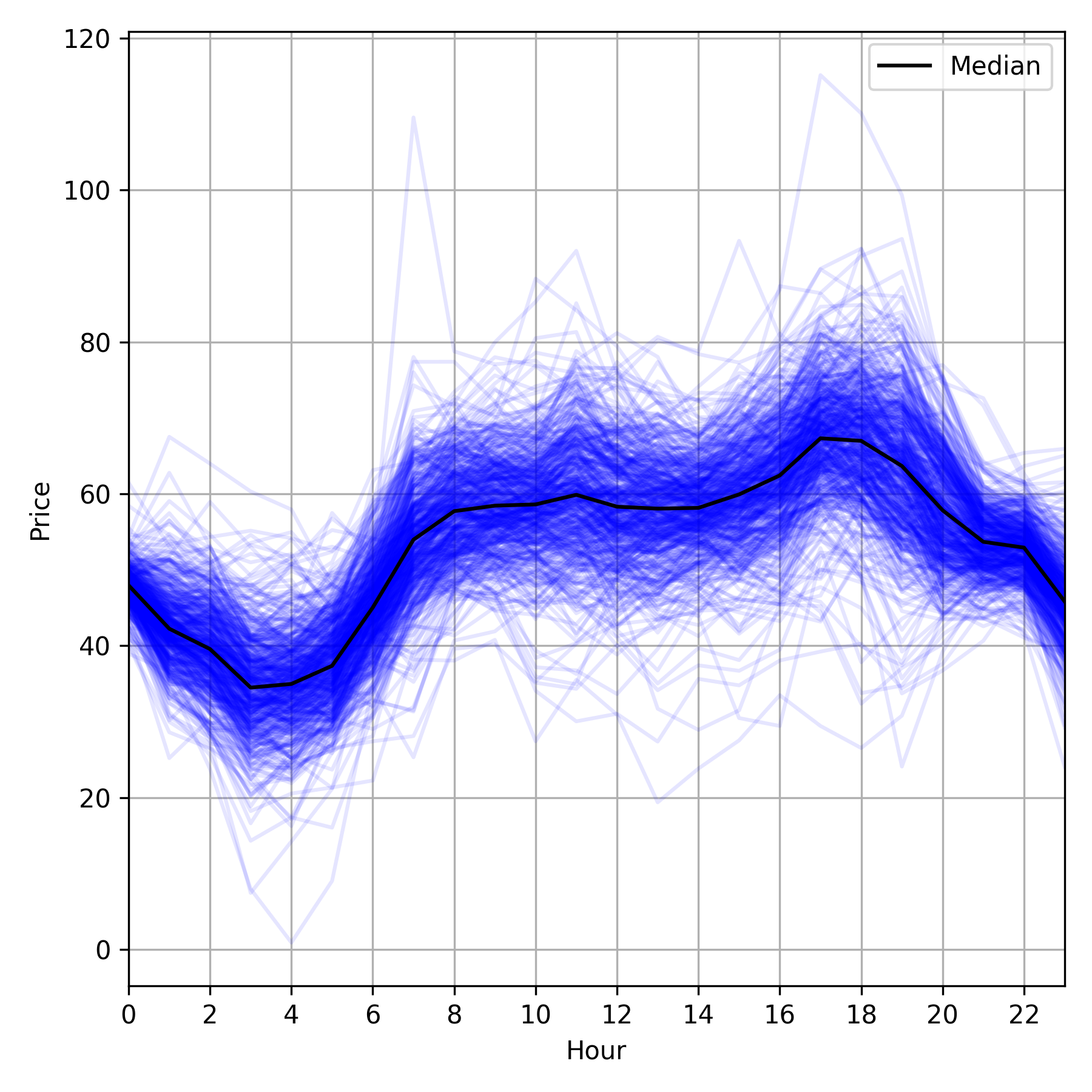}};

  \node[boxwithheader, right=of b1] (b2) {};
  \node[header, anchor=north west] at (b2.north west)
    {Calculate Profit distributions for all $s$ and $b$.};

  \path let \p1 = (b2.north), \p2 = (b2.south) in
    coordinate (center2) at ($(b2.south)!0.5!(b2.north)+(0,-\headerheight/2)$);
  \node[anchor=center] at (center2)
    {\includegraphics[width=\figwidth, keepaspectratio]{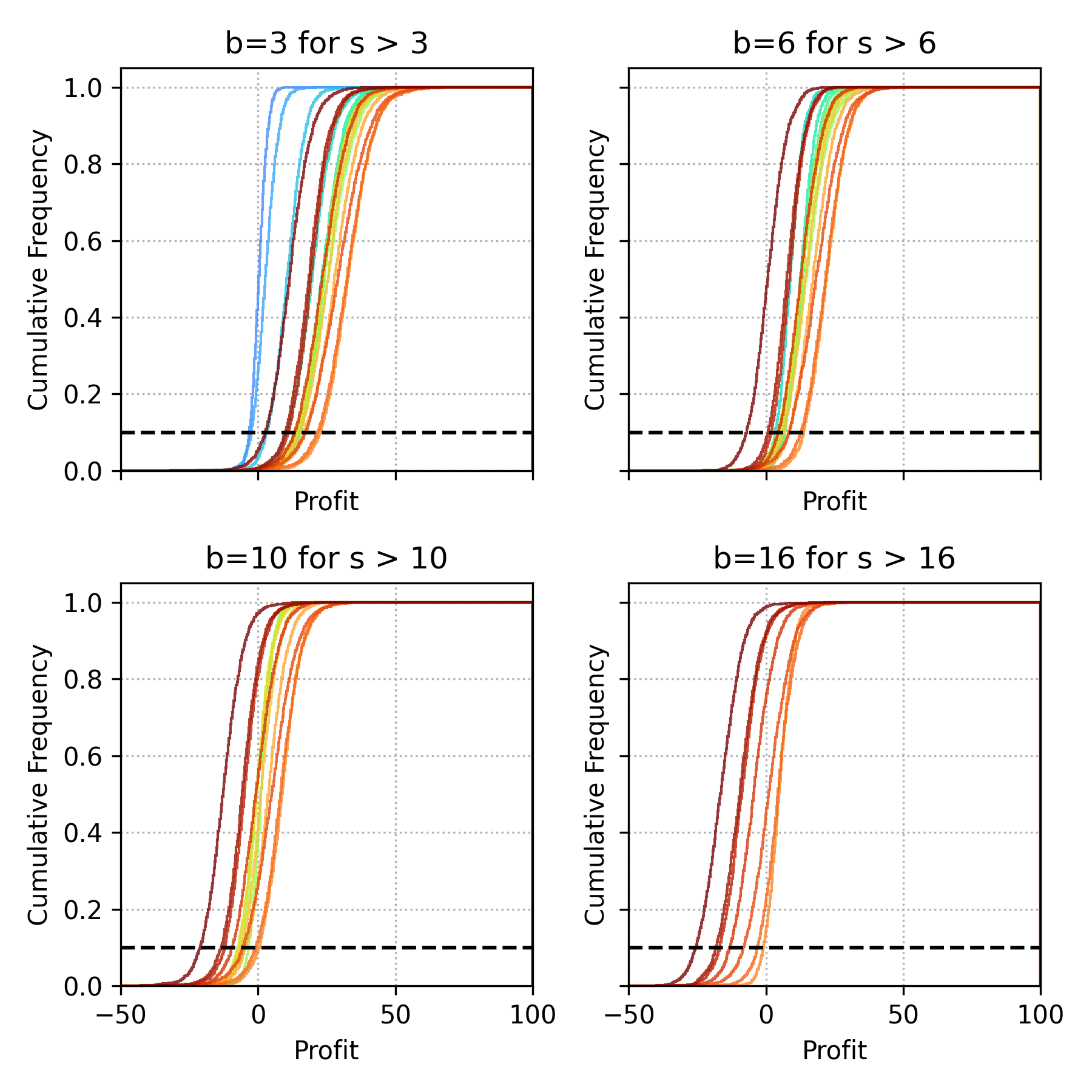}};

  \node[boxwithheader, right=of b2] (b3) {};
  \node[header, anchor=north west] at (b3.north west)
    {Evaluate risk measure on all $b, s$ pairs.};

  \path let \p1 = (b3.north), \p2 = (b3.south) in
    coordinate (center3) at ($(b3.south)!0.5!(b3.north)+(0,-\headerheight/2)$);
  \node[anchor=center] at (center3)
    {\includegraphics[width=\figwidth, keepaspectratio]{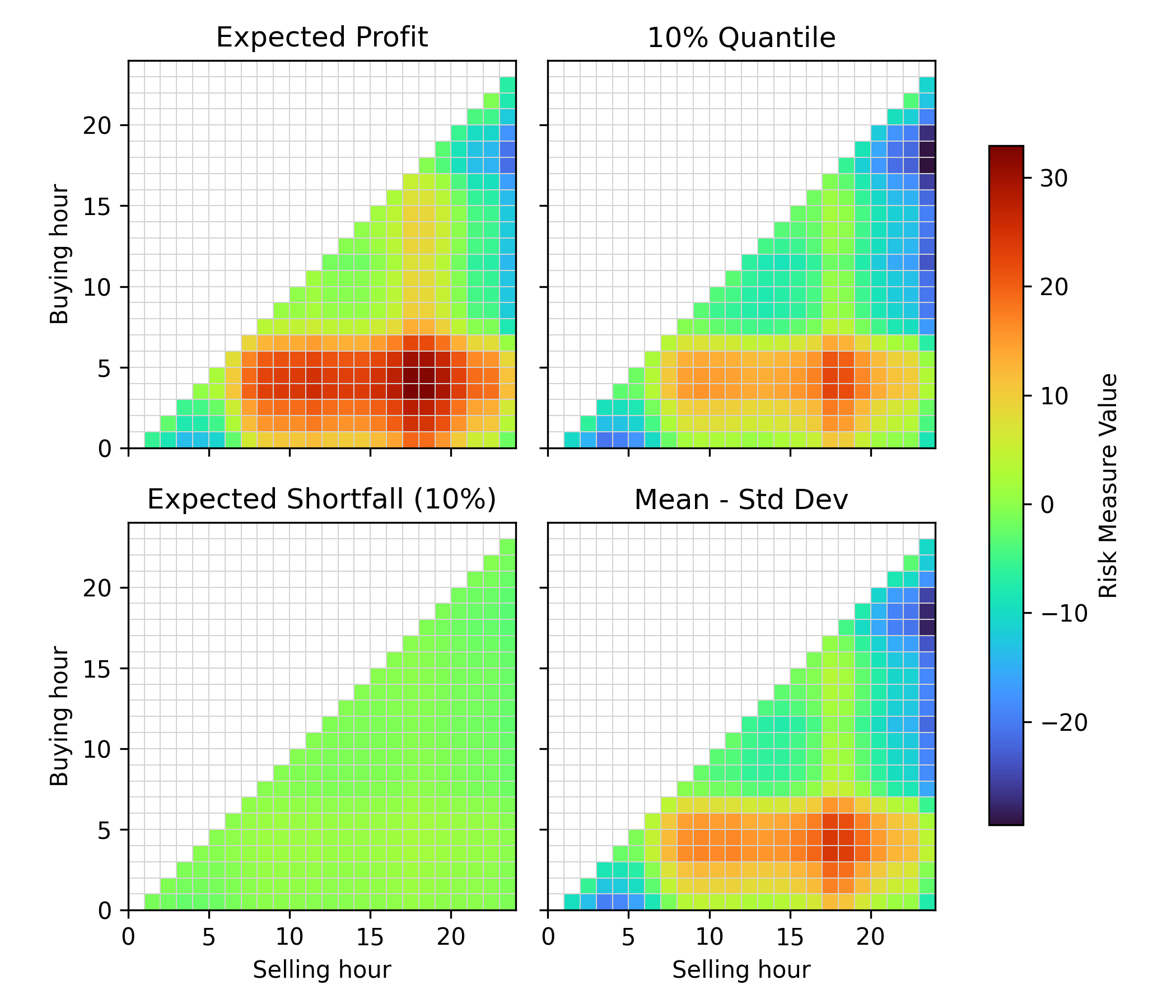}};

\end{tikzpicture}

%% file: tables/forecast_scores.tex
\begin{tabular}{l*{12}{R{1.4cm}}}
\toprule
 & MAE & $\sqrt{\text{RMSE}}$ & CRPS & $\text{VS}_{0.5}$ & $\text{VS}_{1.0}$ & DSS & ES & Brier & RPS & KS & MPD & MHD \\
\midrule
Climatology & {\cellcolor[HTML]{000004}} \color[HTML]{F1F1F1} 69.766 & {\cellcolor[HTML]{000004}} \color[HTML]{F1F1F1} 112.770 & {\cellcolor[HTML]{000004}} \color[HTML]{F1F1F1} 56.318 & {\cellcolor[HTML]{000004}} \color[HTML]{F1F1F1} 9.072 & {\cellcolor[HTML]{000004}} \color[HTML]{F1F1F1} 1694.479 & {\cellcolor[HTML]{2F1163}} \color[HTML]{F1F1F1} 176.766 & {\cellcolor[HTML]{000004}} \color[HTML]{F1F1F1} 291.149 & {\cellcolor[HTML]{451077}} \color[HTML]{F1F1F1} 0.931 & {\cellcolor[HTML]{000004}} \color[HTML]{F1F1F1} 2.907 & {\cellcolor[HTML]{000004}} \color[HTML]{F1F1F1} 0.831 & {\cellcolor[HTML]{000004}} \color[HTML]{F1F1F1} 29.641 & {\cellcolor[HTML]{000004}} \color[HTML]{F1F1F1} 11.117 \\
Naive-BS & {\cellcolor[HTML]{8C2981}} \color[HTML]{F1F1F1} 34.487 & {\cellcolor[HTML]{802582}} \color[HTML]{F1F1F1} 61.676 & {\cellcolor[HTML]{7E2482}} \color[HTML]{F1F1F1} 29.235 & {\cellcolor[HTML]{732081}} \color[HTML]{F1F1F1} 5.416 & {\cellcolor[HTML]{5F187F}} \color[HTML]{F1F1F1} 1119.829 & {\cellcolor[HTML]{000004}} \color[HTML]{F1F1F1} 204.941 & {\cellcolor[HTML]{762181}} \color[HTML]{F1F1F1} 160.579 & {\cellcolor[HTML]{000004}} \color[HTML]{F1F1F1} 0.986 & {\cellcolor[HTML]{0E0B2B}} \color[HTML]{F1F1F1} 2.690 & {\cellcolor[HTML]{0C0926}} \color[HTML]{F1F1F1} 0.807 & {\cellcolor[HTML]{390F6E}} \color[HTML]{F1F1F1} 21.772 & {\cellcolor[HTML]{08071E}} \color[HTML]{F1F1F1} 10.484 \\
LEAR-BS & {\cellcolor[HTML]{FEB078}} \color[HTML]{000000} 16.222 & {\cellcolor[HTML]{FEAE77}} \color[HTML]{000000} 30.211 & {\cellcolor[HTML]{FC8961}} \color[HTML]{F1F1F1} 13.530 & {\cellcolor[HTML]{FB835F}} \color[HTML]{F1F1F1} 3.020 & {\cellcolor[HTML]{FB8560}} \color[HTML]{F1F1F1} 602.827 & {\cellcolor[HTML]{241253}} \color[HTML]{F1F1F1} 180.588 & {\cellcolor[HTML]{FA7F5E}} \color[HTML]{F1F1F1} 79.210 & {\cellcolor[HTML]{400F74}} \color[HTML]{F1F1F1} 0.934 & {\cellcolor[HTML]{CF4070}} \color[HTML]{F1F1F1} 1.745 & {\cellcolor[HTML]{D3436E}} \color[HTML]{F1F1F1} 0.693 & {\cellcolor[HTML]{E85362}} \color[HTML]{F1F1F1} 10.441 & {\cellcolor[HTML]{D9466B}} \color[HTML]{F1F1F1} 6.304 \\
LEAR-N(0, $\Sigma$) & {\cellcolor[HTML]{FEAE77}} \color[HTML]{000000} 16.339 & {\cellcolor[HTML]{FEAA74}} \color[HTML]{000000} 30.357 & {\cellcolor[HTML]{FC9065}} \color[HTML]{000000} 13.220 & {\cellcolor[HTML]{FA7F5E}} \color[HTML]{F1F1F1} 3.049 & {\cellcolor[HTML]{FA7D5E}} \color[HTML]{F1F1F1} 611.727 & {\cellcolor[HTML]{20114B}} \color[HTML]{F1F1F1} 182.419 & {\cellcolor[HTML]{FB8761}} \color[HTML]{F1F1F1} 77.292 & {\cellcolor[HTML]{962C80}} \color[HTML]{F1F1F1} 0.893 & {\cellcolor[HTML]{D6456C}} \color[HTML]{F1F1F1} 1.721 & {\cellcolor[HTML]{D9466B}} \color[HTML]{F1F1F1} 0.690 & {\cellcolor[HTML]{E75263}} \color[HTML]{F1F1F1} 10.537 & {\cellcolor[HTML]{D8456C}} \color[HTML]{F1F1F1} 6.322 \\
DLENAR-IND & {\cellcolor[HTML]{FCFDBF}} \color[HTML]{000000} 14.740 & {\cellcolor[HTML]{FCFDBF}} \color[HTML]{000000} 27.713 & {\cellcolor[HTML]{FCFDBF}} \color[HTML]{000000} 10.769 & {\cellcolor[HTML]{F2645C}} \color[HTML]{F1F1F1} 3.273 & {\cellcolor[HTML]{EE5B5E}} \color[HTML]{F1F1F1} 666.584 & {\cellcolor[HTML]{832681}} \color[HTML]{F1F1F1} 150.644 & {\cellcolor[HTML]{FECF92}} \color[HTML]{000000} 65.028 & {\cellcolor[HTML]{D6456C}} \color[HTML]{F1F1F1} 0.872 & {\cellcolor[HTML]{F7705C}} \color[HTML]{F1F1F1} 1.594 & {\cellcolor[HTML]{F05F5E}} \color[HTML]{F1F1F1} 0.681 & {\cellcolor[HTML]{FCFDBF}} \color[HTML]{000000} 7.420 & {\cellcolor[HTML]{FCFDBF}} \color[HTML]{000000} 5.312 \\
DLENAR-DEP & {\cellcolor[HTML]{FCFDBF}} \color[HTML]{000000} 14.740 & {\cellcolor[HTML]{FCFDBF}} \color[HTML]{000000} 27.713 & {\cellcolor[HTML]{FCFDBF}} \color[HTML]{000000} 10.769 & {\cellcolor[HTML]{FCECAE}} \color[HTML]{000000} 2.595 & {\cellcolor[HTML]{FDEBAC}} \color[HTML]{000000} 527.446 & \bfseries {\cellcolor[HTML]{FCFDBF}} \color[HTML]{000000} 116.299 & {\cellcolor[HTML]{FCFBBD}} \color[HTML]{000000} 62.730 & {\cellcolor[HTML]{FDE3A5}} \color[HTML]{000000} 0.848 & \bfseries {\cellcolor[HTML]{FCFDBF}} \color[HTML]{000000} 1.465 & \bfseries {\cellcolor[HTML]{FCFDBF}} \color[HTML]{000000} 0.662 & {\cellcolor[HTML]{FCFDBF}} \color[HTML]{000000} 7.420 & {\cellcolor[HTML]{FCFDBF}} \color[HTML]{000000} 5.312 \\
DLENAR-DWD & {\cellcolor[HTML]{FCFDBF}} \color[HTML]{000000} 14.740 & \bfseries {\cellcolor[HTML]{FCFDBF}} \color[HTML]{000000} 27.713 & {\cellcolor[HTML]{FCFDBF}} \color[HTML]{000000} 10.769 & \bfseries {\cellcolor[HTML]{FCFDBF}} \color[HTML]{000000} 2.587 & \bfseries {\cellcolor[HTML]{FCFDBF}} \color[HTML]{000000} 525.592 & {\cellcolor[HTML]{FDE3A5}} \color[HTML]{000000} 116.599 & \bfseries {\cellcolor[HTML]{FCFDBF}} \color[HTML]{000000} 62.721 & \bfseries {\cellcolor[HTML]{FCFDBF}} \color[HTML]{000000} 0.847 & {\cellcolor[HTML]{FCF6B8}} \color[HTML]{000000} 1.466 & {\cellcolor[HTML]{FCF2B4}} \color[HTML]{000000} 0.662 & {\cellcolor[HTML]{FCFDBF}} \color[HTML]{000000} 7.420 & {\cellcolor[HTML]{FCFDBF}} \color[HTML]{000000} 5.312 \\
\bottomrule
\end{tabular}

%% file: tables/top_k_brier.tex
\begin{tabular}{lrrrrrrrrrrrrrrrr}
\toprule
 & \multicolumn{4}{l}{Low-$k$} & \multicolumn{4}{l}{High-$k$} & \multicolumn{4}{l}{Low-High-$k$} & \multicolumn{4}{l}{BESS-$k$} \\
 & $k=1$ & $k=2$ & $k=4$ & $k=8$ & $k=1$ & $k=2$ & $k=4$ & $k=8$ & $k=1$ & $k=2$ & $k=4$ & $k=8$ & $k=1$ & $k=2$ & $k=4$ & $k=8$ \\
\midrule
Climatology & {\cellcolor[HTML]{792282}} \color[HTML]{F1F1F1} 0.88 & {\cellcolor[HTML]{701F81}} \color[HTML]{F1F1F1} 0.89 & {\cellcolor[HTML]{641A80}} \color[HTML]{F1F1F1} 0.90 & {\cellcolor[HTML]{4F127B}} \color[HTML]{F1F1F1} 0.92 & {\cellcolor[HTML]{942C80}} \color[HTML]{F1F1F1} 0.86 & {\cellcolor[HTML]{842681}} \color[HTML]{F1F1F1} 0.87 & {\cellcolor[HTML]{6B1D81}} \color[HTML]{F1F1F1} 0.89 & {\cellcolor[HTML]{51127C}} \color[HTML]{F1F1F1} 0.92 & {\cellcolor[HTML]{882781}} \color[HTML]{F1F1F1} 0.87 & {\cellcolor[HTML]{7B2382}} \color[HTML]{F1F1F1} 0.88 & {\cellcolor[HTML]{671B80}} \color[HTML]{F1F1F1} 0.90 & {\cellcolor[HTML]{4F127B}} \color[HTML]{F1F1F1} 0.92 & {\cellcolor[HTML]{782281}} \color[HTML]{F1F1F1} 0.88 & {\cellcolor[HTML]{6B1D81}} \color[HTML]{F1F1F1} 0.89 & {\cellcolor[HTML]{5D177F}} \color[HTML]{F1F1F1} 0.91 & {\cellcolor[HTML]{471078}} \color[HTML]{F1F1F1} 0.93 \\
Naive-BS & {\cellcolor[HTML]{000004}} \color[HTML]{F1F1F1} 1.01 & {\cellcolor[HTML]{07061C}} \color[HTML]{F1F1F1} 0.99 & {\cellcolor[HTML]{090720}} \color[HTML]{F1F1F1} 0.99 & {\cellcolor[HTML]{090720}} \color[HTML]{F1F1F1} 0.99 & {\cellcolor[HTML]{8E2A81}} \color[HTML]{F1F1F1} 0.86 & {\cellcolor[HTML]{5C167F}} \color[HTML]{F1F1F1} 0.91 & {\cellcolor[HTML]{341069}} \color[HTML]{F1F1F1} 0.94 & {\cellcolor[HTML]{1C1044}} \color[HTML]{F1F1F1} 0.97 & {\cellcolor[HTML]{3B0F70}} \color[HTML]{F1F1F1} 0.94 & {\cellcolor[HTML]{2C115F}} \color[HTML]{F1F1F1} 0.95 & {\cellcolor[HTML]{1C1044}} \color[HTML]{F1F1F1} 0.97 & {\cellcolor[HTML]{120D31}} \color[HTML]{F1F1F1} 0.98 & {\cellcolor[HTML]{0D0A29}} \color[HTML]{F1F1F1} 0.98 & {\cellcolor[HTML]{110C2F}} \color[HTML]{F1F1F1} 0.98 & {\cellcolor[HTML]{0E0B2B}} \color[HTML]{F1F1F1} 0.98 & {\cellcolor[HTML]{0A0822}} \color[HTML]{F1F1F1} 0.99 \\
LEAR-BS & {\cellcolor[HTML]{AA337D}} \color[HTML]{F1F1F1} 0.84 & {\cellcolor[HTML]{8E2A81}} \color[HTML]{F1F1F1} 0.86 & {\cellcolor[HTML]{721F81}} \color[HTML]{F1F1F1} 0.89 & {\cellcolor[HTML]{59157E}} \color[HTML]{F1F1F1} 0.91 & {\cellcolor[HTML]{FD9869}} \color[HTML]{000000} 0.72 & {\cellcolor[HTML]{DC4869}} \color[HTML]{F1F1F1} 0.79 & {\cellcolor[HTML]{902A81}} \color[HTML]{F1F1F1} 0.86 & {\cellcolor[HTML]{57157E}} \color[HTML]{F1F1F1} 0.91 & {\cellcolor[HTML]{E75263}} \color[HTML]{F1F1F1} 0.78 & {\cellcolor[HTML]{B5367A}} \color[HTML]{F1F1F1} 0.82 & {\cellcolor[HTML]{802582}} \color[HTML]{F1F1F1} 0.87 & {\cellcolor[HTML]{59157E}} \color[HTML]{F1F1F1} 0.91 & {\cellcolor[HTML]{9C2E7F}} \color[HTML]{F1F1F1} 0.85 & {\cellcolor[HTML]{802582}} \color[HTML]{F1F1F1} 0.87 & {\cellcolor[HTML]{621980}} \color[HTML]{F1F1F1} 0.90 & {\cellcolor[HTML]{400F74}} \color[HTML]{F1F1F1} 0.93 \\
LEAR-N(0, $\Sigma$) & {\cellcolor[HTML]{D0416F}} \color[HTML]{F1F1F1} 0.80 & {\cellcolor[HTML]{BA3878}} \color[HTML]{F1F1F1} 0.82 & {\cellcolor[HTML]{9B2E7F}} \color[HTML]{F1F1F1} 0.85 & {\cellcolor[HTML]{802582}} \color[HTML]{F1F1F1} 0.87 & {\cellcolor[HTML]{FEB47B}} \color[HTML]{000000} 0.70 & {\cellcolor[HTML]{F3655C}} \color[HTML]{F1F1F1} 0.76 & {\cellcolor[HTML]{BA3878}} \color[HTML]{F1F1F1} 0.82 & {\cellcolor[HTML]{842681}} \color[HTML]{F1F1F1} 0.87 & {\cellcolor[HTML]{F7705C}} \color[HTML]{F1F1F1} 0.75 & {\cellcolor[HTML]{DB476A}} \color[HTML]{F1F1F1} 0.79 & {\cellcolor[HTML]{AA337D}} \color[HTML]{F1F1F1} 0.84 & {\cellcolor[HTML]{812581}} \color[HTML]{F1F1F1} 0.87 & {\cellcolor[HTML]{C53C74}} \color[HTML]{F1F1F1} 0.81 & {\cellcolor[HTML]{AD347C}} \color[HTML]{F1F1F1} 0.83 & {\cellcolor[HTML]{8C2981}} \color[HTML]{F1F1F1} 0.86 & {\cellcolor[HTML]{6D1D81}} \color[HTML]{F1F1F1} 0.89 \\
DLENAR-IND & {\cellcolor[HTML]{F2625D}} \color[HTML]{F1F1F1} 0.76 & {\cellcolor[HTML]{DC4869}} \color[HTML]{F1F1F1} 0.79 & {\cellcolor[HTML]{B73779}} \color[HTML]{F1F1F1} 0.82 & {\cellcolor[HTML]{962C80}} \color[HTML]{F1F1F1} 0.85 & {\cellcolor[HTML]{FEB77E}} \color[HTML]{000000} 0.69 & {\cellcolor[HTML]{F8765C}} \color[HTML]{F1F1F1} 0.74 & {\cellcolor[HTML]{CD4071}} \color[HTML]{F1F1F1} 0.80 & {\cellcolor[HTML]{9B2E7F}} \color[HTML]{F1F1F1} 0.85 & {\cellcolor[HTML]{FC8C63}} \color[HTML]{F1F1F1} 0.73 & {\cellcolor[HTML]{EE5B5E}} \color[HTML]{F1F1F1} 0.77 & {\cellcolor[HTML]{C23B75}} \color[HTML]{F1F1F1} 0.81 & {\cellcolor[HTML]{982D80}} \color[HTML]{F1F1F1} 0.85 & {\cellcolor[HTML]{E55064}} \color[HTML]{F1F1F1} 0.78 & {\cellcolor[HTML]{CC3F71}} \color[HTML]{F1F1F1} 0.80 & {\cellcolor[HTML]{A8327D}} \color[HTML]{F1F1F1} 0.84 & {\cellcolor[HTML]{862781}} \color[HTML]{F1F1F1} 0.87 \\
DLENAR-DEP & {\cellcolor[HTML]{FC8C63}} \color[HTML]{F1F1F1} 0.73 & {\cellcolor[HTML]{F2625D}} \color[HTML]{F1F1F1} 0.76 & {\cellcolor[HTML]{D6456C}} \color[HTML]{F1F1F1} 0.79 & {\cellcolor[HTML]{B73779}} \color[HTML]{F1F1F1} 0.82 & {\cellcolor[HTML]{FCFDBF}} \color[HTML]{000000} 0.64 & {\cellcolor[HTML]{FEB078}} \color[HTML]{000000} 0.70 & {\cellcolor[HTML]{EE5B5E}} \color[HTML]{F1F1F1} 0.77 & {\cellcolor[HTML]{BA3878}} \color[HTML]{F1F1F1} 0.82 & {\cellcolor[HTML]{FEC68A}} \color[HTML]{000000} 0.68 & {\cellcolor[HTML]{FC8961}} \color[HTML]{F1F1F1} 0.73 & {\cellcolor[HTML]{E34E65}} \color[HTML]{F1F1F1} 0.78 & {\cellcolor[HTML]{B83779}} \color[HTML]{F1F1F1} 0.82 & {\cellcolor[HTML]{F8765C}} \color[HTML]{F1F1F1} 0.74 & {\cellcolor[HTML]{E85362}} \color[HTML]{F1F1F1} 0.77 & {\cellcolor[HTML]{C73D73}} \color[HTML]{F1F1F1} 0.81 & {\cellcolor[HTML]{A02F7F}} \color[HTML]{F1F1F1} 0.84 \\
DLENAR-DWD & {\cellcolor[HTML]{FC8C63}} \color[HTML]{F1F1F1} 0.73 & {\cellcolor[HTML]{F2645C}} \color[HTML]{F1F1F1} 0.76 & {\cellcolor[HTML]{D6456C}} \color[HTML]{F1F1F1} 0.79 & {\cellcolor[HTML]{B83779}} \color[HTML]{F1F1F1} 0.82 & {\cellcolor[HTML]{FCFDBF}} \color[HTML]{000000} 0.64 & {\cellcolor[HTML]{FEB27A}} \color[HTML]{000000} 0.70 & {\cellcolor[HTML]{EE5B5E}} \color[HTML]{F1F1F1} 0.77 & {\cellcolor[HTML]{BA3878}} \color[HTML]{F1F1F1} 0.82 & {\cellcolor[HTML]{FEC68A}} \color[HTML]{000000} 0.68 & {\cellcolor[HTML]{FC8A62}} \color[HTML]{F1F1F1} 0.73 & {\cellcolor[HTML]{E44F64}} \color[HTML]{F1F1F1} 0.78 & {\cellcolor[HTML]{B83779}} \color[HTML]{F1F1F1} 0.82 & {\cellcolor[HTML]{F9785D}} \color[HTML]{F1F1F1} 0.74 & {\cellcolor[HTML]{E85362}} \color[HTML]{F1F1F1} 0.77 & {\cellcolor[HTML]{C73D73}} \color[HTML]{F1F1F1} 0.81 & {\cellcolor[HTML]{A02F7F}} \color[HTML]{F1F1F1} 0.84 \\
\bottomrule
\end{tabular}

%% file: tables/exceedance_value_at_risk_wide.tex
\begin{tabular}{lrrrrrrrrrrrrrrrrrr}
\toprule
Alpha & \multicolumn{6}{c}{$\alpha=0.5$} & \multicolumn{6}{c}{$\alpha=0.75$} & \multicolumn{6}{c}{$\alpha=0.9$} \\
Duration & \multicolumn{2}{c}{$d=1$} & \multicolumn{2}{c}{$d=2$} & \multicolumn{2}{c}{$d=4$} & \multicolumn{2}{c}{$d=1$} & \multicolumn{2}{c}{$d=2$} & \multicolumn{2}{c}{$d=4$} & \multicolumn{2}{c}{$d=1$} & \multicolumn{2}{c}{$d=2$} & \multicolumn{2}{c}{$d=4$} \\
Cycles & $c=1$ & $c=2$ & $c=1$ & $c=2$ & $c=1$ & $c=2$ & $c=1$ & $c=2$ & $c=1$ & $c=2$ & $c=1$ & $c=2$ & $c=1$ & $c=2$ & $c=1$ & $c=2$ & $c=1$ & $c=2$ \\
\midrule
Climatology & {\cellcolor[HTML]{892881}} \color[HTML]{F1F1F1} -0.20 & {\cellcolor[HTML]{471078}} \color[HTML]{F1F1F1} -0.25 & {\cellcolor[HTML]{862781}} \color[HTML]{F1F1F1} -0.20 & {\cellcolor[HTML]{4A1079}} \color[HTML]{F1F1F1} -0.25 & {\cellcolor[HTML]{792282}} \color[HTML]{F1F1F1} -0.21 & {\cellcolor[HTML]{5A167E}} \color[HTML]{F1F1F1} -0.24 & {\cellcolor[HTML]{F4695C}} \color[HTML]{F1F1F1} -0.11 & {\cellcolor[HTML]{E55064}} \color[HTML]{F1F1F1} -0.12 & {\cellcolor[HTML]{F66C5C}} \color[HTML]{F1F1F1} -0.10 & {\cellcolor[HTML]{E85362}} \color[HTML]{F1F1F1} -0.12 & {\cellcolor[HTML]{F66C5C}} \color[HTML]{F1F1F1} -0.10 & {\cellcolor[HTML]{F3655C}} \color[HTML]{F1F1F1} -0.11 & {\cellcolor[HTML]{FECD90}} \color[HTML]{000000} -0.03 & {\cellcolor[HTML]{FEC68A}} \color[HTML]{000000} -0.04 & {\cellcolor[HTML]{FECD90}} \color[HTML]{000000} -0.03 & {\cellcolor[HTML]{FEC68A}} \color[HTML]{000000} -0.04 & {\cellcolor[HTML]{FED194}} \color[HTML]{000000} -0.03 & {\cellcolor[HTML]{FED194}} \color[HTML]{000000} -0.03 \\
Naive-BS & {\cellcolor[HTML]{A02F7F}} \color[HTML]{F1F1F1} 0.18 & {\cellcolor[HTML]{BC3978}} \color[HTML]{F1F1F1} 0.16 & {\cellcolor[HTML]{B0357B}} \color[HTML]{F1F1F1} 0.17 & {\cellcolor[HTML]{C23B75}} \color[HTML]{F1F1F1} 0.16 & {\cellcolor[HTML]{C53C74}} \color[HTML]{F1F1F1} 0.15 & {\cellcolor[HTML]{DE4968}} \color[HTML]{F1F1F1} 0.13 & {\cellcolor[HTML]{02020D}} \color[HTML]{F1F1F1} 0.32 & {\cellcolor[HTML]{0A0822}} \color[HTML]{F1F1F1} 0.31 & {\cellcolor[HTML]{050416}} \color[HTML]{F1F1F1} 0.32 & {\cellcolor[HTML]{130D34}} \color[HTML]{F1F1F1} 0.30 & {\cellcolor[HTML]{20114B}} \color[HTML]{F1F1F1} 0.29 & {\cellcolor[HTML]{400F74}} \color[HTML]{F1F1F1} 0.26 & {\cellcolor[HTML]{000004}} \color[HTML]{F1F1F1} 0.33 & {\cellcolor[HTML]{000004}} \color[HTML]{F1F1F1} 0.33 & {\cellcolor[HTML]{010106}} \color[HTML]{F1F1F1} 0.33 & {\cellcolor[HTML]{06051A}} \color[HTML]{F1F1F1} 0.31 & {\cellcolor[HTML]{0C0926}} \color[HTML]{F1F1F1} 0.31 & {\cellcolor[HTML]{110C2F}} \color[HTML]{F1F1F1} 0.30 \\
LEAR-BS & {\cellcolor[HTML]{FEB77E}} \color[HTML]{000000} 0.05 & {\cellcolor[HTML]{FEB77E}} \color[HTML]{000000} 0.05 & {\cellcolor[HTML]{FEBB81}} \color[HTML]{000000} 0.05 & {\cellcolor[HTML]{FEBF84}} \color[HTML]{000000} 0.05 & {\cellcolor[HTML]{FEC68A}} \color[HTML]{000000} 0.04 & {\cellcolor[HTML]{FEC68A}} \color[HTML]{000000} 0.04 & {\cellcolor[HTML]{9C2E7F}} \color[HTML]{F1F1F1} 0.18 & {\cellcolor[HTML]{A3307E}} \color[HTML]{F1F1F1} 0.18 & {\cellcolor[HTML]{A3307E}} \color[HTML]{F1F1F1} 0.18 & {\cellcolor[HTML]{A3307E}} \color[HTML]{F1F1F1} 0.18 & {\cellcolor[HTML]{A6317D}} \color[HTML]{F1F1F1} 0.18 & {\cellcolor[HTML]{B0357B}} \color[HTML]{F1F1F1} 0.17 & {\cellcolor[HTML]{762181}} \color[HTML]{F1F1F1} 0.22 & {\cellcolor[HTML]{641A80}} \color[HTML]{F1F1F1} 0.23 & {\cellcolor[HTML]{732081}} \color[HTML]{F1F1F1} 0.22 & {\cellcolor[HTML]{701F81}} \color[HTML]{F1F1F1} 0.22 & {\cellcolor[HTML]{762181}} \color[HTML]{F1F1F1} 0.22 & {\cellcolor[HTML]{732081}} \color[HTML]{F1F1F1} 0.22 \\
LEAR-N(0, $\Sigma$) & {\cellcolor[HTML]{FEA973}} \color[HTML]{000000} 0.06 & {\cellcolor[HTML]{FE9D6C}} \color[HTML]{000000} 0.07 & {\cellcolor[HTML]{FEB078}} \color[HTML]{000000} 0.06 & {\cellcolor[HTML]{FEA973}} \color[HTML]{000000} 0.06 & {\cellcolor[HTML]{FE9D6C}} \color[HTML]{000000} 0.07 & {\cellcolor[HTML]{FEA571}} \color[HTML]{000000} 0.06 & {\cellcolor[HTML]{E34E65}} \color[HTML]{F1F1F1} 0.13 & {\cellcolor[HTML]{DB476A}} \color[HTML]{F1F1F1} 0.14 & {\cellcolor[HTML]{E85362}} \color[HTML]{F1F1F1} 0.12 & {\cellcolor[HTML]{E55064}} \color[HTML]{F1F1F1} 0.12 & {\cellcolor[HTML]{EA5661}} \color[HTML]{F1F1F1} 0.12 & {\cellcolor[HTML]{EE5B5E}} \color[HTML]{F1F1F1} 0.12 & {\cellcolor[HTML]{E85362}} \color[HTML]{F1F1F1} 0.12 & {\cellcolor[HTML]{DE4968}} \color[HTML]{F1F1F1} 0.13 & {\cellcolor[HTML]{F2625D}} \color[HTML]{F1F1F1} 0.11 & {\cellcolor[HTML]{EA5661}} \color[HTML]{F1F1F1} 0.12 & {\cellcolor[HTML]{EC5860}} \color[HTML]{F1F1F1} 0.12 & {\cellcolor[HTML]{EC5860}} \color[HTML]{F1F1F1} 0.12 \\
DLENAR-IND & {\cellcolor[HTML]{FB835F}} \color[HTML]{F1F1F1} 0.09 & {\cellcolor[HTML]{FD9668}} \color[HTML]{000000} 0.07 & {\cellcolor[HTML]{FC8A62}} \color[HTML]{F1F1F1} 0.08 & {\cellcolor[HTML]{FE9D6C}} \color[HTML]{000000} 0.07 & {\cellcolor[HTML]{FC8A62}} \color[HTML]{F1F1F1} 0.08 & {\cellcolor[HTML]{FE9D6C}} \color[HTML]{000000} 0.07 & {\cellcolor[HTML]{D5446D}} \color[HTML]{F1F1F1} 0.14 & {\cellcolor[HTML]{F05F5E}} \color[HTML]{F1F1F1} 0.11 & {\cellcolor[HTML]{D8456C}} \color[HTML]{F1F1F1} 0.14 & {\cellcolor[HTML]{EE5B5E}} \color[HTML]{F1F1F1} 0.11 & {\cellcolor[HTML]{C53C74}} \color[HTML]{F1F1F1} 0.15 & {\cellcolor[HTML]{D8456C}} \color[HTML]{F1F1F1} 0.14 & {\cellcolor[HTML]{D8456C}} \color[HTML]{F1F1F1} 0.14 & {\cellcolor[HTML]{F66C5C}} \color[HTML]{F1F1F1} 0.10 & {\cellcolor[HTML]{D5446D}} \color[HTML]{F1F1F1} 0.14 & {\cellcolor[HTML]{F66C5C}} \color[HTML]{F1F1F1} 0.10 & {\cellcolor[HTML]{CC3F71}} \color[HTML]{F1F1F1} 0.15 & {\cellcolor[HTML]{E85362}} \color[HTML]{F1F1F1} 0.12 \\
DLENAR-DEP & {\cellcolor[HTML]{F97B5D}} \color[HTML]{F1F1F1} 0.09 & {\cellcolor[HTML]{FB835F}} \color[HTML]{F1F1F1} 0.09 & {\cellcolor[HTML]{FB8761}} \color[HTML]{F1F1F1} 0.08 & {\cellcolor[HTML]{FC8E64}} \color[HTML]{000000} 0.08 & {\cellcolor[HTML]{FC8E64}} \color[HTML]{000000} 0.08 & {\cellcolor[HTML]{FE9D6C}} \color[HTML]{000000} 0.07 & {\cellcolor[HTML]{E85362}} \color[HTML]{F1F1F1} 0.12 & {\cellcolor[HTML]{F2625D}} \color[HTML]{F1F1F1} 0.11 & {\cellcolor[HTML]{EE5B5E}} \color[HTML]{F1F1F1} 0.12 & {\cellcolor[HTML]{F66C5C}} \color[HTML]{F1F1F1} 0.10 & {\cellcolor[HTML]{F05F5E}} \color[HTML]{F1F1F1} 0.11 & {\cellcolor[HTML]{F3655C}} \color[HTML]{F1F1F1} 0.11 & {\cellcolor[HTML]{F9785D}} \color[HTML]{F1F1F1} 0.09 & {\cellcolor[HTML]{FB8761}} \color[HTML]{F1F1F1} 0.08 & {\cellcolor[HTML]{FB8761}} \color[HTML]{F1F1F1} 0.08 & {\cellcolor[HTML]{FD9A6A}} \color[HTML]{000000} 0.07 & {\cellcolor[HTML]{FD9A6A}} \color[HTML]{000000} 0.07 & {\cellcolor[HTML]{FEA571}} \color[HTML]{000000} 0.06 \\
DLENAR-DWD & {\cellcolor[HTML]{FA7F5E}} \color[HTML]{F1F1F1} 0.09 & {\cellcolor[HTML]{FB835F}} \color[HTML]{F1F1F1} 0.09 & {\cellcolor[HTML]{FB8761}} \color[HTML]{F1F1F1} 0.08 & {\cellcolor[HTML]{FD9266}} \color[HTML]{000000} 0.08 & {\cellcolor[HTML]{FD9266}} \color[HTML]{000000} 0.08 & {\cellcolor[HTML]{FE9D6C}} \color[HTML]{000000} 0.07 & {\cellcolor[HTML]{E85362}} \color[HTML]{F1F1F1} 0.12 & {\cellcolor[HTML]{F2625D}} \color[HTML]{F1F1F1} 0.11 & {\cellcolor[HTML]{EA5661}} \color[HTML]{F1F1F1} 0.12 & {\cellcolor[HTML]{F8745C}} \color[HTML]{F1F1F1} 0.10 & {\cellcolor[HTML]{EC5860}} \color[HTML]{F1F1F1} 0.12 & {\cellcolor[HTML]{F2625D}} \color[HTML]{F1F1F1} 0.11 & {\cellcolor[HTML]{F9785D}} \color[HTML]{F1F1F1} 0.09 & {\cellcolor[HTML]{FB8761}} \color[HTML]{F1F1F1} 0.08 & {\cellcolor[HTML]{FB835F}} \color[HTML]{F1F1F1} 0.09 & {\cellcolor[HTML]{FD9668}} \color[HTML]{000000} 0.07 & {\cellcolor[HTML]{FE9D6C}} \color[HTML]{000000} 0.07 & {\cellcolor[HTML]{FEA571}} \color[HTML]{000000} 0.06 \\
\bottomrule
\end{tabular}